\documentclass[aps,prx,reprint,superscriptaddress,nofootinbib,longbibliography,amsmath,amssymb]{revtex4-2}

\usepackage{amsthm}
\usepackage{braket}
\usepackage{xcolor,graphicx}
\usepackage{bm,bbm}
\usepackage{booktabs}
\usepackage{enumitem}
\usepackage[colorlinks=true,linkcolor=blue!55!black,citecolor=blue!55!black,
            urlcolor=blue!55!black]{hyperref}
\usepackage{orcidlink} 

\newcommand{\ev}[1]{\langle #1 \rangle}
\newcommand{\norm}[1]{\lVert #1 \rVert}
\newcommand{\abs}[1]{\lvert #1 \rvert}

\newcommand{\Tr}{\operatorname{Tr}}
\newcommand{\tr}{\operatorname{tr}}
\newcommand{\id}{\mathbbm{1}}
\newcommand{\Id}{\mathbb{I}}

\newcommand{\half}{\tfrac{1}{2}}
\newcommand{\eps}{\varepsilon}
\newcommand{\Emap}{\mathcal{E}}

\newcommand{\Svec}{\mathbf{S}}
\newcommand{\Stot}{S_{\mathrm{tot}}}
\newcommand{\mH}{\mathbf{H}}            
\newcommand{\mS}{\mathbf{S}}            
\newcommand{\mV}{\mathbf{V}}            
\newcommand{\mVeff}{\mathbf{V}_{\mathrm{eff}}}
\newcommand{\heff}{H_{\mathrm{eff}}}
\newcommand{\Neel}{N\'eel}
\newcommand{\fold}{\ket{g_L\otimes g_L}}
\newcommand{\pLCU}{p_{\mathrm{LCU}}}
\newcommand{\pQSVT}{p_{\mathrm{QSVT}}}
\DeclareMathOperator{\PREP}{PREPARE}
\DeclareMathOperator{\SEL}{SELECT}

\DeclareMathOperator{\vecop}{vec}
\DeclareMathOperator{\sgn}{sgn}
\newcommand{\bigO}{\mathcal{O}}

\theoremstyle{plain}
\newtheorem{theorem}{Theorem}[section]
\newtheorem{lemma}[theorem]{Lemma}
\newtheorem{proposition}[theorem]{Proposition}
\newtheorem{corollary}[theorem]{Corollary}

\theoremstyle{definition}
\newtheorem{assumption}[theorem]{Assumption}
\newtheorem{definition}[theorem]{Definition}

\theoremstyle{remark}
\newtheorem{remark}[theorem]{Remark}

\begin{document}

\title{Quantum Folding in Chebyshev-Based Quantum Subspace Diagonalization}

\author{Seung Park\,\orcidlink{0009-0007-8520-4123}}
\affiliation{Institute for Convergence Research and Education in Advanced Technology, Yonsei University, Seoul 03722,
Republic of Korea}
\affiliation{Department of Quantum Information, Graduate School, Yonsei University, Seoul 03722, Republic of Korea}

\author{Dongni Chen}
\affiliation{Research Center for Advanced Optics and Photoelectronics, Department of Physics, College of Science, Shantou University, Shantou 515063, China}

\author{Stefano Chesi}
\affiliation{Beijing Computational Science Research Center, Beijing 100193, People's Republic of China}
\affiliation{Department of Physics, Beijing Normal University, Beijing 100875, People's Republic of China}

\author{Kyunghyun Baek}
\email{k.baek@yonsei.ac.kr}
\affiliation{Institute for Convergence Research and Education in Advanced Technology, Yonsei University, Seoul 03722,
Republic of Korea}
\affiliation{Department of Quantum Information, Graduate School, Yonsei University, Seoul 03722, Republic of Korea}

\author{Mahn-Soo Choi\,\orcidlink{0000-0001-6901-5233}}
\email{choims@korea.ac.kr}
\affiliation{Department of Physics, Korea University, Seoul 02841, Republic of Korea}

\date{\today}

\begin{abstract}
Quantum subspace diagonalization (QSD) with a Krylov space spanned by Chebyshev
polynomials of the Hamiltonian is a natural ground-state method for
fault-tolerant quantum computers, but every guarantee it offers---convergence
rate, noise floor, and the cost of turning its classical output back into a
physical state---is controlled by the overlap of the reference state with the
true ground state, which decays exponentially with system size for any product
state and nontrivial Hamiltonian. We propose \emph{quantum folding}: for a system governed by a local (i.e., finite-range)
Hamiltonian, seed the QSD with the product of the ground states of its
subsystems, which differs from the target only through the interface terms.
We prove a nonasymptotic error bound for the Chebyshev QSD that is linear in
the measurement noise, by transplanting Kirby's effective-Hamiltonian analysis
to the Chebyshev--Krylov basis, and identify the block-encoding rescaling of the
Hamiltonian as a parameter of the guarantee. Exact simulations of
antiferromagnetic Heisenberg chains show that the folded reference state
reduces the Krylov dimension for a given accuracy severalfold and lowers the
error under noise by two to five orders of magnitude relative to the \Neel{}
state. We show how the Ritz vector is prepared as a physical state and that
the ground-state generation probability obeys a tight ceiling set by the overlap, so
that recursive folding prepares ground states at polynomial cost. Finally, we
compute the fold fidelity itself: a slow power law for the gapless
spin-$\half$ chain, a size-independent constant for the gapped spin-$1$
chain, and an exact closed form converging to $\tfrac14$ for the AKLT chain.
These results establish a scalable QSD framework that avoids the
exponentially small ground-state overlaps of conventional product-state
initialization.
\end{abstract}

\maketitle

\section{Introduction}
\label{sec:intro}

Quantum subspace diagonalization (QSD), also known as the quantum Krylov
method, estimates the ground-state energy of a Hamiltonian $H$ by generating a
low-dimensional Krylov space on a quantum computer, measuring the projection of
$H$ onto it, and solving the resulting generalized eigenvalue problem
classically \cite{Motta24a,Yoshioka25a}. The approach needs no variational optimization
and no phase estimation, and its error analysis is by now well developed:
Epperly, Lin, and Nakatsukasa \cite{Epperly22a} gave a rigorous treatment of
the real-time-evolution variant, including the thresholding that regularizes
the ill-conditioned overlap matrix; Kirby, Motta, and Mezzacapo
\cite{Kirby23a} introduced the variant studied here, in which the Krylov space
is spanned by Chebyshev polynomials $T_k(H)\ket{\psi_0}$ of the Hamiltonian
and is realized exactly---without Trotter or time-discretization error---by
the qubitization walk operator of a block encoding of $H$ \cite{Low19a}; and
Kirby \cite{Kirby24a} developed an effective-Hamiltonian technique that yields
error bounds linear in the measurement noise for the real-time variant.

All of these guarantees share one dependence: they degrade as the inverse of
the overlap $\abs{\gamma_0}^2=\abs{\braket{g|\psi_0}}^2$ between the reference
state $\ket{\psi_0}$ and the true ground state $\ket{g}$. The overlap sets the
prefactor of the exponential convergence in the Krylov dimension, it sets the
noise floor, and---less often remarked---it sets the scale of the ground-state
signal in the measured data, below which no thresholding can recover it. For
any fixed product state, such as the \Neel{} state of an antiferromagnet, the
overlap with a correlated ground state decays exponentially with the system
size, and with it the method becomes impractical. The choice of reference state
is therefore not a detail of the implementation but the quantity that decides
whether QSD scales.

This article proposes and analyzes a reference state that does not suffer
from this decay. For a system governed by a \emph{local} (i.e., finite-range) Hamiltonian, partitioned
into subsystems $A$ and $B$, the \emph{folded state} $\ket{g_A}\otimes\ket{g_B}$
is the product of the ground states of the parts. It is the exact ground state
of the full Hamiltonian with the interface terms removed, so its overlap with
the true ground state is governed by the physics of the interface rather than
by the bulk: for a one-dimensional chain cut into halves the interface is a
single bond, and the overlap is size-independent for gapped chains and decays
only as a slow power law for gapless ones. \emph{Quantum folding} is the use of
this state as the QSD reference, and, recursively, the ladder
$L\to2L\to4L\to\cdots$ in which the QSD output at each size is turned into a
physical state and used to seed the next. This recursive construction allows for a scalable QSD framework that circumvents the exponentially small ground-state overlaps for fixed product-state initializations. We develop the idea in four steps.

\begin{enumerate}
\item \textbf{Error bound} (Section~\ref{sec:ChebyshevQSD},
  Appendix~\ref{app:ChebyshevQSD}). We carry out the extension proposed in the
  conclusion of \cite{Kirby24a}: the effective-Hamiltonian analysis, which is
  agnostic to the Krylov basis, is transplanted to the Chebyshev--Krylov space.
  The result is an upper bound on the signed ground-state energy error that is
  linear in the spectral-norm noise of the projected matrix pair and converges
  exponentially in the Krylov dimension $D$. The one new ingredient is a
  Chebyshev low-energy filter whose coefficient vector has $\bigO(1)$
  Euclidean norm in the Chebyshev--Krylov basis, by Chebyshev--Parseval orthogonality;
  the bound holds for any rescaling of the Hamiltonian, and when the rescaling
  aligns the ground state with the edge of the Chebyshev interval the
  convergence exponent gains the square root of the gap familiar from the
  Lanczos analysis \cite{Saad80a}.
\item \textbf{Folded reference state} (Section~\ref{sec:FoldedQSD}). Exact
  simulations of open antiferromagnetic Heisenberg chains ($S=\tfrac12$ up to
  $2L=24$ sites, $S=1$ up to $2L=12$) show that the folded state reduces the
  Krylov dimension needed for a given accuracy severalfold and lowers the
  error under measurement noise by two to five orders of magnitude relative to
  the \Neel{} state, with the advantage growing with system size. The same
  simulations expose a technical point that affects any Chebyshev-based QSD:
  the block-encoding normalization of the Hamiltonian decides whether the
  ground state sits at the edge or in the interior of the Chebyshev interval,
  and the two placements differ by orders of magnitude in resolving power
  (Section~\ref{sec:fold-normalization}, Appendix~\ref{app:RescalingFactor}).
\item \textbf{State preparation} (Section~\ref{sec:StatePreparation}). The
  Ritz vector returned by QSD is a classical coefficient vector in a
  nonorthogonal basis. We show that it is prepared as a physical state by a
  linear combination of walk-operator powers or by the quantum singular value
  transformation, at the depth of the deepest QSD measurement circuit; that
  the thresholding which stabilizes the energy also floors the postselection
  probability; that the ground-state generation probability obeys the tight ceiling
  $p\,F\le\abs{\gamma_0}^2$, which a certified filter saturates; and that the
  folding ladder consequently prepares $\ket{g_L}$ at a cost polynomial in
  $L$, with an exponent set by the asymptotic fold overlap.
\item \textbf{Fold fidelity} (Section~\ref{sec:Fidelity},
  Appendix~\ref{app:AKLT}). Everything above rests on the overlap
  $F(L)=\abs{\braket{g_{2L}|g_L\otimes g_L}}^2$, which we determine by using the density-matrix
renormalization group (DMRG) and,
  for the Affleck-Kennedy-Lieb-Tasaki (AKLT) chain, in closed form. Its behavior is dictated by the Anderson
  orthogonality catastrophe \cite{Anderson67a} in the ``cut-and-glue''
  geometry of Dubail and St\'ephan \cite{Dubail11a}: a slow power law
  $L^{-0.107}$ for the gapless spin-$\half$ chain, saturation at
  $F_\infty\simeq0.22$ for the gapped spin-$1$ Heisenberg chain, and
  $F(L)=\bigl(1+3(-1/3)^L\bigr)^2/\bigl(4(1+3\cdot9^{-L})\bigr)\to\tfrac14$
  exactly for the AKLT chain \cite{Affleck87a}. These results explain why
  folding is effective and quantify the resulting advantage.
\end{enumerate}

The idea of solving a large quantum problem through smaller ones has been
pursued in several distinct forms, and it is useful to delineate how quantum
folding differs from them. The methods commonly referred to as ``quantum divide
and conquer'' \cite{Bravyi16a,Peng20a,Ayral21a} decompose a \emph{quantum
circuit} into fragments that fit on a smaller device, cutting wires or
gates and reconstructing the output of the full circuit from the outputs of the
fragments at a classical cost exponential in the number of cuts; the object
being divided is the circuit, and the target is the simulation of a fixed
computation with fewer qubits. 

Hadfield and Papageorgiou \cite{Hadfield18a}
apply a divide-and-conquer strategy to \emph{Hamiltonian simulation}, splitting
the Hamiltonian into groups of terms, simulating each group with the algorithm
best suited to it, and recombining the partial evolutions through splitting
formulas. Closer in spirit to the present work, the deep variational quantum
eigensolver of Fujii \emph{et al.}\ \cite{Fujii22a} divides a lattice into
subsystems, solves each by VQE, and then constructs an effective Hamiltonian
for the inter-subsystem couplings in a reduced basis built from the subsystem
solutions, which is solved by a further VQE; recursive versions of the quantum
approximate optimization algorithm \cite{Bravyi20a} likewise reduce a problem
step by step by eliminating variables on the basis of the solution of a smaller
instance. 

Quantum folding shares with the last two the use of subsystem ground
states as building blocks, but it differs in what is done with them: the
subsystem ground states are not used to truncate the Hilbert space or to define
an effective model, but are combined into a single product state that serves
as the \emph{reference state} of an exact, nonvariational method (QSD) on the
full system, whose error bounds are then controlled by the overlap of that
state with the true ground state. The approximation is therefore confined to
the Krylov projection and to the measurement noise, both of which are
quantified by the bounds of Section~\ref{sec:ChebyshevQSD}, and the interface
between the subsystems is treated exactly rather than perturbatively or in a
reduced basis.

The construction applies to any Hamiltonian with local structure. Its
advantage is naturally largest when the interface is small compared with the
subsystems it joins, as in weakly coupled spin clusters, molecular aggregates,
and other assemblies of strongly correlated units, where the inter-subsystem
interactions modify the ground state without reorganizing its local structure.
Weak coupling is not required, however: the uniformly coupled spin chains
studied below are the more demanding case, in which the bond across the cut is
as strong as any other, and the overlap is nevertheless controlled by the
interface because the interface consists of a single bond.

Throughout we use spin chains as the test bed, but nothing in the construction
is specific to one dimension, to two parts, or to spins
(Section~\ref{sec:fold-general}). Most of the numerical results were obtained
with a Wolfram Language package built on the Q3 framework \cite{Choi22a}, which
exposes the normalization choice discussed in
Section~\ref{sec:fold-normalization} as an option; the fidelity data of
Section~\ref{sec:Fidelity} were obtained with DMRG \cite{Hauschild18a}.

\section{Chebyshev Polynomial-Based Quantum Subspace Diagonalization}
\label{sec:ChebyshevQSD}

Quantum Krylov, or quantum subspace diagonalization (QSD), methods estimate the
ground-state energy $E_0$ of a Hamiltonian $H$ by projecting $H$ onto a
low-dimensional subspace generated on a quantum device and solving the resulting
generalized eigenvalue problem classically; see \cite{Motta24a} for a review.
Epperly, Lin, and Nakatsukasa \cite{Epperly22a} gave the first rigorous analysis
of the real-time-evolution variant, with a noise term scaling sublinearly, as
$\eta^{1/(1+\alpha)}$ with $\alpha\in[0,\tfrac12]$, in the noise rate $\eta$.
Kirby, Motta, and Mezzacapo \cite{Kirby23a} introduced the Krylov space
generated by Chebyshev polynomials of the Hamiltonian, 
which hereafter we call the \emph{Chebyshev--Krylov space},
constructed exactly by
qubitization \cite{Low19a}, and adapted the analysis of \cite{Epperly22a} to it.
Kirby \cite{Kirby24a} then developed an \emph{effective-Hamiltonian} technique
for the real-time QSD that yields an error bound linear in the noise, and
in his conclusion left the construction ``for polynomials rather
than complex exponentials'' as an open extension. 
This section summarizes the outcome of
carrying out that extension for the Chebyshev--Krylov space: a bound with the linear noise scaling
of \cite{Kirby24a} and the square-root-of-gap convergence exponent of the
classical Chebyshev/Lanczos analysis \cite{Saad80a}. Complete statements and
proofs are given in Appendix~\ref{app:ChebyshevQSD}; here we set up the
notation used throughout the article and state the results in the form needed
in Sections~\ref{sec:FoldedQSD}--\ref{sec:Fidelity}.

\subsection{The Chebyshev--Krylov space}
\label{sec:qsd-setup}

Let $H$ be a Hermitian operator on an $N$-dimensional Hilbert space with
eigenpairs $(E_k,\ket{E_k})$, $E_0\le E_1\le\cdots\le E_{N-1}$, and spectral gap
$\Delta:=E_1-E_0$. Let
\begin{equation}
  \ket{\psi_0}=\sum_{k=0}^{N-1}\gamma_k\ket{E_k},\qquad \abs{\gamma_0}^2>0,
  \label{eq:qsd-psi0}
\end{equation}
be a normalized reference state; $\abs{\gamma_0}^2=\abs{\braket{E_0|\psi_0}}^2$
is the \emph{initial overlap}. Given real numbers $E_{\min}'\le E_0$ and
$E_{\max}'\ge E_{N-1}$, set the estimated spectral width $\Omega:=E_{\max}'-E_{\min}'$ and rescale the Hamiltonian by it,
\begin{equation}
  h:=\frac{2H-(E_{\min}'+E_{\max}')\,\id}{\Omega},
  \label{eq:qsd-rescale}
\end{equation}
where $\id$ denotes the identity operator,
so that the spectrum of $h$ lies in $[-1,1]$,
\begin{equation}
  x_k:=\frac{2E_k-(E_{\min}'+E_{\max}')}{\Omega}\in[-1,1].
  \label{eq:qsd-rescale-spectrum}
\end{equation}
The ground eigenvalue sits at
$x_0=-1+\nu$ with \emph{edge slack} $\nu:=2(E_0-E_{\min}')/\Omega\in[0,2)$, which
measures how tightly the \textit{a priori} lower bound $E_{\min}'$ locates $E_0$. The
most common choice in practice is the symmetric window
$E_{\max}'=-E_{\min}'=\lambda$, i.e.\ $h=H/\lambda$, with $\lambda$ the
subnormalization of a block encoding of $H$ (e.g., for $H=\sum_i c_iP_i$ a linear
combination of Pauli operators, $\lambda=\sum_i\abs{c_i}$); in that case $x_0$
generally lies in the \emph{interior} of $[-1,1]$, and the distinction between
edge and interior placement of $x_0$ will recur in
Sections~\ref{sec:qsd-filter} and~\ref{sec:prep-normalization}.

The Chebyshev--Krylov basis of dimension $D$ is
\begin{equation}
  \ket{v_j}:=T_j(h)\ket{\psi_0},\qquad j=0,\dots,D-1,
  \label{eq:qsd-basis}
\end{equation}
with $T_j$ the Chebyshev polynomial of the first kind. Since
$\{T_0,\dots,T_{D-1}\}$ spans the polynomials of degree $\le D-1$, the span of
\eqref{eq:qsd-basis} is the classical Krylov space
\begin{equation}
  \mathcal K_D(H,\ket{\psi_0}) := \operatorname{span}\{H^j\ket{\psi_0}\}_{j<D}
\end{equation} 
for
\emph{any} affine rescaling: the method is Lanczos in a nonorthogonal basis
\cite{Kirby23a}, and its noiseless Ritz values do not depend on
$(E_{\min}',E_{\max}')$. Writing $\mV:=[\ket{v_0}\cdots\ket{v_{D-1}}]$, the
projected matrix pair is
\begin{subequations}\label{eq:qsd-pair}
\begin{align}
  \mH&:=\mV^\dagger H\mV, \\
  \mS&:=\mV^\dagger\mV,
\end{align}
\end{subequations}
and the ground-state energy estimate $\widetilde E_0$ is the least eigenvalue
of the (regularized) generalized eigenvalue problem $\mH c=\widetilde E\,\mS c$,
whose eigenvalues $\widetilde E_0\le\widetilde E_1\le\cdots$ are the Ritz
values. It is useful to expand each Chebyshev--Krylov basis vector in the eigenbasis of $H$, which gives the factorization
\begin{equation}
  \mV=\Psi\,\Gamma\,\Phi,
  \label{eq:qsd-factorization}
\end{equation}
where $\Psi:=[\ket{E_0}\cdots\ket{E_{N-1}}]$ is the matrix of eigenvectors of
$H$, $\Gamma:=\operatorname{diag}(\gamma_k)$, and $\Phi_{kj}:=T_j(x_k)$,
so that $\mS=\Phi^\dagger\Gamma^\dagger\Gamma\Phi$; the Chebyshev--Vandermonde
matrix $\Phi$ plays the role of the phase matrix of the real-time QSD.
Because $\mS$ is a Gram matrix, the squared norm of any vector
$\sum_jc_j\ket{v_j}=\mV c$ in the Krylov space is the quadratic form
$c^\dagger\mS c$; this identity is used repeatedly in
Section~\ref{sec:StatePreparation}.

\subsection{Quantum implementation: block encoding, walk operator, moments}
\label{sec:qsd-implementation}

Following \cite{Kirby23a}, assume a block encoding $(U,G)$ of the rescaled Hamiltonian $h$: 
a unitary $U$ on auxiliary and system registers with $U^2=\id$ and a state
$\ket{G}_a=G\ket{0}_a$ such that
$(\bra{G}_a\otimes\id_s)\,U\,(\ket{G}_a\otimes\id_s)=h$. With the reflection
$R:=(2\ket{G}_a\!\bra{G}_a-\id_a)\otimes\id_s$, the walk operator
$W:=RU$ satisfies the qubitization identity \cite{Low19a,Kirby23a}
\begin{equation}
  (\bra{G}_a\otimes\id_s)\,W^k\,(\ket{G}_a\otimes\id_s)=T_k(h),
  \qquad k=0,1,2,\dots,
  \label{eq:qsd-walk}
\end{equation}
so $W^k$ is a block encoding of $T_k(h)$, exactly and without Trotter or
time-discretization error. The product formula
$T_jT_k=\tfrac12(T_{j+k}+T_{\abs{j-k}})$ and the recurrence
$hT_k=\tfrac12(T_{k+1}+T_{\abs{k-1}})$ reduce every matrix element of the pair
\eqref{eq:qsd-pair} to the \emph{Chebyshev moments}
\begin{equation}
  \mu_\ell:=\bra{\psi_0}T_\ell(h)\ket{\psi_0},\qquad \ell=0,\dots,2D-1,
  \label{eq:qsd-moments}
\end{equation}
namely $\mS_{jk}=\tfrac12(\mu_{j+k}+\mu_{\abs{j-k}})$ and
\begin{multline}
  \mH_{jk}=\frac{\Omega}{8}\Bigl(\mu_{j+k+1}+\mu_{\abs{j+k-1}}
   +\mu_{\abs{j-k}+1}+\mu_{\abs{\,\abs{j-k}-1\,}}\Bigr) \\
   +\frac{E_{\min}'+E_{\max}'}{2}\,\mS_{jk},
  \label{eq:qsd-Hmoments}
\end{multline}
which for the symmetric window reduces to
$\mH_{jk}=\tfrac{\lambda}{4}(\mu_{j+k+1}+\mu_{\abs{j+k-1}}+\mu_{\abs{j-k+1}}
+\mu_{\abs{j-k-1}})$. Thus $2D$ Chebyshev moments ($\mu_0=1$ needs no
measurement) determine the entire $D\times D$ pair---a Toeplitz-plus-Hankel
structure that is the analogue of the Toeplitz economy of the real-time method.
On the device, $\mu_\ell$ is estimated by preparing
$W^{\lfloor\ell/2\rfloor}(\ket{G}_a\otimes\ket{\psi_0})$ and measuring the
expectation value of $R$ (even $\ell$) or of $U$ (odd $\ell$): the
unitarity of $W$ lets the bra and the ket each absorb half of the polynomial,
so the deepest circuit contains only $\lfloor(2D-1)/2\rfloor=D-1$ walk steps
\cite{Kirby23a}. The maximal number of sequential walk-operator applications therefore grows linearly with the Krylov dimension $D$.

\subsection{Noise and regularization}
\label{sec:qsd-noise}

Let $(\mH',\mS')$ be Hermitian estimates of $(\mH,\mS)$ with spectral-norm
errors
\begin{subequations}\label{eq:qsd-noise}
\begin{align}
  \eta_H&:=\norm{\mH'-\mH}, \\
  \eta_S&:=\norm{\mS'-\mS},
\end{align}
\end{subequations}
of any origin (finite sampling, hardware noise), and let
\begin{equation}
  \chi:=\eta_H+\eta_S\norm{H}
  \label{eq:qsd-chi}
\end{equation}
denote the combined noise strength. Since $\mS$ becomes
exponentially ill-conditioned as $D$ grows, the generalized eigenvalue problem
is regularized by \emph{thresholding} \cite{Epperly22a,Kirby23a,Kirby24a}: with
$\Pi'$ the orthogonal projector onto the eigenspaces of $\mS'$ with eigenvalues
exceeding a threshold $\epsilon>0$, one solves the pair
$(\mH'',\mS''):=(\Pi'\mH'\Pi',\Pi'\mS'\Pi')$ on the range of $\Pi'$, and
$\widetilde E_0$ denotes its least eigenvalue. If the moments are each measured
to accuracy $\varsigma$, entrywise bounds convert to $\eta_S\le D\varsigma$ and
$\eta_H\le D\bar\lambda\varsigma$, with 
\begin{equation}
\bar\lambda:=\max(E_{\max}',-E_{\min}')\ge\norm{H};
\end{equation}
the natural threshold is then $\epsilon=\Theta(D\varsigma)$. In the numerical work of
Sections~\ref{sec:FoldedQSD} and~\ref{sec:StatePreparation} the moment noise is
modeled as independent Gaussian perturbations of width $\sigma$ (respectively
$\eta$) on every $\mu_\ell$, $\ell\ge1$, from which \emph{both} $\mH'$ and
$\mS'$ are assembled; the resulting matrix noise is strongly correlated, living
in the $2D$-dimensional space of moment perturbations rather than in the
$\bigO(D^2)$-dimensional space of independent matrix perturbations. This
shared-moment structure is unique to the Chebyshev (single-reference-state)
method and has consequences for state preparation
(Section~\ref{sec:prep-numerics}).

\subsection{Chebyshev low-energy filters}
\label{sec:qsd-filter}

The effective-Hamiltonian technique of \cite{Kirby24a} models generic noise in
$(\mH,\mS)$ \emph{exactly}, at the one point of the subspace that matters, as an
unperturbed Rayleigh--Ritz problem for an effective Hamiltonian $\heff$ in an
effective Krylov space (Appendix~\ref{app:effective}). The technique is agnostic
to the Krylov basis; the only basis-specific ingredient is the construction of
an approximate ground state \emph{inside} the subspace, i.e.\ a polynomial
filter $q=\sum_{j<D}c_jT_j$ of degree at most $D-1$ with $q(x_0)=1$ that is
small, say $\le\varepsilon_D$, on the spectrum above an energy window $\delta$
and---crucially for the noise analysis---has a coefficient vector of
controlled Euclidean norm,
\begin{equation}
  \norm{c}_2^2\le 2B^2,
  \label{eq:qsd-parseval}
\end{equation}
where $B:=\max_{x\in[-1,1]}\abs{q(x)}$; this bound follows for \emph{any} such polynomial from Chebyshev--Parseval
orthogonality (Appendix~\ref{app:filter}). 
The magnitude of $B$ controls how strongly noise and thresholding errors are amplified in the energy bound (Appendix~\ref{app:main}). Thus a small suppression factor $\varepsilon_D$ alone is not sufficient; the magnitude of $B$ must be controlled as well. Property \eqref{eq:qsd-parseval} is
the polynomial counterpart of the Fourier--Parseval bound $\norm{c}\le1$ for
the trigonometric filter of \cite{Epperly22a,Kirby24a}, and it is special to
the Chebyshev--Krylov basis: the same filter state $q(h)\ket{\psi_0}$
expressed in the monomial Krylov basis $h^j\ket{\psi_0}$ has exponentially
large coefficients, one linear-algebraic face of the ill-conditioning of the
power basis. The remaining question is how large the
filter becomes on the spectrally \emph{empty} part of $[-1,1]$, and two filters
answer it differently.

The \emph{two-sided} filter $q_{\mathrm{2s}}(x)=p^*\bigl((x-x_0)^2/4\bigr)$, built
from the minimax residual polynomial $p^*$ of degree $\lfloor(D-1)/2\rfloor$
\cite{Kirby23a}, suppresses the spectrum on both sides of $x_0$ and is
therefore bounded by $1$ on all of $[-1,1]$ wherever $x_0$ sits: $B=1$
unconditionally, at the cost of spending two Krylov vectors per unit of degree
and of a suppression $\varepsilon_D\le2(1+\delta/\Omega)^{-\lfloor(D-1)/2\rfloor}$
that is linear in the gap (Lemma~\ref{lem:twosided}; the closed forms quoted
here are upper bounds that become informative once $D$ exceeds the
onset of exponential convergence, $D\gtrsim\Omega/\delta$ or
$\sqrt{\Omega/\delta}$ respectively). The \emph{one-sided}
filter $q_{\mathrm{1s}}(x)=T_{D-1}(f(x))/T_{D-1}(f(x_0))$, with $f$ the affine map
taking $[x_0+2\delta/\Omega,1]$ onto $[-1,1]$, uses the full degree on the one
side where the spectrum lies and achieves
$\varepsilon_D\le2(1+2\sqrt{\delta/\Omega})^{-(D-1)}$, the square-root
acceleration of Chebyshev filters---but it is uncontrolled on the spectrally
empty interval $[-1,x_0)$, and its supremum there governs $B$
(Lemma~\ref{lem:onesided}). If $\nu=0$ (tight lower edge,
Assumption~\ref{ass:edge}) then $B=1$; more generally $B=\bigO(1)$ as long as
$E_0-E_{\min}'\lesssim\sqrt{\delta\Omega}/D$ (Remark~\ref{rem:slack}), whereas
with the slack $\nu=\Theta(1)$ of a block-encoding normalization fixed in
advance (Section~\ref{sec:fold-normalization}) $B$ grows exponentially in $D$
and the one-sided filter is not applicable. The real-time filter needs no such
distinction because complex exponentials are unimodular on the whole real
line; this is the essential structural difference between the two settings,
and it resides entirely in the conditioning of the representation, not in the
subspace. Edge alignment improves exactly the convergence exponent, from linear
to square-root dependence on the gap, and nothing else.

\subsection{Energy error bounds}
\label{sec:qsd-bounds}

The main result (Theorem~\ref{thm:general}) holds for any rescaling of the
Hamiltonian. Define
\begin{subequations}\label{eq:qsd-constants}
\begin{align}
  \abs{\gamma_0'}^2&:=\abs{\gamma_0}^2-4(\epsilon+\eta_S), \\
  \Delta'&:=\Delta-\frac{2\chi}{\abs{\gamma_0'}^2}, \\
  \zeta&:=4D(\epsilon+\eta_S).
\end{align}
\end{subequations}
Then, for any $0<\delta<\delta'<\norm{H}$ such that
$2\chi/\abs{\gamma_0'}^2<\delta'-\delta$, $\epsilon\ge\eta_S$, and
$\abs{\gamma_0'}^2>0$,
\begin{multline}
  \widetilde E_0-E_0\ \le\ 
  \delta'\,\mathbbm{1}(\delta'>\Delta')
  +\frac{2\chi}{\abs{\gamma_0'}^2} \\
  +\frac{6\norm{H}}{\abs{\gamma_0'}^2}\biggl(
    \frac{2\chi}{\delta'-\delta}+\zeta \\
    +8\Bigl(1+\frac{\delta}{\Omega}\Bigr)^{-2\lfloor(D-1)/2\rfloor}\biggr).
  \label{eq:qsd-main}
\end{multline}
Under the tight-lower-edge Assumption~\ref{ass:edge} the one-sided filter
replaces the convergence term by $8(1+2\sqrt{\delta/\Omega})^{-2(D-1)}$
(Theorem~\ref{thm:edge}); with a finite slack the same theorem holds with the
constants $2\chi$, $4(\epsilon+\eta_S)$, $4D(\epsilon+\eta_S)$ multiplied by
$B^2$. In the corollaries below, $\rho_D$ denotes the convergence factor at
the window $\delta=\Delta/2$, namely $(1+\Delta/(2\Omega))^{-2\lfloor(D-1)/2\rfloor}$
in general and $(1+\sqrt{2\Delta/\Omega})^{-2(D-1)}$ when edge-aligned. Three
specializations are of particular interest:

\begin{itemize}
\item \emph{Noiseless} ($\eta_H=\eta_S=0$, $\epsilon\to0^+$; $\delta=\Delta/2$,
  $\delta'\uparrow\Delta$):
  \begin{equation}
    0\le\widetilde E_0-E_0\le
    \frac{48\norm{H}}{\abs{\gamma_0}^2}\,\rho_D,
    \label{eq:qsd-noiseless}
  \end{equation}
  the analogue of the Kaniel--Paige--Saad bound for Lanczos \cite{Saad80a},
  with the classical linear-in-gap exponent in general and the square-root
  exponent when edge-aligned. That the two differ even without noise is not in
  conflict with the normalization-independence of the exact Ritz values
  (Section~\ref{sec:qsd-setup}): both are upper bounds, and the one-sided
  filter provides a tighter bound when it is available.

\item \emph{Simplified noisy bound} (Corollary~\ref{cor:asymptotic};
  $\epsilon=\Theta(\eta_S)$, $\delta=\Delta/2$, $\delta'=3\Delta/4$, admissible
  whenever $8\chi<\Delta\abs{\gamma_0'}^2$, which guarantees $\delta'\le\Delta'$),
  with the dimensionless noise rate $\eta:=\max(\eta_S,\eta_H/\norm{H})$:
  \begin{equation}
    \widetilde E_0-E_0\le
    \bigO\!\left(\frac{\norm{H}}{\abs{\gamma_0'}^2}\left[
      \Bigl(\frac{\norm{H}}{\Delta}+D\Bigr)\eta
      +\rho_D\right]\right).
    \label{eq:qsd-simplified}
  \end{equation}
\item \emph{Machine-level bound} (Corollary~\ref{cor:machine}): if every moment
  is estimated to accuracy $\varsigma$ (e.g.\ $\varsigma=\bigO(N_s^{-1/2})$ for $N_s$ shots per
  moment) and $\epsilon=D\varsigma$, then
  \begin{equation}
    \frac{\widetilde E_0-E_0}{\bar\lambda}\le
    \bigO\!\left(\frac{1}{\abs{\gamma_0''}^2}\left[
      \Bigl(\frac{\bar\lambda}{\Delta}+D\Bigr)D\varsigma
      +\rho_D\right]\right)
    \label{eq:qsd-machine}
  \end{equation}
  with 
  \begin{math}
    \abs{\gamma_0''}^2:=\abs{\gamma_0}^2-8D\varsigma.
  \end{math}
  Both sides are dimensionless, energies being measured in the unit $\bar\lambda$
  at which the walk operator represents $H$; $\varsigma$ is the quantity directly
  tied to device characteristics, which makes \eqref{eq:qsd-machine} the form
  most convenient for resource estimation.
\end{itemize}

Compared with the adaptation of \cite{Epperly22a} to the Chebyshev--Krylov space in
\cite[Thm.~1]{Kirby23a}, which is built on the same two-sided filter and gives
a noise term $\bigO((D^4\eta)^{1/(1+\alpha)})$ with the same linear-in-gap
convergence term, the bound \eqref{eq:qsd-main} improves the noise scaling to
$\bigO(D\eta)$---linear. Compared with the result for the real-time QSD
\cite{Kirby24a}, \eqref{eq:qsd-main} is its exact analogue in the Chebyshev
space, with the same noise scaling and the same linear-in-gap exponent (the
real-time angle $E\,dt$ is linear in the energy). Edge alignment
(Theorem~\ref{thm:edge}) then gains the square root, at the cost of the
block-encoding access model and of the tight-lower-edge assumption. To reach a suppression $e^{-L}$ of the
excited-state contamination, the real-time analysis needs $D\sim LR/(\pi\delta)$ with the exact spectral range $R$,
whereas \eqref{eq:qsd-main} needs $D\sim\tfrac14L\sqrt{\Omega/\delta}$, a quadratic
improvement in $\Omega/\delta$ in the small-gap regime where these methods are most
stressed. Relative to \cite{Kirby23a} this is an improvement of
\emph{analysis}, not of algorithm: the circuits and the classical
post-processing are unchanged. A finer accounting of the three separable
ingredients (base of the exponential, exponent per matrix dimension, depth
accounting) is given in Appendix~\ref{app:discussion}.

\subsection{The role of the initial overlap}
\label{sec:qsd-overlap}

Every term of \eqref{eq:qsd-main}--\eqref{eq:qsd-machine} carries the prefactor
$1/\abs{\gamma_0'}^2\approx1/\abs{\gamma_0}^2$, and the bounds are void unless
\begin{equation}
  \abs{\gamma_0}^2\ >\ 4B^2(\epsilon+\eta_S)\ \sim\ 8D\varsigma ,
  \label{eq:qsd-signal}
\end{equation}
i.e.\ unless the ground-state signal in the moments exceeds the noise floor of
the thresholded pair. The reference state thus controls the method in three
ways at once: the prefactor of the exponential convergence, the noise floor, and
the very resolvability of the ground state in the measured data. A reference
state whose overlap decays exponentially with system size---as does that of any
fixed product state such as the \Neel{} state---forces an exponential shot
budget through \eqref{eq:qsd-signal} long before the Krylov dimension becomes
an issue. Section~\ref{sec:FoldedQSD} introduces the folded reference state,
whose overlap stays $\bigO(1)$ (or decays only as a slow power law) with system
size, and Section~\ref{sec:Fidelity} explains why.

\section{Quantum Folding in Quantum Subspace Diagonalization}
\label{sec:FoldedQSD}

\subsection{Quantum folding for local Hamiltonians}
\label{sec:fold-general}

Consider a quantum many-body system on a lattice (or, more generally, a graph)
$\Lambda$ governed by a \emph{local} Hamiltonian, $H_\Lambda=\sum_{X\subset\Lambda}
h_X$, where each term $h_X$ is supported on a set $X$ of bounded diameter and
bounded norm. Partition the sites into two disjoint subsystems,
$\Lambda=A\cup B$. Every term is then supported entirely in $A$, entirely in
$B$, or straddles the interface, so that
\begin{equation}
  H_\Lambda=H_A\otimes\id_B+\id_A\otimes H_B+H_{\partial},
  \label{eq:fold-general-split}
\end{equation}
with
\begin{equation}
  H_\partial:=\sum_{X\cap A\ne\emptyset,\ X\cap B\ne\emptyset}h_X ,
  \label{eq:fold-general-split-interface}
\end{equation}
where $H_A$ and $H_B$ are the Hamiltonians of the two subsystems with open
boundaries at the cut and the \emph{interface term} $H_\partial$ collects the
terms that couple them. Let $\ket{g_A}$ and $\ket{g_B}$ denote the ground states
of $H_A$ and $H_B$. The \emph{folded state}
\begin{equation}
  \ket{g_A\otimes g_B}:=\ket{g_A}\otimes\ket{g_B}
  \label{eq:fold-general-folded}
\end{equation}
is the exact ground state of $H_\Lambda-H_\partial$, i.e.\ of the full system
with the interface removed. \emph{Quantum folding} is the use of this state as
the reference state $\ket{\psi_0}$ of the QSD computation on $\Lambda$: the
ground-state problem on $\Lambda$ is seeded by the ground states of its parts,
prepared in parallel (or recursively, by folding the parts themselves;
Section~\ref{sec:prep-ladder}). Nothing in the construction is specific to one
dimension, to two parts, or to spins: with $\Lambda=A_1\cup\cdots\cup A_m$ the
reference is $\bigotimes_j\ket{g_{A_j}}$, and the subsystems may be fermionic
or bosonic modes as long as $H_\Lambda$ is local in the sense above.

The essential property of the construction is that the folded state differs
from an eigenstate of $H_\Lambda$ only by the interface term. Its ground-state overlap
$\abs{\gamma_0}^2=\abs{\braket{g_\Lambda|g_A\otimes g_B}}^2$ is therefore
governed by the physics of the cut rather than by the bulk: heuristically,
$1-\abs{\gamma_0}^2$ is controlled by $\norm{H_\partial}$ measured in units of
the relevant excitation gap, and for gapped systems Lieb--Robinson-type
arguments confine the disturbance to within a correlation length of the
interface (see Section~\ref{sec:fid-criteria}).\footnote{Throughout this work the ground state is assumed to be nondegenerate. A degenerate ground manifold does not affect the ground-state energy returned by a (classical or quantum) Krylov method, but resolving individual ground-state vectors within the manifold requires additional procedures. When the degeneracy is induced by a symmetry, the folded reference state inherits the quantum numbers of its factors and therefore lies in a definite symmetry sector, so that the QSD resolves the lowest state of that sector; if the subsystems themselves have degenerate ground manifolds, as for edge modes, the appropriate member must be selected (sector-resolved folding, Section~\ref{sec:fid-aklt}).}
By contrast, any fixed product
state---the \Neel{} state, a mean-field state, a computational-basis
state---differs from the correlated ground state \emph{everywhere}, and its
overlap decays exponentially in the volume $\abs\Lambda$. The interface of a
bipartition of a $d$-dimensional system of linear size $L$ has
$\abs{\partial}\sim L^{d-1}$ terms, so folding replaces an overlap decaying in
the volume $L^d$ by one decaying at worst in the \emph{area} $L^{d-1}$. In one
dimension the interface is a single bond, $\abs\partial=\bigO(1)$, and the
volume dependence is removed entirely: the overlap is size-independent for gapped chains and
decays only as a slow power law for gapless ones, with an exponent fixed by
boundary scaling dimensions (Section~\ref{sec:Fidelity}). The one-dimensional
case is therefore the one in which the method is most effective and can be
analyzed most precisely, and we use it for all demonstrations in this article.

\subsection{Spin chains as the test bed}
\label{sec:fold-state}

For demonstrations we consider open antiferromagnetic Heisenberg chains of
$2L$ spins of size $S$,
\begin{equation}
  H_{2L}=J\sum_{i=1}^{2L-1}\Svec_i\cdot\Svec_{i+1},\qquad J>0,
  \label{eq:fold-ham}
\end{equation}
together with, in Section~\ref{sec:Fidelity}, the AKLT chain
\cite{Affleck87a,Affleck88a} for integer $S$. We denote by $\ket{g_n}$ the
ground state of the $n$-site chain and cut the chain into two halves of $L$
sites each, $A=\{1,\dots,L\}$ and $B=\{L+1,\dots,2L\}$. The interface term is
the single bond $H_\partial=J\,\Svec_L\cdot\Svec_{L+1}$, so
\eqref{eq:fold-general-split} reads
\begin{equation}
  H_{2L}=H_L\otimes\id+\id\otimes H_L+J\,\Svec_L\cdot\Svec_{L+1},
  \label{eq:fold-split}
\end{equation}
and the folded state is
\begin{equation}
  \fold=\ket{g_L}\otimes\ket{g_L}.
  \label{eq:fold-folded}
\end{equation}
It can be prepared by running the ground-state routine (typically, 
adiabatic or variational) on the half-size problem twice in parallel.
In Section~\ref{sec:StatePreparation} we propose and analyze a general
preparation method that works in tandem with the Chebyshev-based QSD, in
particular recursively, by repeated doubling (Section~\ref{sec:prep-ladder}).
In this section we take it as given and quantify the advantage it provides in the Chebyshev-based
QSD of the full chain, by exact numerical simulation. The conventional
alternative is the \Neel{} state $\ket{S,-S,S,-S,\dots}$.

Recall from Section~\ref{sec:qsd-overlap} that both the prefactor of the
noiseless convergence and the noise floor of QSD scale inversely with the
initial overlap $\abs{\gamma_0}^2=\abs{\braket{g_{2L}|\psi_0}}^2$; in the
schematic form of \eqref{eq:qsd-simplified},
\begin{equation}
  \widetilde E_0-E_0\ \lesssim\
  \underbrace{\frac{\norm{H}}{\abs{\gamma_0}^2}\,
    e^{-cD\kappa}}_{\text{noiseless Rayleigh--Ritz}}
  \;+\;
  \underbrace{\bigO\!\left(\frac{\eta}{\abs{\gamma_0}^2}\right)}_{\text{noise floor}},
  \label{eq:fold-bound}
\end{equation}
with $\kappa=\sqrt{\Delta/\lambda}$ when the rescaling places the ground state
at the edge of the Chebyshev interval (Theorem~\ref{thm:edge}) and
$\kappa=\Delta/\lambda$ when it lies in the interior
(Theorem~\ref{thm:general}); see Section~\ref{sec:fold-normalization}. The
edge-aligned convergence is the one usually quoted for Chebyshev methods, but only the
interior one is available with a block encoding whose normalization is fixed
in advance. In either regime a
reference state with larger ground-state overlap converges faster \emph{and}
tolerates proportionally more noise. Three properties of the folded
state make it the natural handle on $\abs{\gamma_0}^2$.

\begin{figure*}[t]
\centering
\includegraphics[width=0.48\textwidth]{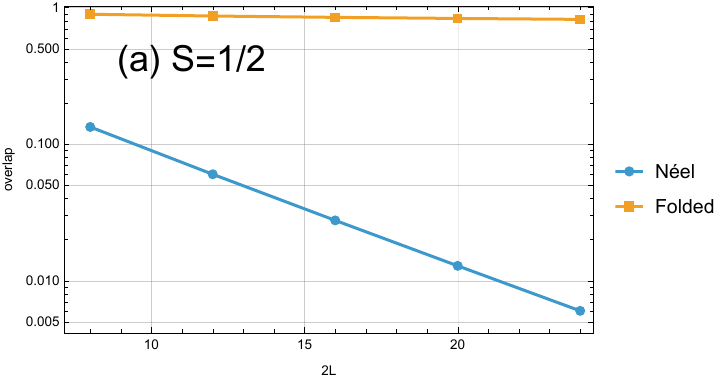}\hfill
\includegraphics[width=0.48\textwidth]{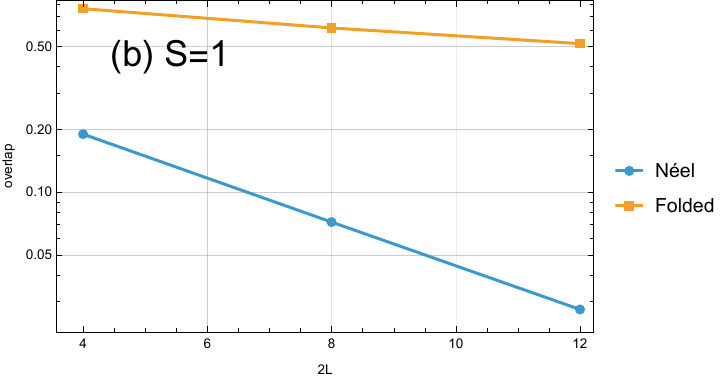}
\caption{Ground-state overlap of the two reference states versus system size.
The \Neel{} overlap decays exponentially, while the folded-state overlap stays
$\bigO(1)$.}
\label{fig:fold-overlap}
\end{figure*}

\paragraph{Size-independent overlap.}
By \eqref{eq:fold-split}, $\fold$ differs from an eigenstate of $H_{2L}$ by a
single-bond perturbation, so its residual weight $1-\abs{\gamma_0}^2$ is
governed by the local physics of the cut rather than accumulating extensively.
For the gapped integer-$S$ chains the overlap converges exponentially to a
nonzero constant; for the gapless spin-$\half$ chain it decays only as a very
slow power law, $\abs{\gamma_0}^2\sim L^{-0.11}$, so that it remains above
$0.5$ up to $L\sim10^3$ and above $0.4$ at $L\sim10^4$
(Section~\ref{sec:Fidelity}). In contrast, the overlap of the
\Neel{} state, or of any fixed product state, decays exponentially,
$\abs{\braket{g_{2L}|\text{\Neel}}}^2\sim e^{-\alpha\cdot2L}$.
Figure~\ref{fig:fold-overlap} quantifies this
for the sizes simulated here; more detailed analysis of the folded state is discussed in Section~\ref{sec:Fidelity}.

\paragraph{Low-energy concentration.}
The energy expectation is
$\ev{H_{2L}}_{\mathrm{fold}}=2\eps_L+J\ev{\Svec_L\cdot\Svec_{L+1}}_{\mathrm{fold}}$,
with $\eps_L$ the half-chain ground energy, so it misses $E_0$ by an
$L$-independent $\bigO(J)$ amount: for $S=\half$, $2L=24$ we find
$\ev{H}_{\mathrm{fold}}-E_0=0.170J$ versus $4.70J$ for the \Neel{} state. The
residual weight of the folded state is concentrated on low-lying excitations,
which accelerates the Rayleigh--Ritz convergence beyond what the overlap alone
suggests.

\paragraph{Exact quantum numbers.}
$\fold$ carries total $S^z=0$ and, being a product of two singlets (for even
$L$), is an \emph{exact total-spin singlet}. For the open $S=1$ chain this
property is essential: the Kennedy edge-state quasi-degeneracy \cite{Kennedy90a} places a
triplet only $0.097J$ above the singlet ground state at $2L=12$, and any triplet
weight in the reference state (as carried by the \Neel{} state) must be
filtered out through this tiny gap. The folded state has strictly zero weight
there, so its convergence is governed by the much larger gap ($0.79J$ at
$2L=12$) to the next state it actually couples to. The same mechanism---the
relevant gap is the gap of the symmetry-resolved spectrum, not of the full
spectrum---reappears in the fidelity analysis of Section~\ref{sec:fid-half}.

\subsection{Numerical setup}
\label{sec:fold-setup}

We simulate the algorithm of Section~\ref{sec:ChebyshevQSD} exactly (i.e.\ with
exact moments \eqref{eq:qsd-moments}, so that hardware noise can be modeled in
a controlled way), working in the conserved $S^z_{\mathrm{tot}}=0$ sector:
dimension $\binom{24}{12}=2{,}704{,}156$ for $S=\half$, $2L=24$, and $73{,}789$
for $S=1$, $2L=12$. The moments are generated by the three-term recursion
$\ket{v_{k+1}}=2h\ket{v_k}-\ket{v_{k-1}}$ with the symmetric normalization
$h=H/\lambda$, where $\lambda$ is the block-encoding subnormalization that is
available on hardware, namely the sum of the bond-operator norms
\begin{equation}
  \lambda_{\mathrm{BE}}=\sum_{i=1}^{2L-1}\norm{J\,\Svec_i\cdot\Svec_{i+1}}
  =(2L-1)\,J\,S(S+1),
  \label{eq:fold-lambdaBE}
\end{equation}
which reduces to the Pauli $1$-norm $3(2L-1)J/4$ for $S=\half$. This choice
guarantees $\operatorname{spec}(h)\subseteq[-1,1]$ a priori but places the
ground state in the \emph{interior} of the interval, $x_0=E_0/\lambda_{\mathrm{BE}}
\approx-0.61$ ($S=\half$, $2L=24$) and $\approx-0.71$ ($S=1$, $2L=12$), i.e.\ with
a sizable edge slack $\nu=1-\abs{E_0}/\lambda_{\mathrm{BE}}\approx0.39$ and
$0.29$ in the sense of Section~\ref{sec:qsd-setup}. This choice is not
innocuous: its substantial cost, and why it does not overturn the conclusions
of this section, are explained in Section~\ref{sec:fold-normalization} and
quantified in Appendix~\ref{app:RescalingFactor}. The alternative used in classical
benchmarking, the \emph{edge-aligned} normalization
\begin{equation}
  \lambda_{\mathrm{edge}}=1.000001\times\max(\abs{E_{\min}},\abs{E_{\max}})
  \label{eq:fold-lambdaedge}
\end{equation}
($=10.45J$ and $15.67J$ for the two models),
places the ground state at the lower edge $x_0\simeq-1$ but presumes knowledge
of $\abs{E_0}$, the very target of the computation; it is used in
Appendix~\ref{app:RescalingFactor} for comparison. Statistical (shot) noise is modeled, as in
\cite{Kirby23a,Kirby24a}, by adding independent Gaussian perturbations of width
$\sigma$ to every measured moment $\mu_\ell$, $\ell\ge1$. The generalized
eigenvalue problem is thresholded with $\epsilon=6\sigma\sqrt D$ (and
$\epsilon=10^{-12}$ in the noiseless case); results are medians over $40$ noise
realizations, with interquartile ranges shown as error bars. All energies are quoted
in units of $J$.

\subsubsection{Dependence on block-encoding rescaling}
\label{sec:fold-normalization}

Before turning to the results, we remark on a technical
point that affects the QSD convergence in Sections~\ref{sec:fold-noiseless}
and~\ref{sec:fold-noise}. One may be tempted to regard the rescaling $\lambda$ as
immaterial for the QSD energy: the Chebyshev--Krylov space coincides with the
power Krylov space $\mathcal K_D(H,\ket{\psi_0})$ for every affine rescaling
(Section~\ref{sec:qsd-setup}), so in exact arithmetic the Ritz values do not
depend on $\lambda$ at all. This invariance of the exact subspace is, however,
the only part of the argument that survives. Substituting $x=\cos\theta$, the
measured moments \eqref{eq:qsd-moments} become trigonometric moments
$\mu_k=\sum_i\abs{\gamma_i}^2\cos(k\theta_i)$ with $\theta_i=\arccos(E_i/\lambda)$,
and what $D$ noisy, thresholded moments can \emph{resolve} is set by gaps in
the angle $\theta$, not in the energy \cite{Epperly22a}. Since
$d\theta/dE=[\lambda\sin\theta]^{-1}$ diverges at the edges of $[-1,1]$, the
placement of the ground state decides the resolving power: with
$\hat\Delta=\Delta/\lambda$,
\begin{equation}
  \theta_0-\theta_1\simeq
  \begin{cases}
    \sqrt{2\hat\Delta}, & x_0=-1\ \text{(edge-aligned)},\\[3pt]
    \hat\Delta/\sin\theta_0, & \abs{x_0}<1\ \text{(interior)},
  \end{cases}
  \label{eq:fold-thetagap}
\end{equation}
i.e.\ edge alignment converts the gap into its square root---the same square
root that separates the exponent $(1+2\sqrt{\delta/\Omega})^{-2(D-1)}$ of
Theorem~\ref{thm:edge}, valid under the tight-lower-edge
Assumption~\ref{ass:edge}, from the linear-in-gap exponent
$(1+\delta/\Omega)^{-2\lfloor(D-1)/2\rfloor}$ of Theorem~\ref{thm:general},
valid for any placement of $x_0$ (and matching the rates of
\cite[Thm.~1]{Kirby23a} and \cite[Thm.~4]{Kirby24a}). 
For the chains studied here the angular gap shrinks by a factor of about
$15$ ($S=\half$, $2L=24$) and $18$ ($S=1$, $2L=12$) in passing from
$\lambda_{\mathrm{edge}}$ to $\lambda_{\mathrm{BE}}$, and the ratio grows with
system size (Appendix~\ref{app:rescale-scaling}).

The consequences, quantified in Appendix~\ref{app:RescalingFactor}
(Figures~\ref{fig:app-scale} and~\ref{fig:app-thetagap}), are significant: under
$\lambda_{\mathrm{BE}}$ the noiseless convergence slope flattens by roughly an
order of magnitude and the noise floors rise by two to three orders of
magnitude relative to the edge-aligned convention, with sublinear rather than
linear dependence on the noise. No published guarantee is violated---the
bounds of \cite{Kirby23a,Kirby24a} hold at either normalization but promise
only the interior convergence, which the algorithm saturates at
$\lambda_{\mathrm{BE}}$, while edge alignment beats them---and a spectral shift
cannot restore edge alignment, since it inflates the LCU norm by the same
amount (Appendix~\ref{app:rescale-noshift}). Because $\lambda_{\mathrm{edge}}$
presumes knowledge of $\abs{E_0}$, the results below are reported for the
implementable $\lambda_{\mathrm{BE}}$; they therefore describe the physical
algorithm rather than its idealized benchmark, and any comparison with
Chebyshev-based QSD results in the literature must first establish which
convention was used. The folded-state advantage, as we shall see, is robust to the
choice and in fact larger under $\lambda_{\mathrm{BE}}$. The state-preparation
stage depends on the same parameter for a different reason
(Section~\ref{sec:prep-normalization}).

\subsection{Noiseless convergence}
\label{sec:fold-noiseless}

\begin{figure*}[t]
\centering
\includegraphics[width=\textwidth]{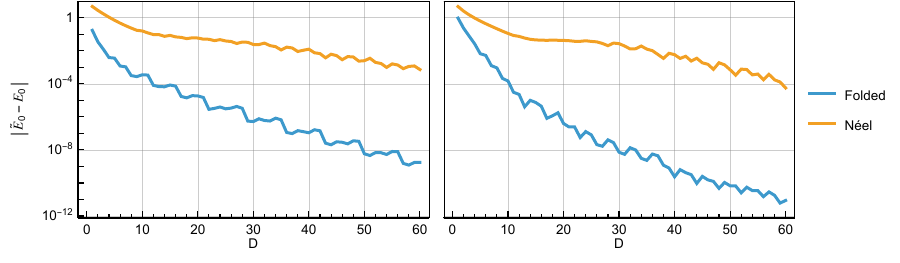}
\caption{Noiseless convergence of the ground-energy error with the Krylov
dimension $D$ for the folded and \Neel{} reference states. Left: $S=1/2$,
$2L=24$. Right: $S=1$, $2L=12$.}
\label{fig:fold-noiseless}
\end{figure*}

Figure~\ref{fig:fold-noiseless} shows the energy error $\abs{\widetilde E_0-E_0}$
versus the Krylov dimension $D$ without noise. Both states exhibit the
exponential convergence \eqref{eq:qsd-noiseless}, but with markedly
different prefactors and, for $S=1$, different effective gaps:
\begin{itemize}
\item For $S=\half$, $2L=24$, the folded state reaches an error of $10^{-4}$ at
  $D\approx12$, $10^{-6}$ at $D\approx27$, and $10^{-8}$ at $D\approx47$, whereas
  the \Neel{} state reaches $10^{-2}$ only at $D\approx8$ and $10^{-3}$ only at
  $D\approx53$, and is still at $\sim5\times10^{-4}$ at $D=60$. At fixed
  $D=20$ the two errors differ by about three orders of magnitude, and at
  $D=60$ by about five.
\item For $S=1$, $2L=12$, the folded state reaches $10^{-4}$ at $D\approx13$,
  $10^{-6}$ at $D\approx28$, and $\sim10^{-11}$ at $D=60$. The \Neel{} curve
  shows a pronounced shoulder at $\sim10^{-2}$ for $D\approx10$--$30$, where its
  triplet weight must be resolved through the Kennedy quasi-degeneracy, and
  reaches $10^{-4}$ only at $D=60$.
\end{itemize}
Since the maximal circuit depth scales linearly with $D$
(Section~\ref{sec:qsd-implementation}), the folded state reduces the circuit
depth required for any target accuracy severalfold---by a factor of four to
five at the accuracy $10^{-4}$ for both models---before noise is even
considered. The staircase shape of the folded curves, with plateaus every
$\sim8$--$10$ Krylov vectors, is a fingerprint of the interior normalization
\eqref{eq:fold-lambdaBE}: the spectrum of $h$ occupies only a subinterval of
$[-1,1]$, on which the Krylov basis is more strongly ill-conditioned than
on the full interval, so that the thresholding at $\epsilon=10^{-12}$ discards
new Krylov directions in batches. The Chebyshev--Krylov space itself is unchanged by the
normalization (Section~\ref{sec:qsd-setup}); what changes is how much of it
survives a finite threshold.

For the Chebyshev-based QSD, the folded reference state $\fold$ thus converts
the exponentially shrinking \Neel{} overlap into a size-independent (or nearly
so) overlap. Consequently, as verified here for $S=\half$ chains up to $2L=24$
and $S=1$ chains up to $2L=12$, the Krylov dimension---hence the maximal
circuit depth---needed for a given accuracy is reduced severalfold, with
intermediate-$D$ errors smaller by three to five orders of magnitude; for open
$S=1$ chains the exact singlet character of the folded state additionally
bypasses the Kennedy edge quasi-degeneracy.

\subsection{Noise robustness}
\label{sec:fold-noise}

\begin{figure*}[t]
\centering
\includegraphics[width=\textwidth]{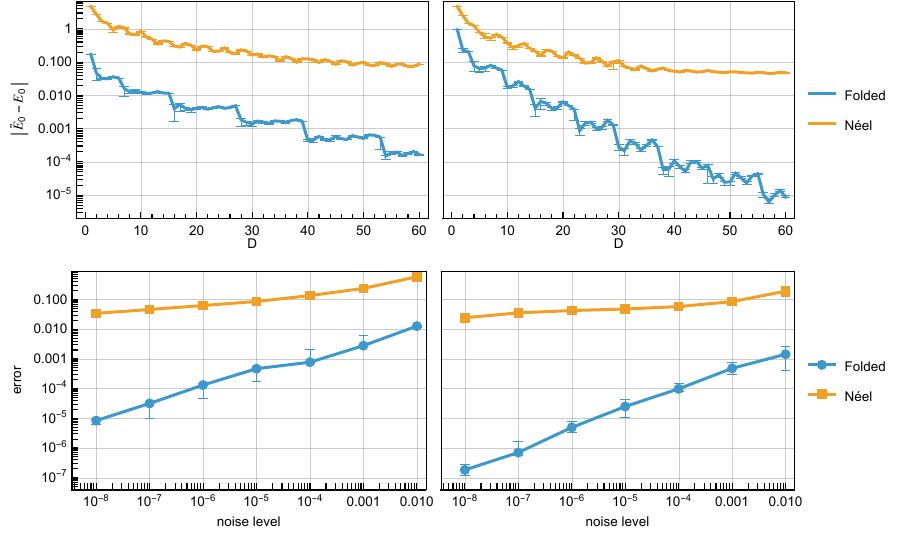}
\caption{Noise robustness. Top row: energy error versus $D$ at fixed moment
noise $\sigma=10^{-5}$. Bottom row: converged error (median over $D\in[45,60]$)
versus $\sigma$. In both rows the plotted values are medians over 40 noise
realizations and the error bars mark the interquartile range. Left: $S=1/2$,
$2L=24$. Right: $S=1$, $2L=12$.}
\label{fig:fold-noise}
\end{figure*}

Figure~\ref{fig:fold-noise} (top row) shows the noisy convergence at
$\sigma=10^{-5}$. The \Neel{} error saturates by $D\approx40$ at a floor of
$\sim0.09$ ($S=\half$) and $\sim0.05$ ($S=1$)---i.e.\ it never gets below a
few percent of $J$---whereas the folded error keeps decreasing through the
largest Krylov dimension simulated, reaching $\sim2\times10^{-4}$ ($S=\half$)
and $\sim10^{-5}$ ($S=1$) at $D=60$, three to four orders of magnitude below
the \Neel{} floor.

The bottom row shows the converged error (median over $D\in[45,60]$) as a
function of $\sigma$. For the folded state the error grows monotonically with
the noise, roughly as $\sigma^{1/2}$ for $S=\half$ (from $\sim10^{-5}$ at
$\sigma=10^{-8}$ to $\sim10^{-2}$ at $\sigma=10^{-2}$) and as $\sigma^{2/3}$ for
$S=1$ (from $\sim2\times10^{-7}$ to $\sim10^{-3}$). For the \Neel{} state the
error is nearly independent of the noise level, varying only between $\sim0.03$
and $\sim0.5$ over six decades of $\sigma$. At every noise level the folded
error is therefore smaller by two to five orders of magnitude, with the
advantage growing as the noise decreases; conversely, the \Neel{} reference does
not reach the folded-state accuracy at \emph{any} noise level in the scanned
range.

Two features of these curves deserve comment, because they differ from the
linear-in-noise behavior predicted by \eqref{eq:qsd-simplified}. First, the
sublinear growth of the folded error and the flattened noiseless slopes of
Figure~\ref{fig:fold-noiseless} have a common origin in the interior
normalization \eqref{eq:fold-lambdaBE}: the tight-lower-edge assumption behind
the square-root exponent (Assumption~\ref{ass:edge}) is violated by a slack
$\nu\approx0.3$--$0.4$, and the applicable guarantee is the interior bound of
Theorem~\ref{thm:general}, whose convergence term
$(1+\delta/\Omega)^{-2\lfloor(D-1)/2\rfloor}$ is not small at any $D\le60$ for
these chains ($\delta/\Omega\sim2\times10^{-3}$). The simulated range is
therefore not the noise-dominated regime of the bound but the regime in which
thresholding decides what is resolved; there the error is set by the retained
subspace, and its dependence on the noise is the sublinear, threshold-driven
$\bigO(\epsilon^{1/(1+\alpha)})$ scaling analyzed in \cite{Epperly22a,Kirby23a},
which with $\epsilon=6\sigma\sqrt D$ is a fractional power of $\sigma$, as
observed. Appendix~\ref{app:RescalingFactor}
(Figure~\ref{fig:app-scale}) shows that the edge-aligned normalization
\eqref{eq:fold-lambdaedge} restores both the square-root convergence and
floors linear in $\sigma$, by the mechanism of Section~\ref{sec:fold-normalization}. Second, the near $\sigma$-independence of the \Neel{} error
reflects the resolvability condition \eqref{eq:qsd-signal}: with
$\abs{\gamma_0}^2\approx6\times10^{-3}$ ($S=\half$) and $0.03$ ($S=1$), the
thresholded subspace retains only the few directions in which the tiny
ground-state signal is resolved, and the error is dominated by the
Rayleigh--Ritz error of that truncated subspace rather than by $\sigma$.
Lowering the noise helps only insofar as it lowers $\epsilon$ and admits one
more direction, hence the staircase in $D$ and the slow drift in $\sigma$.

Both benefits---faster noiseless convergence and a far lower error under
noise---trace back to the same fact: the folded reference state converts the
exponentially shrinking \Neel{} overlap into a size-independent (or nearly so)
overlap. Since the overlap advantage of the folded state \emph{grows}
exponentially with system size (Figure~\ref{fig:fold-overlap}), both benefits
become more pronounced with increasing system size. This motivates recursive folding---using the
QSD output at size $L$ to seed size $2L$---as a ground-state preparation
strategy, which requires turning the classical QSD output back into a physical
state; this is the subject of Section~\ref{sec:StatePreparation}. Two further
remarks connect to Section~\ref{sec:Fidelity}. First, for periodic chains the
folded state breaks two bonds instead of one; for the gapped $S=1$ models the
asymptotic overlap is found to be nearly the same in both geometries,
while for the gapless $S=\half$ chain the two-cut geometry decays with a larger
power-law exponent. Second, for the AKLT chain the short correlation length
($\xi=1/\ln3$) makes the overlap converge to its asymptote within a few sites,
but the asymptote itself is fixed by the recoupling of the edge spins at the
junction and equals $\abs{\gamma_0}^2\to\tfrac14$ exactly---smaller than the
spin-$\half$ Heisenberg values in Figure~\ref{fig:fold-overlap} at these sizes,
but strictly independent of $L$.

\section{Ground-State Preparation}
\label{sec:StatePreparation}

Chebyshev-based QSD returns a \emph{classical} description of an approximate
ground state: the lowest eigenpair $(\widetilde E_0,c)$ of the thresholded
generalized eigenvalue problem $\mH c=\widetilde E\,\mS c$ defines the Ritz vector
\begin{equation}
  \ket{\Psi}\ \propto\ \sum_{k=0}^{D-1}c_k\,T_k(h)\ket{\psi_0}=A\ket{\psi_0},
  \label{eq:prep-ritz}
\end{equation}
where for later use we have defined
\begin{equation}
  A:=\sum_{k=0}^{D-1}c_kT_k(h)=p(h)
\end{equation}
with $p(x)=\sum_kc_kT_k(x)$ a polynomial of degree $D-1$. To use this state as
(half of) the folded reference state $\fold$ of the next larger system, the
Ritz vector must be prepared as a physical state on the quantum computer. This
section examines why the linear-combination-of-unitaries (LCU) technique is the
natural construction in this setting, why the quantum singular value
transformation (QSVT) is its optimized variant, how the conditioning of $\mS$
and the thresholding that regularizes it control the postselection cost of
both, and how the cost of the resulting \emph{folding ladder}
$L_0\to2L_0\to\cdots\to2^kL_0$ scales with system size. Throughout, $h=H/\lambda$
is normalized to $[-1,1]$ as in Section~\ref{sec:qsd-implementation}, and $\ket{g}$
denotes the exact ground state of the system being prepared; the overlap
$\abs{\braket{g_{2L}|g_L\otimes g_L}}^2$ of the prepared folded state with the
ground state of the doubled system is studied in Section~\ref{sec:Fidelity}.

\subsection{The state-preparation problem}
\label{sec:prep-problem}

The task is: given the classical data $(c,\mS)$ and the same block encoding
$(U,G)$ used in the QSD run, prepare the normalized state
$\ket{\Psi}=A\ket{\psi_0}/\norm{A\ket{\psi_0}}$. Since the Krylov basis is not
orthogonal, the physical norm of the Ritz vector is a quadratic form in the Gram
matrix,
\begin{equation}
  \norm{A\ket{\psi_0}}^2=\sum_{j,k}c_j^*\braket{v_j|v_k}c_k=c^\dagger\mS c,
  \label{eq:prep-gram}
\end{equation}
computable classically from the already-measured $\mS$. The apparent
obstruction is that neither the basis states nor $A$ are unitary. The
resolution is that the qubitization identity \eqref{eq:qsd-walk} already
expresses each basis state as a \emph{postselected block of a power of a single
unitary} $W$: the non-unitarity has been accounted for once, in the
block-encoding construction. Two implementations present themselves.

\subsection{LCU over walk-operator powers}
\label{sec:prep-lcu}

The LCU technique \cite{Berry15a,Childs17a} implements $A=\sum_kc_kW^k$ with
$c_k=e^{i\theta_k}\abs{c_k}$: adjoin a counter register of $\lceil\log_2D\rceil$
qubits and perform $\PREP:\ket0\mapsto\norm{c}_1^{-1/2}\sum_k\sqrt{\abs{c_k}}\ket
k$ on the counter, then $\SEL:=\sum_k\ket k\!\bra k\otimes e^{i\theta_k}W^k$ on
$\ket{G}_a\otimes\ket{\psi_0}$, then $\PREP^\dagger$, and postselect the counter
on $\ket0$ and the block-encoding auxiliaries on $\ket{G}_a$. Because $\SEL$
involves powers of one and the same walk operator, its cost is linear rather
than quadratic in $D$: controlled-$W^{2^j}$ conditioned on bit $j$ of the
counter costs $\sum_j2^j=D-1$ applications of controlled $W$. The preparation
circuit is therefore about as deep as the \emph{longest measurement circuit
already executed} in the QSD run ($D-1$ walk steps,
Section~\ref{sec:qsd-implementation}), plus a $\log_2D$-qubit counter---the
precise sense in which LCU is the natural construction here.

\begin{proposition}[LCU postselection amplitude]
\label{prop:lcu-amp}
The postselection succeeds with amplitude
\begin{equation}
  \sqrt{\pLCU}=\frac{\norm{A\ket{\psi_0}}}{\norm{c}_1}
  =\frac{\sqrt{c^\dagger\mS c}}{\norm{c}_1}.
  \label{eq:prep-lcu-amp}
\end{equation}
\end{proposition}

Equation \eqref{eq:prep-lcu-amp} is invariant under $c\to\alpha c$, as it must
be; under the $\mS$-normalization $c^\dagger\mS c=1$ it reduces to
$1/\norm{c}_1$, but generic eigensolvers return Euclidean-normalized $c$, for
which the factor $\sqrt{c^\dagger\mS c}$ is essential. The postselection probability is
predicted classically from the measured $\mS$ before any preparation is
attempted. Repeat-until-success costs $\bigO(1/\pLCU)$ preparations; amplitude
amplification improves this to $\bigO(1/\sqrt{\pLCU})$ rounds.

\subsection{QSVT with classically computed phase factors}
\label{sec:prep-qsvt}

Since $A=p(h)$ for a known polynomial, one may instead implement $p(h)$
directly by quantum signal processing / QSVT \cite{Gilyen19a,Martyn21a}:
classically convert the Chebyshev coefficients into phase factors (stable
algorithms exist up to very large degrees \cite{Haah19a,Dong21a}) and run the
phased qubitization sequence with $D-1$ queries to $U$ and $U^\dagger$. QSP
requires $\abs{p(x)}\le1$ on $[-1,1]$, so $p$ is rescaled by
$\norm{p}_\infty:=\max_{x\in[-1,1]}\abs{p(x)}$, and implements fixed-parity
polynomials, so $p=p_{\mathrm{even}}+p_{\mathrm{odd}}$ is combined with a
one-qubit LCU. This route needs no counter and no $\PREP$ circuit, and its
success amplitude is normalized by the sup-norm rather than the $\ell_1$-norm,
\begin{equation}
  \sqrt{\pQSVT}=\frac{\norm{p(h)\ket{\psi_0}}}{\norm{p}_\infty}
  =\frac{\sqrt{c^\dagger\mS c}}{\norm{p}_\infty}.
  \label{eq:prep-qsvt-amp}
\end{equation}

\begin{lemma}[Norm inequalities for Chebyshev expansions]
\label{lem:prep-norms}
Let $p(x)=\sum_{k=0}^{D-1}c_kT_k(x)$ and $\ket{\psi_0}=\sum_i\gamma_i\ket{E_i}$.
Then
\begin{equation}
\begin{split}
  \norm{p(h)\ket{\psi_0}}&\le\max_i\abs{p(x_i)}\le\norm{p}_\infty \\
  &\le\norm{c}_1\le\sqrt{2D-1}\,\norm{p}_\infty .
\end{split}
  \label{eq:prep-chain}
\end{equation}
\end{lemma}

\begin{proof}
$\norm{p(h)\ket{\psi_0}}^2=\sum_i\abs{\gamma_i}^2\abs{p(x_i)}^2\le\max_i\abs{p(x_i)}^2$
gives the first two inequalities. With $x=\cos\theta$, $T_k(x)=\cos k\theta$
and $\abs{T_k}\le1$, so $\abs{p(x)}\le\norm{c}_1$ pointwise. Finally, the
Chebyshev orthogonality $\int_{-1}^1T_jT_k\,w\,dx=\pi\delta_{jk}(1+\delta_{j0})/2$
with $w(x)=(1-x^2)^{-1/2}$ gives
\begin{equation}
  \abs{c_0}^2+\frac12\sum_{k=1}^{D-1}\abs{c_k}^2
  =\frac1\pi\int_{-1}^1\abs{p(x)}^2\,w(x)\,dx\le\norm{p}_\infty^2
  \label{eq:prep-parseval}
\end{equation}
(the Chebyshev--Parseval bound \eqref{eq:parseval} of
Appendix~\ref{app:ChebyshevQSD}), and the Cauchy--Schwarz inequality
\begin{equation}
  \norm{c}_1^2\le\bigl(1+2(D-1)\bigr)\Bigl(\abs{c_0}^2+\frac12\sum_{k=1}^{D-1}\abs{c_k}^2\Bigr)
\end{equation}
yields the last inequality.
\end{proof}

\begin{corollary}[QSVT vs.\ LCU]
\label{cor:prep-ratio}
$1\le\sqrt{\pQSVT}/\sqrt{\pLCU}=\norm{c}_1/\norm{p}_\infty\le\sqrt{2D-1}$: QSVT never
has a smaller postselection amplitude than LCU, and can have one larger by up
to a factor $\sim\sqrt{2D}$ ($\sim2D$ in probability).
\end{corollary}

Equality $\norm{p}_\infty=\norm{c}_1$ requires a point where all $\abs{T_k}=1$
simultaneously with all terms in phase, i.e.\ $x=\pm1$ (where
$T_k(\pm1)=(\pm1)^k$): for alternating-sign coefficients the bound saturates at
$x=-1$ and QSVT offers no advantage. ``Cancellation-heavy'' coefficients are
the opposite regime; for random signs $c_k=\pm1$ a Salem--Zygmund bound
\cite{Salem54a} gives $\norm{p}_\infty\sim\sqrt{D\log D}$ while $\norm{c}_1=D$,
so incoherent coefficients generically gain a factor $\sim\sqrt{D/\log D}$ in
amplitude, within a logarithm of the bound of Corollary~\ref{cor:prep-ratio}.

\subsection{Ill-conditioning, gauge freedom, and thresholding}
\label{sec:prep-thresholding}

The eigenvalues of $\mS$ are the squared singular values of $\mV$. The dominant
ones are $\bigO(1)$ ($\mS_{kk}=\norm{T_k(h)\ket{\psi_0}}^2=\bigO(1)$, so $\tr\mS\sim
D$), but the smallest decay rapidly with $D$: once the Krylov space has
captured the spectral content of $\ket{\psi_0}$ resolvable by a degree-$(D-1)$
filter, each new basis vector is nearly a combination of its predecessors. A
small eigenpair $(\lambda_v,v)$ of $\mS$ means precisely that
$\norm{\sum_kv_kT_k(h)\ket{\psi_0}}^2=v^\dagger\mS v=\lambda_v\approx0$: the
polynomial $q_v=\sum_kv_kT_k$ is tiny on the spectrum \emph{weighted by}
$\ket{\psi_0}$ despite having $\bigO(1)$ coefficients. The map $c\mapsto\mV c$ is
therefore nearly many-to-one; shifting $c\to c+\alpha v$ changes the physical
state and the Rayleigh quotient by only $\bigO(\abs\alpha\sqrt{\lambda_v})$ while
changing $\norm{c}_1$ arbitrarily. The energy is blind to these
directions---but the preparation cost is not, because \emph{the LCU cost is
determined by the representation, not by the state} \eqref{eq:prep-lcu-amp}. An eigensolver applied
to the raw noisy pair $(\mH',\mS')$ generically returns a $c$ heavily
contaminated along near-null directions (the noise renders them spuriously
significant): a large, rapidly oscillating coefficient vector representing a
state of modest norm, with $\norm{c}_1\gg\sqrt{c^\dagger\mS c}$ and a collapsed
postselection probability. QSVT softens but does not remove the problem, since
a contaminated $p$ still oscillates to large values inside spectral gaps or
beyond the spectral edges, where $\norm{p}_\infty$ is set.

The thresholding prescription of Section~\ref{sec:qsd-noise} was introduced to
stabilize the \emph{energy} \cite{Epperly22a}: solved directly, the raw noisy
problem admits coefficient vectors with $\mH'c,\mS'c\approx0$, so the noise
perturbs the Rayleigh quotient near $0/0$ and produces spurious eigenvalues,
often below the true ground energy. Its role for the \emph{state-preparation
cost} is the following elementary but essential bound.

\begin{proposition}[Thresholding floors the postselection probabilities]
\label{prop:prep-floor}
Let $c$ solve the thresholded problem with $c^\dagger\mS c=1$, supported on
eigenvectors of $\mS$ with eigenvalues $>\epsilon$. Then
\begin{subequations}\label{eq:prep-floor}
\begin{align}
  \norm{c}_2&\le\frac1{\sqrt\epsilon}, \\
  \norm{c}_1&\le\sqrt D\,\norm{c}_2\le\sqrt{\frac D\epsilon}, \\
  \pQSVT&\ge\pLCU=\frac1{\norm{c}_1^2}\ge\frac\epsilon D .
\end{align}
\end{subequations}
\end{proposition}

\begin{proof}
Expand $c=\sum_{\lambda_i>\epsilon}a_iv_i$ in the eigenbasis of $\mS$, so
$\sum_i\lambda_i\abs{a_i}^2=1$; then $\norm{c}_2^2=\sum_i\abs{a_i}^2\le
\epsilon^{-1}$. Cauchy--Schwarz gives $\norm{c}_1\le\sqrt D\norm{c}_2$,
Proposition~\ref{prop:lcu-amp} gives $\pLCU=\norm{c}_1^{-2}$, and
Corollary~\ref{cor:prep-ratio} gives $\pQSVT\ge\pLCU$.
\end{proof}

Thus the \emph{same} regularization parameter that stabilizes the energy is a
floor on the postselection probability. Nothing of value is lost: the discarded
directions are coefficient patterns whose physical states have norm
$\le\sqrt\epsilon$, so removing them perturbs the Ritz state only at the
$\bigO(\sqrt\epsilon)$ level while removing the unbounded part of the
representation. Thresholding is a \emph{gauge fixing}: among the many
coefficient vectors representing nearly the same state, it selects the
minimal-norm representative supported on the well-resolved subspace. Since for
folding only the overlap with the larger system's ground state matters, the
representation can be optimized further for cost---by \emph{residual
$\ell_1$ gauge fixing}, i.e.\ the small convex program
\begin{equation}
  \min_c\ \norm{c}_1\quad\text{s.t.}\quad
  (c-c_{\mathrm{Ritz}})^\dagger\mS(c-c_{\mathrm{Ritz}})\le\delta^2,
  \label{eq:prep-l1}
\end{equation}
whose radius $\delta$ translates into the overlap guarantee
$\abs{\braket{\Psi_\delta|\Psi}}\ge\sqrt{1-\delta^2}$; by degree truncation,
which shortens the circuit and reduces $\norm{c}_1$ simultaneously; or by
preferring QSVT for cancellation-heavy coefficients.

\begin{remark}[Folding step]
\label{rem:prep-folding}
The prepared state serves as half of $\ket{\psi_0^{(2L)}}=\fold$, and the
larger computation requires only that
$\abs{\braket{g_{2L}|g_L\otimes g_L}}^2$ be at least inverse-polynomial in the
system size, so approximate preparation---truncated or
gauge-optimized $c$, sampling noise in $c$ itself---is tolerated: all such
errors enter only through the overlap of the larger run. The product reference
requires two independent preparations, so the naive per-attempt postselection
probability is $p^2$; with amplitude amplification each copy costs
$\bigO(1/\sqrt p)$ rounds and the two costs add. Both copies reuse the same
$L$-site block encoding and phase factors, and both preparations cost $\sim D$
queries, the same order as the deepest QSD measurement circuit: state
preparation for folding does not change the depth class of the algorithm, and
the new cost is entirely in the postselection/amplification factor, which
Proposition~\ref{prop:prep-floor} ties to the regularization threshold.
\end{remark}

\subsection{Numerical demonstrations}
\label{sec:prep-numerics}

\begin{figure*}[t]
  \centering
  \includegraphics[width=\textwidth]{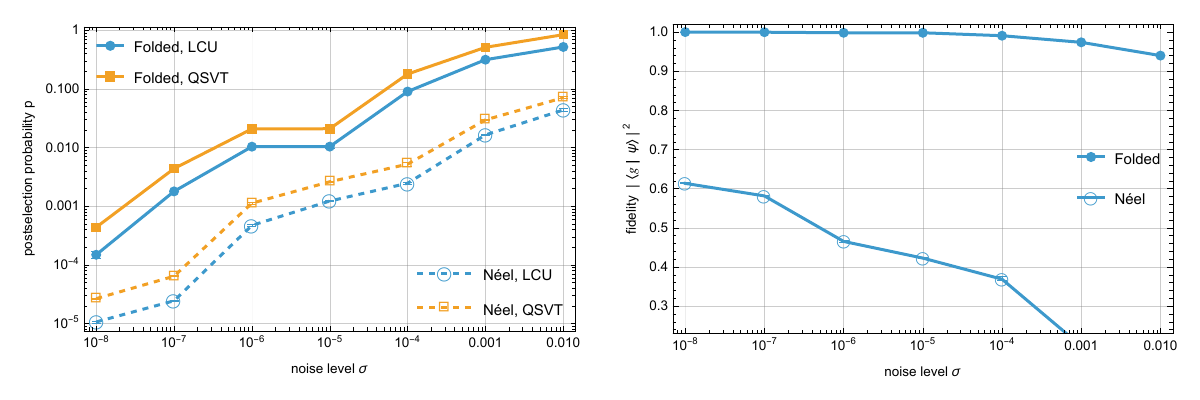}
  \caption{Thresholded pipeline ($\epsilon=6\sigma\sqrt D$, $D=30$) for the
  folded reference $\ket{g_{12}}\otimes\ket{g_{12}}$ (filled symbols) and the
  \Neel{} reference (open symbols) on the $2L=24$ spin-$\half$ Heisenberg chain, versus
  the moment-noise width $\sigma$. Left: postselection probability $p$ for the LCU
  (circles) and QSVT (squares) implementations. Right: fidelity
  $\abs{\braket{g|\Psi}}^2$ of the prepared state with the exact ground
  state. Medians over 50 noise realizations.}
  \label{fig:prep-folded}
\end{figure*}

We illustrate the analysis on the open spin-$\half$ Heisenberg chain normalized
by the block-encoding subnormalization $\lambda_{\mathrm{BE}}=3(2L-1)/4$ (the
Pauli $\ell_1$-norm), with the folded reference $\ket{g_{12}}\otimes\ket{g_{12}}$
on $2L=24$ sites ($\abs{\gamma_0}^2=0.819$) and the \Neel{} reference on the
same chain ($\abs{\gamma_0}^2=0.0060$), both at Krylov dimension $D=30$. Measurement
noise is modeled as in Section~\ref{sec:fold-setup}: i.i.d.\ Gaussian noise of
width $\sigma$ is added to the moments $\mu_k$, $k\ge1$, and both $\mH'$ and
$\mS'$ are assembled from the same noisy moments. The
generalized eigenvalue problem is thresholded at $\epsilon=6\sigma\sqrt D$ as in Section~\ref{sec:fold-setup}, the
postselection probabilities are evaluated with the \emph{true} $\mS$---i.e.\ they
are the physical ones, $\pLCU=c^\dagger\mS c/\norm{c}_1^2$ and
$\pQSVT=c^\dagger\mS c/\norm{p}_\infty^2$---and the fidelity of the prepared
state with the exact ground state is monitored. Figure~\ref{fig:prep-folded}
shows the thresholded pipeline; the raw (unthresholded) pipeline and the
$\ell_1$ gauge fixing of Section~\ref{sec:prep-thresholding} are summarized in
the text.

Four observations summarize the demonstrations.
\emph{(i) Thresholding is the essential step.} Without it, the coefficient
vector returned by the eigensolver is fit to noise in the near-null subspace of
$\mS'$---exactly the gauge pathology of Section~\ref{sec:prep-thresholding}:
in the same setting the raw pipeline's postselection probabilities collapse to
$10^{-10}$--$10^{-6}$ over the scanned noise range, and the state it delivers
has infidelity $\approx1$. Thresholding lifts $p$ by four to eight orders of
magnitude (Figure~\ref{fig:prep-folded}, left), well above the
guaranteed floor $\epsilon/D=6\sigma/\sqrt D$ of
Proposition~\ref{prop:prep-floor}, while the fidelity of the folded
reference tracks the noise level, staying within $10^{-4}$ of unity at
$\sigma=10^{-8}$ and degrading only to $0.93$ at $\sigma=10^{-2}$ (right). The
\Neel{} reference, with an overlap more than a hundred times smaller, is a
factor $10$--$20$ below the folded one in postselection probability at every $\sigma$,
and its prepared state is poor (fidelity $0.6$ at the smallest noise, falling
to $0.06$):
at this Krylov dimension and interior normalization the \Neel{} Ritz vector is
not converged (Section~\ref{sec:fold-noiseless}), which is the
state-preparation face of the resolvability condition \eqref{eq:qsd-signal}.
\emph{(ii) Under the correlated moment noise, QSVT provides little gain.} The ratio
$\pQSVT/\pLCU$ is only $1.6$--$2.7$ for both references, much smaller than the
worst-case $2D-1=59$ of Corollary~\ref{cor:prep-ratio}, and it is
equally small ($1$--$2$) for the raw pipeline, whose collapse QSVT does not remedy.
The structural reason is that, by $T_iT_j=\tfrac12(T_{i+j}+T_{\abs{i-j}})$, the
measured data couple to $c$ only through the Chebyshev coefficients of the
polynomial $p^2$,
\begin{equation}
  c^\dagger\mS'c=\sum_{m=0}^{2D-2}w_m(c)\,\mu_m',
  \label{eq:prep-moment-coupling}
\end{equation}
where the weights $w_m(c)$ are defined by
\begin{equation}
  p(x)^2=\sum_{m=0}^{2D-2}w_m(c)\,T_m(x),
\end{equation}
so noise-dominated near-null directions are constrained at the level of the
polynomial, not the coefficient vector: the contamination inflates
$\norm{p}_\infty$ together with $\norm{c}_1$ instead of producing the
incoherent coefficient patterns that QSVT's sup-norm normalization exploits.
(Under an---unphysical for this algorithm---independent matrix-element noise
model the raw-regime ratio is $\approx3$--$5$.) For the Chebyshev method,
regularization, not the choice of implementation, is the essential step.
\emph{(iii) The thresholded postselection probability has an anomalous noise
dependence, which gauge fixing removes.} In Figure~\ref{fig:prep-folded} $p$
\emph{decreases} as the noise decreases---by three orders of magnitude over
six decades of $\sigma$---because a smaller $\epsilon$ retains more marginal
directions, which inflate $\norm{c}_1$ (and $\norm{p}_\infty$) without
improving the state. Since $\epsilon$ grows with the noise level $\sigma$, a larger $\sigma$ discards the eigenspaces of $\mS'$ with the eigenvalues below $\epsilon$, reducing the dimension of the retained subspace and hence $\norm{c}_1$, which raises the postselection probability at the cost of a weaker filter and the modest fidelity loss seen in the right panel. The residual $\ell_1$ gauge fixing \eqref{eq:prep-l1}
removes this: with $\delta=0.15$ (overlap guarantee $\ge0.989$) the folded
postselection probability is flat at $\pLCU\approx0.8$ across the whole noise range,
i.e.\ at its ceiling $\abs{\gamma_0}^2$ (Section~\ref{sec:prep-ceiling}), and
$\pQSVT/\pLCU\to1.0$--$1.1$, since minimizing $\norm{c}_1$ toward its lower
bound $\norm{p}_\infty$ produces a maximally coherent coefficient vector. The
practical hierarchy is thus: thresholding rescues the computation from
pathology; then \emph{either} QSVT \emph{or} $\ell_1$ gauge fixing harvests
the remaining incoherence---doing both is largely redundant.
\emph{(iv) The fidelity cost of gauge fixing is bounded}, saturating near the
budget $\delta^2$, which is irrelevant in the folding context where only the
reference overlap of the next level matters.

\subsubsection{Dependence on block-encoding rescaling}
\label{sec:prep-normalization}

The demonstrations above depend on one design parameter that has so far been
kept implicit: the rescaling $h=H/\lambda$. It must be committed to
\emph{before} the algorithm runs, and the spectral radius is precisely what the
algorithm does not know---for the antiferromagnetic chain it equals $\abs{E_0}$,
the target itself. The operationally available choice is the block-encoding
subnormalization $\lambda_{\mathrm{BE}}=\sum_i\norm{J\,\Svec_i\cdot\Svec_{i+1}}
=(2L-1)JS(S+1)$ of \eqref{eq:fold-lambdaBE}, which guarantees
$\operatorname{spec}(h)\subseteq[-1,1]$ a priori and places
$x_0=E_0/\lambda_{\mathrm{BE}}$ in the \emph{interior} of $[-1,1]$
($\approx-0.61$ for the chains above), as in Section~\ref{sec:FoldedQSD}. All
figures in this section use it.

Section~\ref{sec:fold-normalization} explained why the QSD \emph{energy}
depends on this choice: everything the algorithm measures lives in the angle
$\theta=\arccos(E/\lambda)$, and interior placement compresses the angular
gap that a Krylov step resolves. The \emph{preparation cost} depends on the
same parameter through a second, independent mechanism. It is a property of
the representation $c$ of the Ritz state rather than of the state, and
$\lambda$ sets where $x_0$ falls in $[-1,1]$, which controls the
\emph{coherence} of the filter coefficients. With the edge-aligned convention
$\lambda\approx\abs{E_0}$, $x_0$ sits at the edge $-1$, where $T_k(-1)=(-1)^k$:
the converged filter's coefficients are alternating-sign coherent,
$\norm{c}_1=\abs{p(-1)}=\norm{p}_\infty$ exactly, the saturation case of
Lemma~\ref{lem:prep-norms} is realized---$\pQSVT=\pLCU$ to all digits and the
$\ell_1$ gauge fixing is inert---and the edge placement partially protects even
the raw pipeline, whose failures become a discrete, bimodal per-realization
event, since a spurious Ritz value can displace the true one only by falling
below $-1$. With $\lambda_{\mathrm{BE}}$ the coefficients are incoherent, and
the modest QSVT advantage and $\ell_1$ slack of Figure~\ref{fig:prep-folded}
appear. A third effect concerns the admissibility of filters for QSVT and is
met in Section~\ref{sec:prep-ceiling}: the one-sided Chebyshev amplifier with
its $\sqrt{\hat\Delta}$ acceleration is bounded on $[-1,1]$ only when $x_0$ is
at the edge, so in the interior the certified filter costs
$D\sim\hat\Delta^{-1}$ rather than $\hat\Delta^{-1/2}$.

The two dependences are thus of different kinds and should not be conflated.
For the energy, $\lambda$ acts through the \emph{measurement}: it sets the
angular resolution of the moments, a property of the data that no
post-processing can recover. For the preparation, $\lambda$ acts through the
\emph{representation}: it sets how a given state is written in the Chebyshev
basis, which thresholding, $\ell_1$ gauge fixing, or the certified filter can
partly repair. Both lead to the same conclusion---edge alignment is favorable on both
fronts, with square-root-enhanced resolution for the energy and coherent
coefficients with $\pQSVT=\pLCU$ for the preparation---and on its
unavailability, since the spectral radius is the computation's target. The
implementable $\lambda_{\mathrm{BE}}$ incurs the interior penalty on both fronts,
and all figures in this section are computed in that regime. Both mechanisms are
collected in Appendix~\ref{app:RescalingFactor}.

\subsection{Generation probability, ceiling, and the certified filter}
\label{sec:prep-ceiling}

The figure of merit for state preparation is not the postselection probability alone:
a successful postselection onto the wrong state is of no use. The relevant
quantity is the \emph{ground-state generation probability}, the per-attempt
probability that the postselection succeeds \emph{and} the delivered state is
the ground state. Recall that in the Chebyshev-based QSD every entry of
$\mH$ and $\mS$ is a real linear combination of the moments $\mu_\ell$
\eqref{eq:qsd-Hmoments}, which are real as expectation values of Hermitian
operators, so that both matrices are real symmetric; the coefficient vector $c$
of the thresholded generalized eigenvalue problem, and hence the polynomial
$p(x)$, can therefore be taken real, as we do throughout. The phases of the
overlaps $\gamma_i=\braket{E_i|\psi_0}$ play no role in what follows, and we
write $\gamma_i^2$ for $\abs{\gamma_i}^2$. With $\ket{\psi_0}=\sum_i\gamma_i\ket{E_i}$ and the
QSVT-normalized filter $A=p(h)/\norm{p}_\infty$, we have
\begin{equation}
  \pQSVT=\frac{\sum_i\gamma_i^2p(x_i)^2}{\norm{p}_\infty^2},
\end{equation}
and
\begin{equation}
  F=\frac{\gamma_0^2p(x_0)^2}{\sum_i\gamma_i^2p(x_i)^2},
\end{equation}
and the spectral sum cancels in the product:
\begin{equation}
  \pQSVT\,F=\gamma_0^2\,\frac{p(x_0)^2}{\norm{p}_\infty^2}
  =\abs{\braket{g|A|\psi_0}}^2=\gamma_0^2\,\zeta_{\mathrm{f}},
  \label{eq:prep-production}
\end{equation}
with
\begin{equation}
  \zeta_{\mathrm{f}}:=\frac{p(x_0)^2}{\norm{p}_\infty^2}\in[0,1].
\end{equation}
The factorization separates the two factors according to what determines
them: the overlap $\gamma_0^2$ is fixed by the physics---a property of the
Hamiltonian and the reference state on which the algorithm has no
influence---while the \emph{filter efficiency} $\zeta_{\mathrm f}$ is a
purely classical property of the polynomial, the fraction of its maximal
response that is concentrated at the target energy. The ceiling
$\pQSVT F\le\gamma_0^2$ is trivial and \emph{tight}: $\zeta_{\mathrm f}\to1$
exactly when $\abs p$ attains its supremum at $x_0$. For LCU the same
computation gives $\pLCU F=\gamma_0^2p(x_0)^2/\norm{c}_1^2\le\pQSVT F$. In this
language the phenomena of Sections~\ref{sec:prep-thresholding}--%
\ref{sec:prep-numerics} are all $\zeta_{\mathrm f}$-statements: noise
contamination lowers $\zeta_{\mathrm f}$ without lowering $F$ (marginal
retained directions add components to $p$ that are small on the weighted
spectrum but large in spectral gaps and near the edges), and thresholding and
$\ell_1$ gauge fixing are $\zeta_{\mathrm f}$-restoration procedures. The
bound \eqref{eq:prep-production} is a law of the spectral-filter paradigm: it
holds because $A$ commutes with $H$, so a filter can only reweight the
eigencomponents already present. Improving the reference state---which is what
folding \emph{is}---raises the ceiling before the filter is applied. With
amplitude amplification the expected block-encoding count per ground-state copy
is $\bigO(D/\sqrt{\pQSVT})$ up to the residual $1/F$, so the ceiling reappears as
the familiar $1/\gamma_0$ overlap dependence of amplified ground-state
preparation \cite{Lin20a}.

\begin{figure*}[t]
  \centering
  \includegraphics[width=0.98\textwidth]{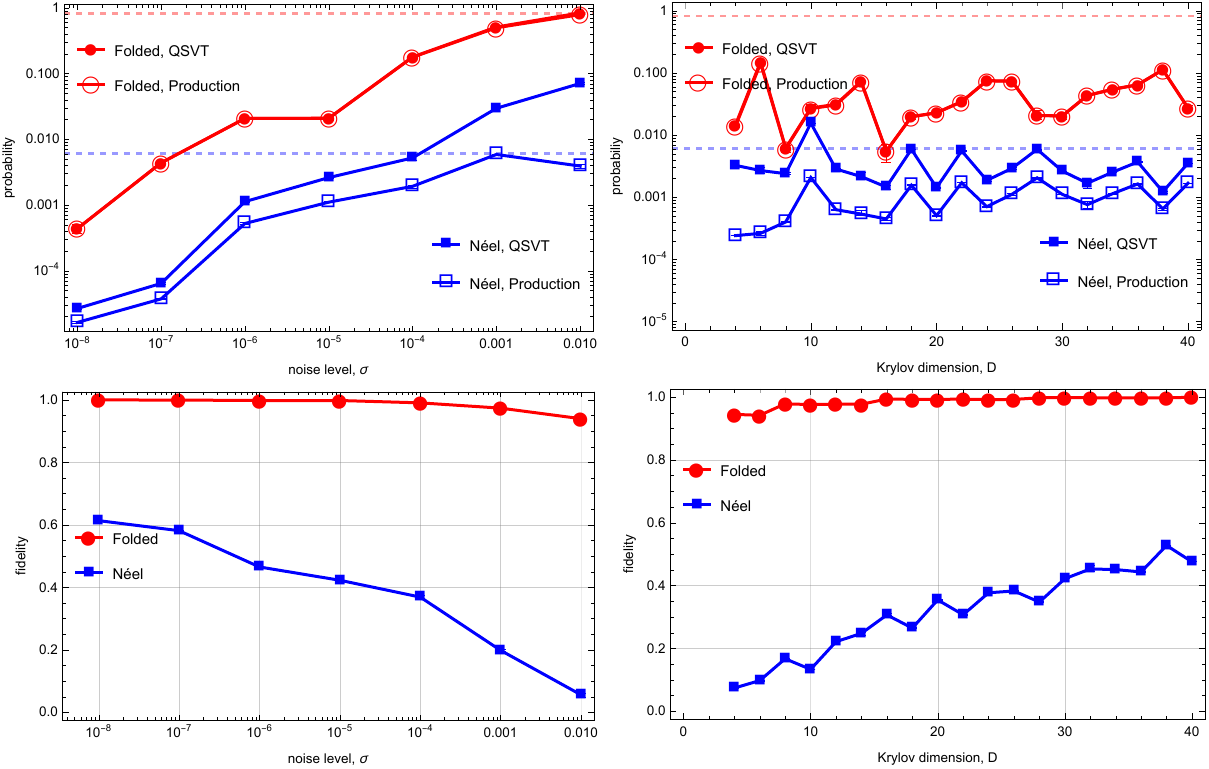}
  \caption{Postselection probability $\pQSVT$ (filled symbols), ground-state
  generation probability $\pQSVT F$ (open symbols), and fidelity $F$ of the prepared
  state, for the thresholded pipeline with the folded (red circles) and \Neel{}
  (blue squares) references of the $2L=24$ spin-$\half$ Heisenberg chain
  ($\gamma_0^2=0.819$ and $0.0060$). Left column: versus the moment-noise
  width $\sigma$ at $D=30$. Right column: versus the Krylov dimension $D$ at
  $\sigma=10^{-5}$. Dashed lines mark each reference's ceiling $\gamma_0^2$.
  The postselection probability may exceed the ceiling (a successful postselection
  onto a poorly converged state); the generation probability never does
  \eqref{eq:prep-production}. Thresholded pipeline, $\epsilon=6\sigma\sqrt D$,
  $\lambda_{\mathrm{BE}}=3(2L-1)/4$; medians over noise realizations.}
  \label{fig:prep-ceiling}
\end{figure*}

Figure~\ref{fig:prep-ceiling} tests the identity. The left column scans the
noise. For the folded reference the fidelity stays above $0.93$ over the whole
range, so postselection and ground-state generation probabilities nearly coincide, and the ceiling is attained
only at strong noise ($\pQSVT F\approx0.8$ at $\sigma=10^{-2}$,
$\zeta_{\mathrm f}\approx1$); at small noise the generation probability sits at
$\sim4\times10^{-4}$ against the same ceiling---$\zeta_{\mathrm f}\lesssim
10^{-3}$, i.e.\ a filter whose weight is almost entirely spent on the
spectrally empty region, even though the state it delivers has fidelity
$0.99$. For the \Neel{} reference the two rates separate
by the fidelity, which falls from $0.6$ to $0.06$ as the noise grows; the
generation probability reaches the ceiling near $\sigma=10^{-3}$ and then
\emph{decreases}, while the postselection probability continues to rise and crosses the
ceiling at $\sigma=10^{-2}$: once the moment noise becomes comparable to the
$\gamma_0^2$-scale ground-state signal, thresholding (correctly, given the
data) discards part of the ground component, and a successful postselection
increasingly delivers the wrong state. This is the state-preparation face of
the resolvability condition \eqref{eq:qsd-signal}; for a reference whose
overlap is below the threshold scale altogether (a random state,
$\gamma_0^2\approx2^{-2L}$) we find the generation probability collapsing by decades
while the postselection probability rises (not shown). The right column scans the
Krylov dimension at fixed noise and shows that the gauge slack is a
$\sigma$-effect, not a $D$-effect: the folded generation probability fluctuates between
$5\times10^{-3}$ and $0.15$ with no systematic trend from $D=4$ to $40$, one to
two decades below its ceiling, even as the fidelity is $\ge0.95$ throughout.
Increasing the Krylov dimension does not close the efficiency gap---the
clearest evidence that the certified filter or the $\ell_1$ gauge fixing is a
distinct, necessary ingredient rather than a by-product of convergence.
For the \Neel{} reference the fidelity improves steadily with $D$ (from
$0.08$ to $0.5$), the postselection probability remains just below the ceiling, and the
generation probability follows the fidelity. In the random-reference case (not shown)
the $D$-scan exposes a further competition, \emph{noise dilution}: the
ground-state information carried by the higher moments is of size
$\sim\gamma_0^2$, and once it falls below the threshold $\epsilon\propto
\sigma\sqrt D$ additional dimensions are detrimental, so that the generation probability
peaks at a state-dependent $D^\ast(\sigma)$ and then degrades. The Krylov
dimension is thus a resource to be tuned against $\sigma$ and $\gamma_0^2$,
not increased indefinitely.

\begin{figure*}[t]
  \centering
  \includegraphics[width=0.48\textwidth]{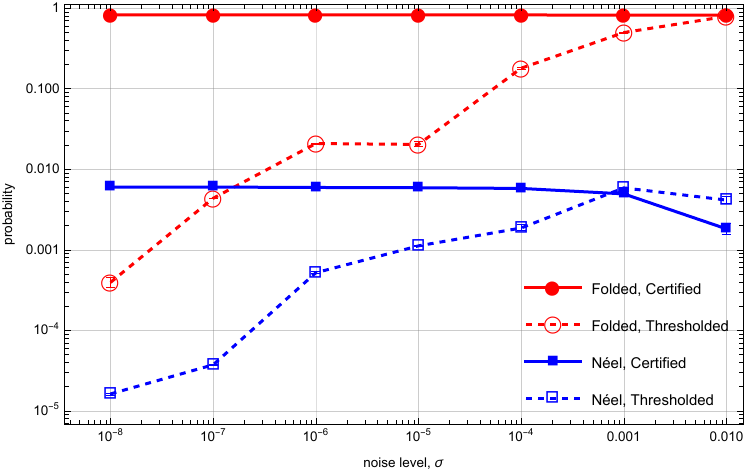}\hfill
  \includegraphics[width=0.48\textwidth]{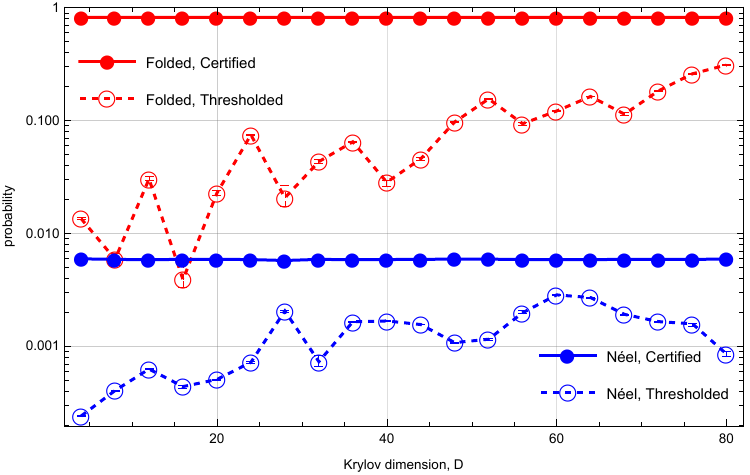}
  \caption{Certified filtering realized numerically for the $2L=24$
  spin-$\half$ Heisenberg chain of Figures~\ref{fig:prep-folded}
  and~\ref{fig:prep-ceiling} ($\gamma_0^2=0.819$ and $0.0060$): the
  closed-form eigenstate filter $R_\ell$ built solely from the two lowest
  \emph{thresholded} Ritz values (center $a=\tilde x_0$, half-width
  $(\tilde x_1-\tilde x_0)/2$), at the same degree budget
  $\ell=\lfloor(D-1)/2\rfloor$ as the Krylov pipelines. Left: generation
  probability versus moment noise at $D=30$; the certified pipeline (solid)
  remains at the ceilings $\gamma_0^2$ for both references over the entire
  noise range, except for the \Neel{} reference at the largest noise, where
  the thresholded Ritz values that define the filter are themselves
  inaccurate, while the thresholded Ritz-coefficient pipeline (dashed) falls
  short at small noise by more than three decades for the folded reference and
  by two and a half for the \Neel{} reference. Right: generation probability
  versus Krylov dimension $D$ at $\sigma=10^{-5}$; the certified pipeline sits
  at the ceiling from the smallest $D$ shown, whereas the thresholded pipeline
  approaches it only slowly for the folded reference, reaching $0.3$ at
  $D=80$, and remains well below it for the \Neel{} reference over the whole
  range. Medians over 30 noise realizations; error bars, where visible, mark
  the interquartile range.}
  \label{fig:prep-certified}
\end{figure*}

\paragraph{The certified filter.}
Logically, \eqref{eq:prep-production} bounds the generation probability only from
above; a matching lower bound $\zeta_{\mathrm f}\ge\mathrm{const}>0$ is needed for the
scaling of $\gamma_0^2$ to control the scaling of the cost, and the thresholded
pipeline alone provides none. The lower bound is supplied constructively:
nothing forces the preparation polynomial to be the noisy Ritz-coefficient
polynomial. QSD's role can be limited to supplying \emph{spectral data}---a
point $\mu$ inside the gap and a width $\hat\Delta$, obtained from the two
lowest thresholded Ritz values in rescaled units,
$\tilde x_k:=\widetilde E_k/\lambda$---and the preparation polynomial is then built
classically with certified properties: (i) $\norm{p}_\infty\le1$, (ii)
$p(x)\ge1-\epsilon$ for $x\le\mu-\hat\Delta/2$, (iii) $\abs{p(x)}\le\epsilon$
for $x\ge\mu+\hat\Delta/2$. The textbook shifted-Chebyshev amplifier
$T_D(\ell(x))/T_D(\ell(x_0))$ (see Appendix~\ref{app:onesided}) achieves the square-root acceleration
$D=\bigO(\hat\Delta^{-1/2}\log\tfrac1\epsilon)$ but violates (i) whenever there is
room below $x_0$: it is admissible only at the \emph{edge}. For an interior
target the correct object is a polynomial approximation to the step
$\tfrac12(1-\sgn(x-\mu))$---erf-based constructions \cite{Lin20a}, or, when a
sharp energy estimate $a\approx x_0$ is available, the eigenstate-filtering
polynomial of \cite{Lin20b},
\begin{equation}
  R_\ell(y;\Delta)=\frac{T_\ell\bigl(-1+2\frac{y^2-\Delta^2}{1-\Delta^2}\bigr)}
  {T_\ell\bigl(-1-2\frac{\Delta^2}{1-\Delta^2}\bigr)}
\end{equation}
with $y:=(x-a)/2$, which satisfies $R_\ell(0)=1$, $\abs{R_\ell}\le1$ on $[-1,1]$, and $\abs{R_\ell(y)}\le
2e^{-\sqrt2\ell\Delta}$ for $\Delta\le\abs y\le1$---at degree
$D=\Theta(\hat\Delta^{-1}\log\tfrac1\epsilon)$, optimal for interior targets
\cite{Eremenko07a}. Choosing $\epsilon=\gamma_0\delta$ with $\delta\le\half$,
conditions (i)--(iii) give $\zeta_{\mathrm f}\ge(1-\epsilon)^2\ge\tfrac14$ and
$F\ge1-4\delta^2$, hence
\begin{equation}
  \pQSVT\,F\ \ge\ \tfrac14\bigl(1-4\delta^2\bigr)\gamma_0^2
  \label{eq:prep-floor-certified}
\end{equation}
uniformly in system size, with the polynomial degree given by
\begin{equation}
  D=\bigO\!\left(\frac1{\hat\Delta}\log\frac1{\gamma_0\delta}\right).
 \end{equation} 
Together with the ceiling, the generation probability is
$\Theta(\gamma_0^2)$---\emph{provided the certified filter (or an equivalent
gauge-fixing step) is part of the protocol}. Figure~\ref{fig:prep-certified}
realizes the construction numerically: fed by the thresholded Ritz values, the
certified pipeline inherits the noise robustness of the energy and remains at the
ceiling across the entire noise range and already at the smallest Krylov
dimensions; the $D=\Theta(\hat\Delta^{-1})$ cost is paid instead in the
fidelity of the delivered state, which converges at the certified rate (not
shown). For folding, where only
the reference overlap of the next level matters, the certified filter is the
preparation tool of choice at small noise. Whenever the certified filter's state
lies in the $\delta$-ball of the Ritz vector it is feasible for the $\ell_1$
program \eqref{eq:prep-l1}, so the $\ell_1$ minimizer inherits the guarantee,
$\zeta_{\ell_1}\ge\zeta_{\mathrm{filter}}(1-\bigO(\delta))$: the empirical
saturation of the ceilings by the gauge-fixed curves is guaranteed by this bound.

\subsection{Scaling with system size: the folding ladder}
\label{sec:prep-ladder}

With the matching lower bound \eqref{eq:prep-floor-certified} in place, the
generation probability is $\Theta(\gamma_0^2)$ up to constants and polynomial degree
overhead, and its system-size scaling is entirely the scaling of
$\gamma_0^2(L)$---and here the two seeding strategies differ qualitatively.

\emph{Product-state seeding.} The overlap of a fixed product state with a
correlated ground state factorizes into per-site fidelity deficits and decays
exponentially, $\gamma_0^2\sim e^{-cL}$, an orthogonality catastrophe of the
trivial, extensive kind (Section~\ref{sec:fid-criteria}). The \Neel{} data of
Figure~\ref{fig:fold-overlap} calibrate $c\approx0.22$--$0.25$ for the
spin-$\half$ chain. Direct preparation at size $L$ then costs $e^{cL/2}$
amplification rounds, and the same exponential appears in the shot count needed
merely to resolve the $\gamma_0^2$-scale signal in the moments
\eqref{eq:qsd-signal}.

\emph{Folded seeding.} The reference $\fold$ differs from $\ket{g_{2L}}$ by a
single missing bond. The fold overlap, or \emph{fold fidelity},
\begin{equation}
  F(L)=\abs{\braket{g_{2L}|g_L\otimes g_L}}^2 ,
  \label{eq:prep-fold-fidelity}
\end{equation}
is studied in Section~\ref{sec:Fidelity}, whose
results we import: for the \emph{gapped} integer-$S$ chains, $F(L)$ approaches
a constant $F_\infty$ with $\bigO(e^{-L/\xi})$ corrections---$F_\infty=\tfrac14$
exactly for the AKLT chain (Appendix~\ref{app:AKLT}) and $F_\infty\approx0.22$
for the spin-$1$ Heisenberg chain; for the \emph{gapless}
spin-$\half$ chain the Anderson orthogonality catastrophe applies in its
mildest form, a single local scatterer, giving the slow power law
$F(L)\simeq b\,L^{-\alpha}$ with $\alpha\approx0.107$ (open chains). One
caveat for open gapped chains with edge modes: the fold overlap is diluted
across the degenerate edge sectors unless the edge states are matched (pinning
fields or symmetry resolution), so $F_\infty$ presumes sector-resolved folding
(Section~\ref{sec:fid-aklt}).

Preparing $\ket{g_{2L}}$ consumes \emph{two} prepared copies of $\ket{g_L}$, so
costs compound down the ladder $L_0\to2L_0\to\cdots\to2^kL_0$. Without
amplification the per-shot probabilities decrease doubly exponentially,
$P_k=F_kP_{k-1}^2$ with $F_k:=F(2^kL_0)$; with amplitude amplification at each level the expected
query count obeys the recursion
\begin{equation}
  \mathcal{N}(2L)\approx\frac1{\sqrt{F(2L)}}\bigl[\,2\mathcal{N}(L)+\bigO\bigl(D(2L)\bigr)\bigr],
  \label{eq:prep-recursion}
\end{equation}
whose growth is governed by $\prod_j2F_j^{-1/2}$. For a gapped chain
($F_j\to F_\infty$),
\begin{equation}
  \mathcal{N}(L)\sim\left(\frac L{L_0}\right)^{1+\frac12\log_2(1/F_\infty)},
  \label{eq:prep-ladder-gapped}
\end{equation}
\emph{polynomial} in $L$ with an exponent measured by the asymptotic fold
overlap: exactly $2$ for the AKLT chain ($F_\infty=\tfrac14$) and $\approx2.1$
for the spin-$1$ Heisenberg chain ($F_\infty\approx0.22$). For the gapless case
$F_j\sim(2^jL_0)^{-\alpha}$ inserts an extra factor $L^{(\alpha/4)\log_2L}$,
formally quasi-polynomial but with $\alpha/4\approx0.027$ indistinguishable
from polynomial at any realistic size, so that the overlaps alone contribute
$\mathcal{N}(L)\sim L^{1+(\alpha/4)\log_2L}$, with a prefactor set by $F(L)\gtrsim0.5$
over the whole practical range.
Folding thus converts an exponential state-preparation problem into an
(effectively) polynomial one, with the exponent supplied by the very overlaps
the QSD ladder measures. The secondary factors scale polynomially, with one
correction to the naive estimates: the extensive normalization
$\lambda_{\mathrm{BE}}\sim L$ keeps $x_0$ at a fixed \emph{interior} point, so
by Section~\ref{sec:prep-ceiling} the certified filter costs
$D\sim\hat\Delta^{-1}$, not $\hat\Delta^{-1/2}$; with
$\hat\Delta\sim\Delta/\lambda_{\mathrm{BE}}$ this gives $D\sim L\,\mathrm{polylog}$
for gapped chains ($\Delta=\bigO(1)$) and $D\sim L^2\,\mathrm{polylog}$ for gapless
ones ($\Delta\sim1/L$). In the recursion \eqref{eq:prep-recursion} the
$D$-term is subdominant for gapped chains, leaving the ladder exponent
unchanged, while for gapless chains the filter degree dominates and
$\mathcal{N}(L)\sim L^2\,\mathrm{polylog}$ times the mild quasi-polynomial overlap factor.
Either way the total remains polynomial---circuit depth, not ceiling. Two
assumptions are implicit and should be treated as part of the protocol: each
level's Ritz vector is gauge-fixed so that $\zeta_{\mathrm f}$ is bounded away from zero (or the
preparation polynomial is replaced by the certified filter) before it seeds the
next level, and, for gapped chains with edge modes, the folding is
sector-resolved.

\section{Initial Fidelity of the Folded State with the Ground State}
\label{sec:Fidelity}

Sections~\ref{sec:FoldedQSD} and~\ref{sec:StatePreparation} rest on a single
quantity: the overlap between the ground state $\ket{g_{2L}}$ of a length-$2L$
chain and the product $\fold$ of two half-chain ground states,
\begin{equation}
  F(L):=\abs{\braket{g_{2L}|g_L\otimes g_L}}^2=\abs{\gamma_0}^2 .
  \label{eq:fid-def}
\end{equation}
(We use the squared-overlap convention throughout, so that $F$ is the
probability that the folded state is found in the ground state; it is the
quantity that enters the bounds of Sections~\ref{sec:ChebyshevQSD}
and~\ref{sec:StatePreparation} directly.)
It sets the convergence prefactor and the noise floor of the folded QSD
(Section~\ref{sec:fold-state}), the ceiling of the generation probability and the
exponent of the folding ladder (Sections~\ref{sec:prep-ceiling}
and~\ref{sec:prep-ladder}). This section determines how $F(L)$ behaves with
$L$ for the three models of interest---the antiferromagnetic Heisenberg chain
for half-integer $S$ (gapless) and integer $S$ (gapped, Haldane phase), and the
AKLT chain---and explains the outcome in terms of the Anderson orthogonality
catastrophe. Throughout, $L$ is even (so that $2L$ is also even), ensuring that
the ground states are nondegenerate singlets except where edge modes of the
Haldane phase introduce a four-fold degeneracy on open chains. Both pure-state
and, for the AKLT chain, mixed-state fidelities are considered.

\subsection{Orthogonality-catastrophe criteria}
\label{sec:fid-criteria}

The numerical results below show that $F(L)$ either saturates or decays,
depending on the spectral-gap (more precisely, the quantum-critical)
properties of the model. The physical interpretation is the Anderson
orthogonality catastrophe (AOC) \cite{Anderson67a}: ground states of two
many-body Hamiltonians differing only by a \emph{local} perturbation can
become orthogonal in the thermodynamic limit. For a Fermi gas scattered by a
localized potential, Anderson found $\abs{\braket{\Psi_0|\Psi_0'}}\sim
N^{-\alpha}$ with $\alpha\propto\sum_\ell(2\ell+1)(\delta_\ell/\pi)^2$,
$\delta_\ell$ the Fermi-level phase shifts. The mechanism is transparent in
the mode picture: the overlap factorizes (exactly for quadratic models,
approximately in general) into a product over modes,
$\braket{\Psi_0|\Psi_0'}=\prod_k\cos\theta_k$, where $\theta_k$ measures how
much mode $k$ is rotated by the perturbation, so
$-\ln\abs{\braket{\Psi_0|\Psi_0'}}^2\approx\sum_k\theta_k^2$.
Everything depends on how this sum scales with $L$:
\begin{enumerate}
  \item $\sum_k\theta_k^2\to\mathrm{const}$: the overlap converges to a
    nonzero constant---no catastrophe;
  \item $\sum_k\theta_k^2\sim\ln L$: power-law decay---the Anderson
    catastrophe proper;
  \item $\sum_k\theta_k^2\sim L$: exponential decay---the trivial
    orthogonality of extensively different states.
\end{enumerate}
Three conditions must hold for the power-law scenario:
\begin{enumerate}
  \item \emph{Gaplessness.} There must be a continuum of excitations of
    arbitrarily low energy. A gap $\Delta>0$ cuts off the infrared modes:
    perturbation theory converges, $\sum_k\theta_k^2$ converges, and a local
    perturbation leaves the overlap finite (rigorously, via
    Lieb--Robinson-based quasi-adiabatic continuation for gapped phases
    \cite{Hastings06a}).
  \item \emph{Locality of the perturbation.} The perturbation must couple to
    $\bigO(1)$ degrees of freedom. An extensive perturbation gives exponential
    decay regardless of the gap, with extensive fidelity susceptibility
    $\chi_F\sim L$; the catastrophe is precisely the marginal case where
    $\chi_F$ diverges only logarithmically. This is the distinction between
    the \Neel{} and the folded reference states of Section~\ref{sec:FoldedQSD}.
  \item \emph{Nonvanishing coupling to the soft modes.} The infrared modes must
    actually feel the perturbation, giving rise to a nonzero phase shift (in
    field-theory terms, a boundary-condition-changing operator of nonzero
    scaling dimension). Symmetries can protect against this: if the
    perturbation commutes with the operators creating the soft modes,
    $\theta_k\equiv0$ for those modes and they do not contribute. The relevant
    gap may therefore differ from the gap of the full spectrum; see the
    spin-$\half$ Heisenberg chain below.
\end{enumerate}
The exponent is universal in the renormalization-group sense: it is fixed by
low-energy data only (phase shifts at the Fermi level; the central charge and
boundary-operator dimensions in conformal field theory). The same physics
underlies the X-ray edge singularity \cite{Nozieres69a}.

The overlap \eqref{eq:fid-def} is of the ``cut-and-glue'' type: the two states
are ground states of Hamiltonians differing by a single bond (two bonds for
periodic chains). Dubail and St\'ephan \cite{Dubail11a} studied the
\emph{logarithmic bipartite fidelity} $\mathcal F:=-\ln
\abs{\braket{g_{2L}|g_L\otimes g_L}}^2$, which is precisely $-\ln F(L)$ in our
notation, and found that it exhibits universal scaling at a quantum critical
point, $\mathcal F\simeq\frac c8\,g\ln L$ with $c$ the central charge and $g$
an $\bigO(1)$ geometry factor (aspect ratio, boundary conditions). Accordingly, the
three model classes realize the three scenarios:
\begin{itemize}
  \item \emph{Half-integer-$S$ Heisenberg (gapless, SU(2)$_1$, $c=1$):} the
    catastrophe occurs and $F$ decays as a universal power of $L$ with an
    exponent set by $c$.
  \item \emph{Integer-$S$ Heisenberg (Haldane phase, gapped):} no bulk
    catastrophe; the overlap converges exponentially to a nonzero constant. The
    constant is \emph{not} $1$: it is governed by the junction mismatch, where
    the dangling edge spin-$\half$'s of $\fold$ must be locked into the bond
    configuration of $\ket{g_{2L}}$, and it depends on which member of the
    quasi-degenerate open-chain ground space is used.
  \item \emph{AKLT:} the exactly solvable reference case---the exact matrix-product structure
    makes the exponential convergence and the junction constant explicit, and
    the fidelity-susceptibility bound $\chi_F=\bigO(1)$ holds uniformly in $L$.
\end{itemize}

\subsection{Numerical results}
\label{sec:fid-numerics}

All numerical results below were obtained by DMRG \cite{White92a,Schollwock11a}
with the TeNPy library \cite{Hauschild18a}. Half-chain bond dimensions ranged
from $\chi=16$ for $L\le12$ up to $\chi=320$ for $L=64$; energies are converged
to $10^{-9}$ or better in all reported cases. Overlaps were computed as inner
products of MPSs after concatenating two half-chain MPSs with a trivial bond
between them.

\subsubsection{Spin-\texorpdfstring{$\half$}{1/2} Heisenberg chain}
\label{sec:fid-half}

\begin{figure}[t]
\centering
\includegraphics[width=\linewidth]{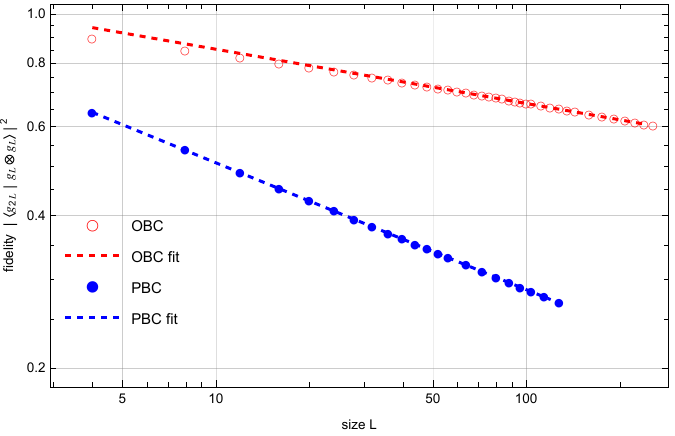}
\caption{Spin-$\half$ Heisenberg antiferromagnet: fold fidelity
$F(L)=\abs{\braket{g_{2L}|g_L\otimes g_L}}^2$ versus $L$ on a log-log scale
for open (OBC) and periodic (PBC) chains, with the power-law fits
\eqref{eq:fid-half-obc} and \eqref{eq:fid-half-pbc} (dashed). The two-cut
periodic geometry decays with a larger exponent, but not twice the open-chain
one.}
\label{fig:fid-half}
\end{figure}

\paragraph{Open boundary conditions.}
With open boundary conditions (OBC), both the full chain ($2L$ sites) and the
two halves ($L$ sites) have unique singlet ground states for even $L$: the
``one cut, no edges'' geometry. We computed $F(L)$ for $L=4$ to $256$ (full
chains up to $512$ sites); the data and the power-law fit are shown in
Figure~\ref{fig:fid-half}. The form that describes all data with $L\ge48$
at close to numerical precision is a pure power law with a small exponent,
\begin{equation}
  F(L)\approx b\,L^{-p},\qquad b=1.09,\quad p=0.107,
  \label{eq:fid-half-obc}
\end{equation}
with root-mean-square residual below $10^{-3}$, an order of magnitude better
than $1/\log L$ and $1/\sqrt L$ fits (not shown). Hence $F(L)\to0$ as
$L\to\infty$, but the decay is exceedingly slow: extrapolating
\eqref{eq:fid-half-obc} gives $F(10^3)\approx0.52$ and $F(10^4)\approx0.41$,
so that the folded overlap $\abs{\gamma_0}^2=F$ of Section~\ref{sec:FoldedQSD}
exceeds $0.5$ up to $L\sim10^3$. With $p$ this small the data cannot strictly
exclude a nonzero $F_\infty$ up to $\sim0.1$, and multiplicative logarithmic
corrections from the marginal SU(2) operator \cite{Affleck89a} are possible on
top of \eqref{eq:fid-half-obc}; the fitted $p$ is therefore an effective
exponent over $L\in[48,256]$.

The slow power-law decay is the Anderson orthogonality catastrophe of
Section~\ref{sec:fid-criteria}: all three conditions are met (gapless bulk,
single-bond perturbation, nonvanishing coupling to the gapless singlet modes),
placing the system in the marginal $\sum_k\theta_k^2\sim\ln L$ scenario. In the
Dubail--St\'ephan language the logarithmic bipartite fidelity grows as
$\mathcal F=p\ln L\approx0.107\ln L$, of the same order as the universal scale
$c/8=0.125$ for $c=1$ (the $\bigO(1)$ geometry factor of the OBC--OBC cut and the
SU(2)$_1$ logarithmic corrections account for the difference). Even though the
cut is an RG-irrelevant \emph{defect}, so that local observables recover their bulk values away from it, the
many-body overlap still decays as $L^{-p}$ with $p$ set by boundary scaling
dimensions.

The relevant gap here is that of the total-spin-resolved spectrum, not of the
full spectrum (condition~3 of Section~\ref{sec:fid-criteria}). Both $\fold$
and $\ket{g_{2L}}$ are total-spin singlets, and the bond perturbation
$\Svec_L\cdot\Svec_{L+1}$ is SU(2)-invariant, so the entire overlap problem
lives inside the $\Stot=0$ sector; the gapless \emph{triplet} (des
Cloizeaux--Pearson) excitations are irrelevant to it. Like the triplet gap,
the singlet-sector gap closes as $1/L$, but it is larger by several factors,
consistent with boundary conformal field theory. This is the fidelity-side
counterpart of the observation in Section~\ref{sec:fold-state} that the folded
state, being an exact singlet, is filtered through the singlet-sector gap
rather than through the smallest gap of the full spectrum.

\paragraph{Periodic boundary conditions.}
With periodic boundary conditions (PBC), both halves and the full chain are
rings with unique singlet ground states: the ``two cuts, no edges
anywhere'' geometry, free of boundary effects. The data and fit are shown in
Figure~\ref{fig:fid-half}. Again a pure power law describes all data at
close to numerical precision,
\begin{equation}
  F(L)\approx b\,L^{-p},\qquad b=0.90,\quad p=0.25,
  \label{eq:fid-half-pbc}
\end{equation}
with root-mean-square residual below $10^{-3}$. The exponent is considerably
larger than the OBC value $p=0.107$, indicating a faster decay with two cuts.
The two cuts are not independent: $F^{\mathrm{PBC}}(L)\ne\bigl(F^{\mathrm{OBC}}(L)\bigr)^2$,
and the exponent is not doubled.

\subsubsection{Spin-1 Heisenberg chain (Haldane phase)}
\label{sec:fid-haldane}

The spin-$1$ Heisenberg chain on a ring has a unique singlet ground state
\cite{White93a} with bulk Haldane gap $\Delta\approx0.41J$ and correlation
length $\xi\approx6$. On an open chain it has the four-fold near-degenerate
edge manifold \cite{Kennedy90a}; the global ground state is the $\Stot=0$
singlet (Lieb--Mattis), which we isolate by the $\Stot^2$ projection described
for AKLT below. We compare two geometries: ``OBC'' is $\abs{\braket{s_{2L}|s_L
\otimes s_L}}^2$ with open boundaries throughout (genuine singlets), and
``PBC'' is $\abs{\braket{g^{\mathrm{PBC}}_{2L}|g^{\mathrm{PBC}}_L\otimes
g^{\mathrm{PBC}}_L}}^2$.

\begin{figure}[t]
\centering
\includegraphics[width=\linewidth]{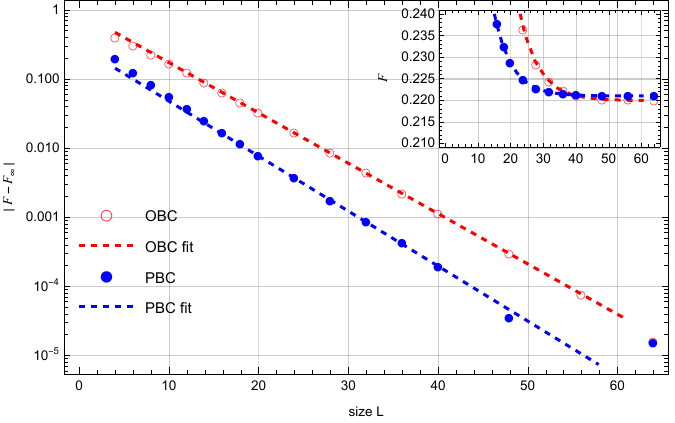}
\caption{Spin-$1$ Heisenberg chain (Haldane phase), OBC vs.\ PBC. Main panel:
finite-size deviations $\abs{F(L)-F_\infty}$ of the fold fidelity on a
logarithmic scale, with the asymptotic values \eqref{eq:fid-haldane-Finf}
obtained independently from exponential fits (dashed). The approximately
linear behavior is consistent with exponentially decaying finite-size
corrections. Inset: $F(L)$ itself for the two geometries, converging toward
values near $0.22$. Unlike the exactly solvable AKLT case, the OBC and PBC
fidelities do not coincide at finite $L$.}
\label{fig:fid-haldane}
\end{figure}

Both geometries are well described by $F(L)=F_\infty+b\,e^{-L/\xi}$
(Figure~\ref{fig:fid-haldane}). Independent fits give
\begin{equation}
  F_\infty^{\mathrm{OBC}}\simeq0.220,\qquad F_\infty^{\mathrm{PBC}}\simeq0.221,
  \label{eq:fid-haldane-Finf}
\end{equation}
with finite-size decay scales $\xi_{\mathrm{OBC}}\simeq6.5$ and
$\xi_{\mathrm{PBC}}\simeq5.2$, consistent with the bulk correlation length.
Despite their different finite-size behavior, both geometries show clear
saturation near $F_\infty\simeq0.22$, a folded overlap that is strictly
independent of $L$ (this is the value used for the ladder exponent in
Section~\ref{sec:prep-ladder}). Both asymptotes lie close to the AKLT value
$\tfrac14$, but the spin-$1$ Heisenberg ground state is not exactly a
valence-bond solid, so its boundary recoupling differs from the exact AKLT
$\tfrac14$ by an $\bigO(1)$ amount that survives the thermodynamic limit. Contrast this with AKLT (Figure~\ref{fig:fid-aklt}), where
OBC and PBC give numerically identical fidelities at every $L$.

\subsubsection{Spin-1 AKLT chain}
\label{sec:fid-aklt}

For clearer insight into the $S=1$ Heisenberg chain we consider the AKLT chain
\cite{Affleck87a,Affleck88a}, which is exactly solvable; the closed-form
fidelity is derived in Appendix~\ref{app:AKLT}. The AKLT model has a four-fold
\emph{exactly} degenerate ground manifold on each open chain (singlet $+$
triplet of the two edge spin-$\half$'s, all at $E=0$), so ``the lowest state
in the $S^z=0$ sector'' is not well defined: that sector contains two
degenerate ground states, the genuine total-spin singlet ($\Stot=0$) and the
$S^z=0$ component of the edge triplet ($\Stot=1$). To obtain a well-defined
pure-state fidelity we isolate the genuine singlet on each chain by computing
the $2\times2$ matrix of $\Stot^2$ within the $S^z=0$ ground space and
selecting the eigenvector with eigenvalue $0$ (numerically the two eigenvalues
come out as $\{0,2\}$ and $\bra s\Stot^2\ket s\sim10^{-15}$). This is the
``sector-resolved folding'' assumed in Section~\ref{sec:prep-ladder}.

\begin{figure*}[t]
\centering
\includegraphics[width=0.48\textwidth]{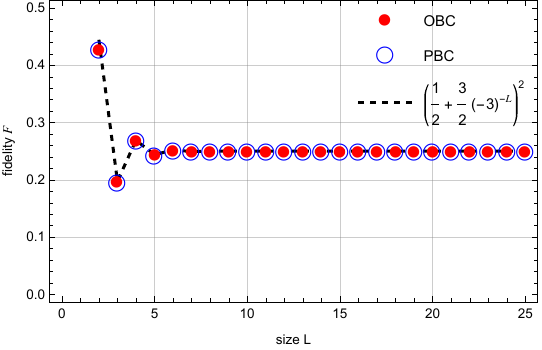}\hfill
\includegraphics[width=0.48\textwidth]{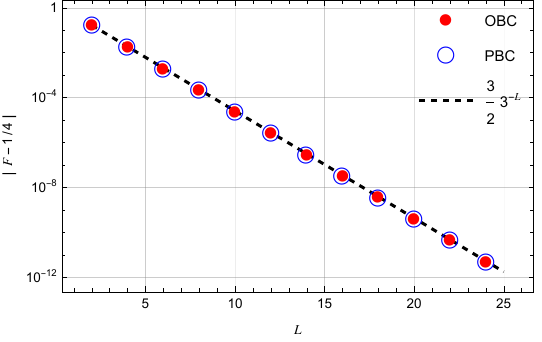}
\caption{Spin-$1$ AKLT chain, OBC vs.\ PBC. ``OBC'' is the genuine
singlet--singlet fidelity $\abs{\braket{s_{2L}|s_L\otimes s_L}}^2$ with open
boundaries throughout; ``PBC'' is $\abs{\braket{g^{\mathrm{PBC}}_{2L}|
g^{\mathrm{PBC}}_L\otimes g^{\mathrm{PBC}}_L}}^2$. Left: the fidelity $F(L)$
approaching $\tfrac14$ from above for even $L$ and from below for odd $L$,
with the squared two-term asymptotics
$\bigl(\tfrac12+\tfrac32(-\tfrac13)^L\bigr)^2$ (dashed). Right: the deviation
$\abs{F-\tfrac14}$ on a logarithmic scale, falling on the leading term
$\tfrac32\,3^{-L}$ of \eqref{eq:fid-aklt-exact} over more than ten orders of
magnitude, i.e.\ decaying at the AKLT correlation length $\xi=1/\ln3$. The
OBC and PBC data \emph{coincide at every $L$} (to all digits computed),
because the exact VBS structure makes the boundary recoupling identical in the
two geometries (Appendix~\ref{app:AKLT-fidelity}).}
\label{fig:fid-aklt}
\end{figure*}

The singlet--singlet fidelity converges \emph{exponentially} to $\tfrac14$ (not
$\tfrac12$) at the AKLT correlation length $\xi=1/\ln3\approx0.91$
(Figure~\ref{fig:fid-aklt}), in agreement with the exact result
\begin{equation}
  F(L)=\frac{\bigl(1+3(-\tfrac13)^L\bigr)^2}{4\bigl(1+3(\tfrac19)^L\bigr)}
      =\tfrac14+\tfrac32\bigl(-\tfrac13\bigr)^L+\tfrac32\bigl(\tfrac19\bigr)^L+\cdots
  \label{eq:fid-aklt-exact}
\end{equation}
of Appendix~\ref{app:AKLT}. With PBC, both halves and the full chain are rings
with unique ground states at $E=0$, and the fidelity is \emph{identical} to the
OBC singlet--singlet value at every $L$: the exact valence-bond structure makes
the boundary recoupling the same in the two geometries (the singlet ground
state of the open chain is the trace closure of the MPS, i.e.\ the ring state).
The limit $\tfrac14$ is the square of a recoupling amplitude $\tfrac12$:
gluing the two halves requires re-pairing the valence bonds at \emph{two}
points (the cut bond and the outer edges, or the two cut bonds of the rings),
each contributing $1/\sqrt2$ to the amplitude. The
energy cost of the cut is likewise $L$-independent: with the AKLT Hamiltonian
written as $H=\sum_i\bigl[\Svec_i\cdot\Svec_{i+1}+\tfrac13(\Svec_i\cdot
\Svec_{i+1})^2+\tfrac23\bigr]=2\sum_iP^{(2)}_{i,i+1}$, each cut bond joins two
uncorrelated, maximally mixed spin-$1$'s for which $\ev{P^{(2)}}=5/9$, so
$\ev{H_{2L}}_{\mathrm{fold}}-E_0=10/9$ for the single cut of the open chain and
$\eps=20/9\approx2.22$ for the two cuts of the rings.

The \emph{mixed-state} fidelity
$F(\rho_L\otimes\rho_L,\rho_{2L})=\bigl(\Tr\sqrt{\sqrt{\rho}\,\sigma\sqrt{\rho}}\bigr)^2$,
with $\rho_N$ the maximally mixed state on the four-dimensional ground manifold
of length $N$, equals $\tfrac14$ \emph{exactly} for every $L$ tested. Thus for
sector-unresolved folding the weight on the ground manifold is the same
$\tfrac14$, but it is spread over the four degenerate sectors of
the doubled chain; only the singlet component seeds the singlet ground state of
the next level, which is why Section~\ref{sec:prep-ladder} presumes
sector-resolved folding for gapped chains with edge modes.

\subsection{Consequences for folding}
\label{sec:fid-consequences}

Table~\ref{tab:fid-summary} collects the asymptotics. They explain the two
empirical facts on which Sections~\ref{sec:FoldedQSD}
and~\ref{sec:StatePreparation} rest.

\begin{table*}[t]
\centering
\small
\begin{tabular}{llll}
\toprule
Model & Spectrum & $F(L)=\abs{\braket{g_{2L}|g_L\otimes g_L}}^2$ & $F_\infty$ \\
\midrule
$S=\half$ Heisenberg, OBC & gapless, $c=1$ & $1.09\,L^{-0.107}$ & $0$ \\
$S=\half$ Heisenberg, PBC & gapless, $c=1$ & $0.90\,L^{-0.25}$ & $0$ \\
$S=1$ Heisenberg, OBC/PBC & gapped, $\Delta\approx0.41J$ & $0.22+b\,e^{-L/\xi}$, $\xi\approx5$--$6.5$ & $\approx0.22$ \\
$S=1$ AKLT, OBC/PBC & gapped, $\xi=1/\ln3$ & $\frac{(1+3(-1/3)^L)^2}{4(1+3\cdot9^{-L})}$ & $\tfrac14$ (exact) \\
\bottomrule
\end{tabular}
\caption{Asymptotic behavior of the fold fidelity for the three model classes;
$F$ is the folded overlap $\abs{\gamma_0}^2$ entering
Sections~\ref{sec:FoldedQSD} and~\ref{sec:StatePreparation}.}
\label{tab:fid-summary}
\end{table*}

First, the folded reference state of Section~\ref{sec:FoldedQSD} has an
$\bigO(1)$ ground-state overlap because the folding perturbation is a single bond:
for gapped chains condition~1 of the AOC fails and the overlap saturates, at a
junction-recoupling value ($\tfrac14$ for AKLT, $\approx0.22$ for the spin-$1$
Heisenberg chain) that is smaller than the finite-size values in
Figure~\ref{fig:fold-overlap} but strictly independent of $L$; for the gapless
spin-$\half$ chain the catastrophe does occur, but in its mildest, single-cut
form, with an exponent $p\approx0.1$ so small that $\abs{\gamma_0}^2$ stays
above $0.5$ for $L\lesssim10^3$. Any extensive perturbation---the difference
between the \Neel{} state and the ground state---puts the system in the
exponential scenario instead, which is the exponential decay of the \Neel{}
curves in Figure~\ref{fig:fold-overlap}. Second, the folding ladder of
Section~\ref{sec:prep-ladder} has a polynomial cost because $\prod_jF_j^{-1/2}$
grows polynomially: the factor $2$ per level contributes $L/L_0$, and the
overlaps contribute $F_\infty^{-k/2}=(L/L_0)^{\frac12\log_2(1/F_\infty)}$ for
gapped chains and the quasi-polynomial $L^{(p/4)\log_2L}$ for the gapless
chain. The exponent of
the ladder is thus fixed by boundary recoupling (gapped) or by boundary scaling
dimensions (gapless), both universal low-energy data, and it is measured by
the very overlaps that the QSD ladder produces.

\section{Conclusion}
\label{sec:Conclusion}

Every guarantee of quantum subspace diagonalization is a statement about the
overlap of the reference state with the ground state, and for any fixed
product state that overlap decays exponentially with the size of the system.
Quantum folding replaces the product state by the product of the ground
states of the subsystems, which is the exact ground state of the full
Hamiltonian with its interface terms removed, so that the overlap is decided
by the physics of the interface rather than by the bulk. For a chain cut into
two halves the interface is a single bond, and we have shown that the
resulting overlap is size-independent for gapped chains, decays only as a slow
power law for gapless ones, and in either case turns the scaling of the QSD
from exponential to polynomial in the system size.

Four results support this conclusion. The Chebyshev QSD admits an error bound
linear in the measurement noise (Theorem~\ref{thm:general}), obtained by
transplanting the effective-Hamiltonian analysis of \cite{Kirby24a} to the
Chebyshev--Krylov basis; the one new ingredient is a Chebyshev low-energy
filter whose coefficient vector has $\bigO(1)$ norm, and the analysis exposes
the block-encoding rescaling of the Hamiltonian as a parameter of the
guarantee, with an edge-aligned convergence rate in $\sqrt{\Delta}$
(Theorem~\ref{thm:edge}) that a fixed block-encoding normalization cannot
reach. In exact simulations of open Heisenberg chains the folded reference
reduces the Krylov dimension needed for a given accuracy severalfold and
lowers the error under noise by two to five orders of magnitude relative to
the \Neel{} state, an advantage that grows with the system size. The Ritz
vector can be turned back into a physical state at the depth of the deepest
QSD measurement circuit, the thresholding that stabilizes the energy also
floors the postselection probability, and the ground-state generation probability
obeys the ceiling $pF\le\abs{\gamma_0}^2$, saturated by a certified filter;
the folding ladder $L_0\to2L_0\to\cdots$ therefore prepares $\ket{g_L}$ at a
cost polynomial in $L$, with an exponent fixed by the asymptotic fold
overlap. Finally, the fold fidelity itself follows the Anderson orthogonality
catastrophe in its mildest, single-cut form: $F\sim L^{-0.107}$ for the
gapless spin-$\half$ chain, $F_\infty\simeq0.22$ for the spin-$1$ Heisenberg
chain, and $F\to\tfrac14$ exactly for the AKLT chain, where the limit is the
square of a valence-bond recoupling amplitude.

Three practical lessons deserve emphasis because they apply to any
Chebyshev-based QSD, folded or not. First, the normalization of the block
encoding is not a mere convention: it fixes where the ground state sits
on the Chebyshev interval, and an interior placement costs a factor
$\sqrt{\Omega/\Delta}$ in resolving power and a factor $\hat\Delta^{-1/2}$ in
the degree of any filter, which no post-processing recovers
(Appendix~\ref{app:RescalingFactor}). Second, the coefficient vector returned
by the eigensolver is defined only up to a gauge in the thresholded near-null
space, and a successful postselection onto the wrong state is of no use; gauge
fixing or the certified filter must be part of the protocol, not an
afterthought. Third, for gapped systems with edge modes the folding must be
sector-resolved, since the fold weight on the ground manifold is otherwise
diluted over degenerate edge sectors.

Several directions follow naturally. Nothing in the construction is specific
to one dimension, to two parts, or to spins (Section~\ref{sec:fold-general}),
but in $d>1$ dimensions the interface of a bipartition carries $L^{d-1}$
terms, so the folded overlap decays in the area rather than the volume; whether
this residual decay can be removed---by folding into many small parts, by
preparing the interface adiabatically, or by seeding with a state that already
carries the interface entanglement---is the question that decides the reach of
the method beyond chains. Fermionic lattice models and molecular fragments are
covered by the same argument and connect folding to the fragment-based
strategies of quantum chemistry. On the analytical side, the fold fidelity of
gapped systems should admit a rigorous $L$-independent lower bound through
Lieb--Robinson-based quasi-adiabatic continuation \cite{Hastings06a}, and the
logarithmic corrections expected for the SU(2)$_1$ chain \cite{Affleck89a}
remain to be resolved numerically. Excited states lie in the interior of the
Chebyshev interval for every normalization, so the two-sided filter of
Theorem~\ref{thm:general} is the relevant tool there, and the folded reference
can be adapted by folding excited states of the parts. Finally, our noise
model---Gaussian noise on the moments---captures shot noise but not coherent
errors in the walk operator; establishing how the linear-in-noise bound and
the folded-state advantage survive under realistic device noise is the natural
next step toward an implementation.

Although we have focused on Chebyshev-based QSD, the need for a ``good'' reference state, i.e., one with sufficiently large initial overlap, is not specific to this method. Similar overlap dependence appears in other ground-state algorithms, including state preparation based on quantum phase estimation \cite{Abrams99a,Lee25a} and eigenstate-filtering approaches based on quantum signal processing or the quantum singular value transformation \cite{Lin20a,Lin20b}, in which the success probability or the resource cost is likewise governed by the initial overlap. Since quantum folding is fundamentally a strategy for constructing a reference state whose difference from the target ground state is confined to the interfaces between subsystems, rather than a procedure specific to the quantum Krylov method, the same idea may also be useful as an initialization strategy for these algorithms. Exploring such extensions may therefore provide a broader route toward scalable quantum ground-state preparation beyond Chebyshev-based QSD.

\begin{acknowledgments}
S.P.\ and K.B.\ were supported by the Yonsei University Research Fund of 2026-22-0212.
K.B.\ was supported by the National Research Foundation of Korea (Grant Nos.~RS-2025-18362970, RS-2026-25519864, RS-2026-25613219), the Korean ARPA-H Project through the Korea Health Industry Development Institute (KHIDI) funded by the Ministry of Health \& Welfare, Korea (RS-2025-25456722), and the Ministry of Trade, Industry, and Energy (MOTIE), Korea, under the project ``Industrial Technology Infrastructure Program'' (RS-2024-00466693). 
D.C.\ acknowledges support from the STU Scientific Research Initiation Grant (Grant No.~NTF25003T).
S.C.\ acknowledges support from the National Science Association
Funds (Grant No.~U2230402) and the National Natural Science Foundation of China (Grant No. 12675011).
M.-S.C.\ was supported by the National Research Foundation of Korea (Grant Nos.~RS-2023-NR068116, 2023R1A2C1005588, RS-2024-00432563, RS-2024-00404854).
\end{acknowledgments}

\appendix
\section{A linear-in-noise error bound for the Chebyshev polynomial-based quantum subspace diagonalization}
\label{app:ChebyshevQSD}

This appendix gives the complete statements and proofs summarized in
Section~\ref{sec:ChebyshevQSD}. We combine the effective-Hamiltonian error
analysis of Kirby \cite{Kirby24a}, developed for quantum Krylov algorithms based
on real-time evolution, with the Krylov space generated by Chebyshev polynomials
of the Hamiltonian \cite{Kirby23a}. The result is a nonasymptotic upper bound on
the signed ground-state energy error of the thresholded method that is
\emph{linear} in the spectral-norm noise rates of the projected matrix pair and
converges exponentially in the Krylov dimension $D$ with rate
$(1+\delta/\Omega)^{-2\lfloor(D-1)/2\rfloor}$ for an arbitrary rescaling of the
Hamiltonian---in particular for the block-encoding normalization available in
practice, which places the ground state in the interior of the Chebyshev
interval (Theorem~\ref{thm:general})---and with the square-root-enhanced rate
$(1+2\sqrt{\delta/\Omega})^{-2(D-1)}$ when the ground state is aligned with
the lower edge (Theorem~\ref{thm:edge}).

\paragraph{What is new relative to \cite{Kirby24a}.}
The effective-Hamiltonian technique of \cite{Kirby24a} is agnostic to the
Krylov basis: it models noise in the projected pair exactly as an unperturbed
Rayleigh--Ritz problem for an effective Hamiltonian, and then needs only an
approximate ground state \emph{inside} the subspace, built as a polynomial
filter applied to the reference state, together with a bound on the
filter's coefficient vector in the basis actually measured. In the real-time
space that filter is a trigonometric polynomial in the phase $E\,dt$, its
coefficient norm is controlled by Fourier--Parseval, $\norm{c}\le1$, and the
phase is linear in the energy for every $E$---there is nothing to choose.
The only genuinely new technical ingredient here is therefore
Section~\ref{app:filter}: Chebyshev low-energy filters that (a) equal $1$ at
the ground energy, (b) are uniformly bounded on the low-energy window, (c) are
exponentially small above the window, and (d) have a coefficient vector of
$\bigO(1)$ Euclidean norm in the \emph{Chebyshev--Krylov} basis, by
Chebyshev--Parseval orthogonality. Property (d) is the polynomial counterpart
of the Fourier--Parseval bound; it holds for any polynomial of degree $<D$ that
is bounded on $[-1,1]$, and would fail in the monomial Krylov basis
$h^j\ket{\psi_0}$.

This ingredient is inherently tied to the block-encoding rescaling of the
Hamiltonian, in a way that has no analogue in the real-time setting. A block
encoding represents $h=(2H-(E_{\min}'+E_{\max}')\id)/\Omega$ with a
normalization fixed \emph{before} the computation, and the Chebyshev
polynomials $T_k(h)$ behave differently on the spectrally occupied part of
$[-1,1]$ and on the spectrally empty part: a filter must be small on the
former but only \emph{bounded} on the latter, and how large it gets there
decides the constant $B$ in property (d). If the normalization places the
ground state at the lower edge of $[-1,1]$ (edge slack $\nu=0$), no empty
interval lies below $x_0$, a one-sided filter using the full degree is
admissible, and the suppression rate is $\sqrt{\delta/\Omega}$ per Krylov step;
if the ground state lies in the interior---as it does for the LCU
normalization available in practice, Section~\ref{sec:fold-normalization}---the
one-sided filter grows exponentially on $[-1,x_0)$, and one must use the
two-sided filter of \cite{Kirby23a}, bounded by $1$ everywhere but with a
rate linear in $\delta/\Omega$ and half the degree. The angular picture of
Appendix~\ref{app:RescalingFactor} shows that this is not an artifact of the
proof: the resolving power of the measured moments themselves is governed by
the same placement. The rescaling is thus not a normalization convention but a
parameter of the guarantee, and the two theorems of this appendix are its two
values.

Everything else in Sections~\ref{app:effective}--\ref{app:main} is a
transcription of \cite{Kirby24a} with constants recomputed; we flag every
point of difference. For the reader's convenience the setting of
Section~\ref{sec:qsd-setup} is restated first.

\subsection{The Chebyshev--Krylov method}\label{app:setup}

\subsubsection{Rescaling and the Krylov basis}

Let $H$ be a Hermitian operator on an $N$-dimensional Hilbert space with eigenpairs $(E_k,\ket{E_k})$, $E_{\mathrm{min}} \equiv E_0 \leq E_1 \leq \cdots \leq E_{N-1} \equiv E_{\mathrm{max}}$, spectral gap $\Delta := E_1 - E_0$, and let
\begin{equation}
\ket{\psi_0} \;=\; \sum_{k=0}^{N-1}\gamma_k \ket{E_k},
\qquad \abs{\gamma_0}^2 > 0,
\end{equation}
be a normalized reference state. Suppose we are given real numbers $E_{\mathrm{min}}' \le E_{\mathrm{min}}$ and $E_{\mathrm{max}}' \ge E_{\mathrm{max}}$; set $\Omega := E_{\mathrm{max}}' - E_{\mathrm{min}}'$ and define the rescaled Hamiltonian
\begin{equation}\label{eq:rescale}
h \;:=\; \frac{2H - (E_{\mathrm{min}}'+E_{\mathrm{max}}')\,\id}{\Omega},
\end{equation}
whose eigenvalues
\begin{equation}
x_k \;:=\; \frac{2E_k - (E_{\mathrm{min}}'+E_{\mathrm{max}}')}{\Omega}
\end{equation}
lie in $[-1,1]$.
The ground eigenvalue sits at $x_0 = -1 + \nu$ with \emph{edge slack}
\begin{equation}\label{eq:slack}
\nu \;:=\; \frac{2(E_0 - E_{\mathrm{min}}')}{\Omega} \;\in\; [0,2),
\end{equation}
which measures the tightness of the lower bound $E_{\mathrm{min}}'$. The Chebyshev--Krylov basis of dimension $D$ is
\begin{equation}\label{eq:basis}
\ket{v_j} \;:=\; T_j(h)\,\ket{\psi_0}, \qquad j = 0,\dots,D-1,
\end{equation}
with $T_j$ the Chebyshev polynomial of the first kind. Since $\{T_0,\dots,T_{D-1}\}$ spans the polynomials of degree $\le D-1$, the span of \eqref{eq:basis} is the classical Krylov space $\mathcal{K}_D(H,\ket{\psi_0})$, and the method is Lanczos in a nonorthogonal basis \cite{Kirby23a}. Writing $\mV := [\,\ket{v_0}\;\cdots\;\ket{v_{D-1}}\,]$, the projected pair is
\begin{subequations}\label{eq:pair}
\begin{align}
\mH &:= \mV^\dagger H\, \mV, \\
\mS &:= \mV^\dagger \mV .
\end{align}
\end{subequations}
It is convenient to record the factorization, used repeatedly below,
\begin{equation}\label{eq:factorization}
\mV = \Psi\,\Gamma\,\Phi,
\end{equation}
where $\Psi := [\,\ket{E_0}\;\cdots\;\ket{E_{N-1}}\,]$ is the matrix of eigenvectors of $H$, $\Gamma := \operatorname{diag}(\gamma_k)$, and $\Phi_{kj} := T_j(x_k)$,
so that $\mS = \Phi^\dagger \Gamma^\dagger \Gamma\, \Phi$; the matrix $\Phi$ is a Chebyshev--Vandermonde matrix and plays the role of the phase matrix of \cite[Eq.~(101)]{Kirby24a}.

\subsubsection{Moment structure and quantum implementation}\label{app:moments}

The product formula $T_jT_k = \tfrac{1}{2}(T_{j+k}+T_{\abs{j-k}})$ and the recurrence $h\,T_k = \tfrac{1}{2}(T_{k+1}+T_{\abs{k-1}})$ reduce all matrix elements to the Chebyshev moments
\begin{equation}\label{eq:moments}
\mu_\ell := \bra{\psi_0} T_\ell(h)\ket{\psi_0}, \qquad \ell = 0,\dots,2D-1 .
\end{equation}
Explicitly, $\mS_{jk} = \tfrac{1}{2}\bigl(\mu_{j+k}+\mu_{\abs{j-k}}\bigr)$, and, for $j,k\ge 1$,
\begin{multline}\label{eq:Hmoments}
\mH_{jk} = \frac{\Omega}{8}\Bigl(\mu_{j+k+1}+\mu_{j+k-1}+\mu_{\abs{j-k}+1}+\mu_{\abs{\,\abs{j-k}-1\,}}\Bigr) \\
+ \frac{E_{\mathrm{min}}'+E_{\mathrm{max}}'}{2}\,\mS_{jk},
\end{multline}
with the edge cases $j=0$ or $k=0$ handled by the convention $T_{-1}:=T_1$. Thus $2D$ scalar estimands determine the entire pair, the analogue of the Toeplitz economy of the real-time method (there, entries depend on $j-k$ only; here, on $j+k$ and $\abs{j-k}$, a Toeplitz-plus-Hankel structure). On a quantum computer the moments \eqref{eq:moments} are estimated exactly, without Trotter error, from powers of the qubitization walk operator as described in Section~\ref{sec:qsd-implementation} \cite{Low19a,Kirby23a}.

\subsubsection{Noise model and regularization}\label{app:noise}

Let $(\mH',\mS')$ be Hermitian estimates of $(\mH,\mS)$, with spectral-norm errors
\begin{subequations}
\begin{align}
\eta_H &:= \norm{\mH'-\mH}, \\
\eta_S &:= \norm{\mS'-\mS},
\end{align}
\end{subequations}
and define the combined noise strength
\begin{equation}
\chi := \eta_H + \eta_S\norm{H} ,
\end{equation}
matching \cite[Eq.~(129)]{Kirby24a}. No assumption is made about the source of the errors (finite sampling, hardware noise, etc.). Given a threshold $\epsilon>0$, let $\Pi'$ be the orthogonal projector onto the eigenspaces of $\mS'$ with eigenvalues exceeding $\epsilon$, put
\begin{subequations}
\begin{align}
\mS'' &:= \Pi'\mS'\Pi', \\
\mH'' &:= \Pi'\mH'\Pi',
\end{align}
\end{subequations}
and let $\widetilde{E}_0$ be the least eigenvalue of the pair $(\mH'',\mS'')$ restricted to the range of $\Pi'$. This is the standard thresholding regularization \cite{Epperly22a,Kirby23a,Kirby24a}. If entrywise error bounds $\max_{jk}\abs{\Delta \mS_{jk}}\le \varsigma$ are more natural (as for shot noise on the moments), one may use $\eta_S \le D\varsigma$ and similarly for $\eta_H$.

Throughout, $\id$ denotes the identity and $\mathbbm{1}(\cdot)$ the indicator function of a condition.

\subsection{Effective Hamiltonian and effective Krylov space}\label{app:effective}

This subsection recapitulates the technique of \cite[Secs.~2 and 4.1]{Kirby24a}, which is agnostic to the choice of Krylov basis; we restate it in our notation, with one constant recomputed (Lemma~\ref{lem:effH}).

\paragraph{Two matrix-perturbation tools.}
The analysis below, and that of Sections~\ref{app:projector}--\ref{app:main},
uses only two classical facts about Hermitian perturbations, both
basis-independent \cite{Bhatia97a}. \emph{Weyl's inequality} controls
eigenvalues: if $A$ and $A+E$ are Hermitian with eigenvalues sorted in
increasing order, then $\abs{\lambda_k(A+E)-\lambda_k(A)}\le\norm{E}$ for every
$k$, with no dependence on gaps or on the dimension. It is used to bound the
shift of $E_0$ and of the gap under $H\to\heff$, and to locate the eigenvalues
of the noisy overlap matrix $\mS'$ on the thresholded-out subspace in
$[-\eta_S,\epsilon]$. The \emph{Davis--Kahan $\sin\Theta$ theorem}
\cite{Davis70a} controls eigenvectors: if $P$ is the spectral projector of $A$
onto eigenvalues in an interval $I$ and $\tilde Q$ the spectral projector of
$A+E$ onto eigenvalues at distance at least $\gamma>0$ from $I$, then
\begin{equation}\label{eq:daviskahan}
\norm{\tilde Q\,P}\;\le\;\frac{\norm{E}}{\gamma},
\end{equation}
i.e.\ a perturbation of size $\norm{E}$ tilts a spectral subspace by at most
an angle $\arcsin(\norm{E}/\gamma)$ away from a subspace that is $\gamma$ away
in the spectrum; unlike Weyl's inequality it requires a gap, and $1/\gamma$ is
exactly the sensitivity of eigenvectors to near-degeneracy. It is used with
$A=H$, $A+E=\heff$, $\gamma=\delta'-\delta$ to convert the filter's small weight
above $E_0+\delta$ in the spectrum of $H$ into small weight above $E_0+\delta'$
in the spectrum of $\heff$ (Theorem~\ref{thm:expectation}), and with
$A=\mS$, $A+E=\mS'$ in the trace bound on the thresholded directions
(Theorem~\ref{thm:projector}). Only the construction of the filter
(Section~\ref{app:filter}) depends on the choice of Krylov basis---trigonometric
polynomials for the real-time space of \cite{Kirby24a} or Chebyshev polynomials
here.

Fix any coordinate vector $c'\ne 0$ in the range of $\Pi'$, i.e.\ $c' = \Pi' c'$. Define the effective Krylov basis
\begin{subequations}\label{eq:effective}
\begin{equation}
\mVeff{} := \sqrt{\frac{c'^\dagger \mS' c'}{c'^\dagger \mS c'}}\;\mV,
\end{equation}
the state
\begin{equation}
\ket{\psi} := \mVeff{} c' ,
\end{equation}
and the effective Hamiltonian
\begin{equation}
\heff := H + \bigl(c'^\dagger \mH'' c' - \bra{\psi}H\ket{\psi}\bigr)\,
\frac{\ket{\psi}\bra{\psi}}{\braket{\psi|\psi}^2} .
\end{equation}
\end{subequations}
By construction \cite[Eqs.~(24)--(27)]{Kirby24a},
\begin{equation}\label{eq:matchpoint}
\frac{c'^\dagger \mH'' c'}{c'^\dagger \mS'' c'} \;=\; \frac{\bra{\psi}\heff\ket{\psi}}{\braket{\psi|\psi}} ,
\end{equation}
so the Rayleigh quotient of the noisy, thresholded pair at $c'$ equals an exact Rayleigh quotient of $\heff$ at $\ket{\psi}$; by the variational principle, $\widetilde{E}_0 \le \bra{\psi}\heff\ket{\psi}/\braket{\psi|\psi}$. Following \cite[Eq.~(28)]{Kirby24a},
\begin{equation}\label{eq:effHraw}
\norm{\heff - H} \;\le\; \frac{\norm{c'}^2\,\eta_H + \norm{c'}^2\,\norm{H}\,\eta_S}{\braket{\psi|\psi}}
\;=\; \frac{\norm{c'}^2}{\braket{\psi|\psi}}\;\chi .
\end{equation}
The ratio $\norm{c'}^2/\braket{\psi|\psi}$ is controlled by the norm bounds of Theorem~\ref{thm:projector} below; anticipating them:

\begin{lemma}\label{lem:effH}
Let $c'$ be the ansatz coordinates constructed in Theorem~\ref{thm:projector} from a filter with coefficient bound $\norm{c}^2 \le 2B^2$, and suppose $\abs{\gamma_0'}^2 := \abs{\gamma_0}^2 - 4B^2(\epsilon + \eta_S) > 0$. Then
\begin{equation}\label{eq:effH}
\norm{\heff - H} \;\le\; \frac{2B^2\chi}{\abs{\gamma_0'}^2} .
\end{equation}
\end{lemma}

\begin{proof}
By the first norm bound in \eqref{eq:normbounds} below, $\braket{\psi|\psi} \ge \norm{c'}^2\bigl(\abs{\gamma_0}^2 - 2B^2(\epsilon+\eta_S)\bigr)/(2B^2) \ge \norm{c'}^2\,\abs{\gamma_0'}^2/(2B^2)$; insert this into \eqref{eq:effHraw}.
\end{proof}

The only difference from \cite[Eq.~(135)]{Kirby24a} is the factor $2B^2$ (there equal to $1$), which originates in the coefficient bound $\norm{c}^2\le 2B^2$ of Lemma~\ref{lem:onesided} replacing the bound $\norm{c}\le 1$ available for the trigonometric filter.

\subsection{Chebyshev low-energy filters}\label{app:filter}

This subsection contains the basis-specific ingredient, replacing the trigonometric filter of \cite[Lemma~3.3]{Epperly22a} used in \cite[Thm.~2]{Kirby24a}. The rest of the analysis (Sections~\ref{app:projector}--\ref{app:main}) uses a filter only through the five properties collected in the following definition, so we first state the interface and then exhibit two filters that satisfy it: a \emph{two-sided} filter, valid for any placement of the ground state in $[-1,1]$, which yields the main theorem, and a \emph{one-sided} filter, valid only when the ground state is aligned with the lower edge, which upgrades the convergence exponent to a square root.

Fix the Krylov dimension $D\ge 3$ and an energy window $\delta>0$, and put
\begin{equation}
\tilde\delta := \frac{2\delta}{\Omega} ;
\end{equation}
we assume throughout that $x_0+\tilde\delta < 1$, i.e.\ that the window ends below the upper edge of the interval.

\begin{definition}[Admissible filter]\label{def:filter}
A real polynomial $q$ of degree at most $D-1$ is an \emph{admissible filter} with suppression $\varepsilon_D\in(0,1]$ and bound $B\ge1$ if
\begin{enumerate}[label=(F\arabic*),ref=(F\arabic*)]
\item\label{F1} $q(x_0)=1$;
\item\label{F2} $\abs{q(x_k)}\le\varepsilon_D$ for every eigenvalue $x_k\ge x_0+\tilde\delta$, i.e.\ $E_k-E_0\ge\delta$;
\item\label{F3} $\abs{q(x_k)}\le1$ for every eigenvalue $x_k\in[x_0,x_0+\tilde\delta)$;
\item\label{F4} $\max_{x\in[-1,1]}\abs{q(x)}\le B$;
\item\label{F5} its coefficient vector $c\in\mathbb R^D$ in the expansion $q=\sum_{j=0}^{D-1}c_jT_j$ satisfies $\norm{c}_2^2\le2B^2$.
\end{enumerate}
We write $\Xi_D:=2\varepsilon_D^2$ for the associated \emph{convergence factor}.
\end{definition}

Property \ref{F5} is a consequence of \ref{F4} for \emph{any} polynomial of degree $\le D-1$: with the Chebyshev weight $w(x)=(1-x^2)^{-1/2}$ and the orthogonality $\int_{-1}^1T_jT_k\,w\,dx=\pi\delta_{jk}(1+\delta_{j0})/2$,
\begin{equation}\label{eq:parseval}
\begin{split}
\frac{\pi}{2}\norm{c}_2^2&\;\le\;\pi c_0^2+\frac{\pi}{2}\sum_{j\ge1}c_j^2
\;=\;\int_{-1}^{1}q(x)^2\,w(x)\,dx \\
&\;\le\;B^2\int_{-1}^1w(x)\,dx\;=\;\pi B^2 .
\end{split}
\end{equation}
This Chebyshev--Parseval bound is the polynomial counterpart of the Fourier--Parseval bound $\norm{c}\le1$ for the trigonometric filter of \cite{Epperly22a,Kirby24a}, and it is special to the Chebyshev--Krylov basis: the same filter state $q(h)\ket{\psi_0}$ expanded in the monomial Krylov basis $h^j\ket{\psi_0}$ has exponentially large coefficients. The whole question is therefore how large $B$ must be, i.e.\ how large the filter gets on the spectrally \emph{empty} part of $[-1,1]$.

\subsubsection{The two-sided filter (any placement of the ground state)}\label{app:twosided}

Let $d:=\lfloor(D-1)/2\rfloor$ and let $p^*_d$ be the minimax residual polynomial of degree $d$ on $[\tilde\delta^2/4,1]$ \cite[Lemma~B.2]{Kirby23a},
\begin{equation}
  p^*_d(y):=\frac{T_d\bigl((1+\tilde\delta^2/4-2y)/(1-\tilde\delta^2/4)\bigr)}
                 {T_d\bigl((1+\tilde\delta^2/4)/(1-\tilde\delta^2/4)\bigr)}
  \label{eq:app-twosided}
\end{equation}
and
\begin{equation}
  q_{\mathrm{2s}}(x):=p^*_d\!\Bigl(\frac{(x-x_0)^2}{4}\Bigr).
\end{equation}
Squaring the distance to $x_0$ makes $q_{\mathrm{2s}}$ suppress the spectrum on \emph{both} sides of $x_0$, which is what allows it to be bounded wherever $x_0$ sits.

\begin{lemma}[Two-sided Chebyshev filter; cf.\ {\cite[App.~B]{Kirby23a}}]\label{lem:twosided}
For any $x_0\in[-1,1]$, $q_{\mathrm{2s}}$ is an admissible filter of degree $2d\le D-1$ with $B=1$ and
\begin{equation}
\begin{split}
\varepsilon_D^{\mathrm{2s}} &:= \frac{1}{T_d\bigl((1+\tilde\delta^2/4)/(1-\tilde\delta^2/4)\bigr)} \\
&\;\le\; 2\Bigl(1+\frac{\tilde\delta}{2}\Bigr)^{-\lfloor(D-1)/2\rfloor}
= 2\Bigl(1+\frac{\delta}{\Omega}\Bigr)^{-\lfloor(D-1)/2\rfloor},
\end{split}
\end{equation}
or equivalently,
\begin{equation}
\Xi_D^{\mathrm{2s}} \le 8\Bigl(1+\frac{\delta}{\Omega}\Bigr)^{-2\lfloor(D-1)/2\rfloor}.
\end{equation}
\end{lemma}

\begin{proof}
$p^*_d(0)=1$ gives \ref{F1}. For $y=(x-x_0)^2/4\in[\tilde\delta^2/4,1]$---which covers every $x\in[-1,1]$ with $\abs{x-x_0}\ge\tilde\delta$---the argument of the numerator $T_d$ lies in $[-1,1]$, so $\abs{p^*_d(y)}\le1/T_d\bigl((1+\tilde\delta^2/4)/(1-\tilde\delta^2/4)\bigr)=\varepsilon_D^{\mathrm{2s}}$; this is \ref{F2}, and $\varepsilon_D^{\mathrm{2s}}<1$ because the argument of the denominator exceeds $1$ and $d\ge1$. The closed-form bound follows from $T_d(\cosh t)\ge\tfrac12e^{dt}$, which gives $\varepsilon_D^{\mathrm{2s}}\le2\bigl((1+\tilde\delta/2)/(1-\tilde\delta/2)\bigr)^{-d}\le2(1+\tilde\delta/2)^{-d}$. For $y\in[0,\tilde\delta^2/4)$ the argument exceeds $1$ but does not exceed that of the denominator, and $T_d$ is increasing on $[1,\infty)$, so $\abs{p^*_d(y)}\le1$; since $(x-x_0)^2/4\in[0,1]$ for all $x,x_0\in[-1,1]$, this gives \ref{F3} and \ref{F4} with $B=1$. \ref{F5} is \eqref{eq:parseval}.
\end{proof}

\subsubsection{The one-sided filter (edge-aligned ground state)}\label{app:onesided}

When no spectrum lies below $x_0$, suppression is needed on one side only, and the full degree $D-1$ can be spent on it. Let $a:=x_0+\tilde\delta$ and let $f(x):=\bigl(2x-(1+a)\bigr)/(1-a)$ be the affine map with $f(a)=-1$, $f(1)=1$. Define
\begin{equation}\label{eq:filter}
q_{\mathrm{1s}}(x) \;:=\; \frac{T_{D-1}\bigl(f(x)\bigr)}{T_{D-1}\bigl(f(x_0)\bigr)}
\end{equation}
and
\begin{equation}
\gamma := \frac{\tilde\delta}{\,1-x_0-\tilde\delta\,}
= \frac{\tilde\delta}{\,2-\nu-\tilde\delta\,}
\;\ge\; \frac{\delta}{\Omega}
\end{equation}
so that $f(x_0) = -(1+2\gamma)$.

\begin{lemma}[One-sided Chebyshev filter]\label{lem:onesided}
The polynomial $q_{\mathrm{1s}}$ has degree $D-1$ and satisfies \ref{F1}--\ref{F3} of Definition~\ref{def:filter} with
\begin{equation}
\begin{split}
\varepsilon_D^{\mathrm{1s}} &:= \frac{1}{T_{D-1}(1+2\gamma)} \\
&\le 2\bigl(1+2\sqrt{\gamma}\bigr)^{-(D-1)}
\le 2\Bigl(1+2\sqrt{\delta/\Omega}\Bigr)^{-(D-1)}
\end{split}
\end{equation}
or equivalently
\begin{equation}
  \Xi_D^{\mathrm{1s}} \le 8\Bigl(1+2\sqrt{\delta/\Omega}\Bigr)^{-2(D-1)},
\end{equation}
and \ref{F4}--\ref{F5} with
\begin{equation}\label{eq:Bbound}
B \;\le\; \frac{T_{D-1}\bigl(1+2\gamma'\bigr)}{T_{D-1}\bigl(1+2\gamma\bigr)},
\end{equation}
where
\begin{equation}
\gamma' := \frac{\nu+\tilde\delta}{2-\nu-\tilde\delta} ;
\end{equation}
in particular $B=1$ if $\nu=0$.
\end{lemma}

\begin{proof}
\ref{F1} is immediate from $T_{D-1}(-y)=(-1)^{D-1}T_{D-1}(y)$. For \ref{F2}, $f$ maps $[a,1]$ onto $[-1,1]$, where $\abs{T_{D-1}}\le 1$; the closed-form bound follows from $T_n(1+2\gamma) = \cosh\bigl(n\operatorname{arccosh}(1+2\gamma)\bigr) \ge \tfrac{1}{2}\bigl(1+2\gamma+2\sqrt{\gamma+\gamma^2}\bigr)^{n} \ge \tfrac12(1+2\sqrt\gamma)^n$. For \ref{F3}, $f$ maps $[x_0,a]$ into $[-(1+2\gamma),-1]$, and $\abs{T_{D-1}(y)}$ is increasing in $\abs{y}$ for $\abs{y}\ge 1$, so the numerator never exceeds the denominator in magnitude. For \ref{F4}, $f(-1) = -(3+a)/(1-a)$ and a short computation gives $\abs{f(-1)} = 1+2\gamma'$ with $\gamma'$ as stated; monotonicity of $\abs{T_{D-1}}$ then bounds $\abs{q_{\mathrm{1s}}}$ on $[-1,x_0]$ by \eqref{eq:Bbound}, while on $[x_0,1]$ properties \ref{F2}--\ref{F3} give $\abs{q_{\mathrm{1s}}}\le1$. When $\nu=0$ the interval $[-1,x_0)$ is empty and $B=1$. \ref{F5} is \eqref{eq:parseval}.
\end{proof}

\begin{remark}[Edge slack]\label{rem:slack}
Property \ref{F4} is where the tightness of the lower spectral bound $E_{\mathrm{min}}'$ enters: the one-sided filter is uncontrolled on the spectrally empty interval $[-1,x_0)$, and its supremum there governs the coefficient norm. Using $\tfrac12 e^{nt}\le T_n(\cosh t)\le e^{nt}$ in \eqref{eq:Bbound},
\begin{equation}
B \;\le\; 2\exp\Bigl((D-1)\bigl[\operatorname{arccosh}(1+2\gamma') - \operatorname{arccosh}(1+2\gamma)\bigr]\Bigr).
\end{equation}
Furthermore, since $\frac{d}{du}\operatorname{arccosh}(1+2u) = \bigl(u(1+u)\bigr)^{-1/2}$ is decreasing while $\gamma'-\gamma = \nu/(2-\nu-\tilde\delta)=\nu/(1-a)$, with $a:=x_0+\tilde\delta<1$ the upper end of the energy window, 
the quantity in the bracket is at most $(\gamma'-\gamma)/\sqrt{\gamma}$, so that $B\le 2e$ whenever
\begin{equation}\label{eq:slackcondition}
\frac{(D-1)\nu}{(1-a)\sqrt{\gamma}} \;\le\; 1 .
\end{equation}
Since $\nu=2(E_0-E_{\mathrm{min}}')/\Omega$ and $\gamma\ge\delta/\Omega$, condition \eqref{eq:slackcondition} holds in particular when
\begin{equation}
  E_0 - E_{\mathrm{min}}' \;\le\; \frac{(1-a)\sqrt{\delta\,\Omega}}{2(D-1)} .
\end{equation}
(In the edge-aligned regime $\nu,\tilde\delta\ll1$ one has $1-a\approx2$ and $\gamma'-\gamma\approx\nu/2$; in general $1-a$ is an $\bigO(1)$ factor, since the window is assumed to end below the upper edge.) Thus the lower bound $E_{\mathrm{min}}'$ must locate $E_0$ to accuracy $\sim\sqrt{\delta\Omega}/D$. A priori bounds of this type are standard inputs to Chebyshev-filter and block-encoding ground-state algorithms \cite{Lin20a}, and the algorithm itself returns an upper estimate of $E_0$ that could refine them; but knowledge of $E_0$ is not the obstruction. The rescaled operator $h$ must be \emph{block encoded}, and a block encoding of $H$ built from an LCU decomposition places the spectrum in $[-1,1]$ with a normalization fixed by the decomposition, not by what one knows about $E_0$; shifting or rescaling $H$ before encoding inflates the LCU norm by the same amount, so the ground state never moves closer to the lower edge than it is at $\lambda_{\mathrm{BE}}$ (Appendix~\ref{app:rescale-noshift}). In the setting of a block encoding fixed in advance the slack is therefore $\nu=\Theta(1)$ regardless of any estimate of $E_0$ (Section~\ref{sec:fold-normalization}), $B$ grows exponentially in $D$, and the one-sided filter is useless. The two-sided filter of Lemma~\ref{lem:twosided} is the instrument for that case; the one-sided filter is the instrument for the edge-aligned case, which we formalize as:
\end{remark}

\begin{assumption}[Tight lower edge]\label{ass:edge}
$\nu = 0$, i.e.\ $E_{\mathrm{min}}' = E_0$, so that $B = 1$ in Lemma~\ref{lem:onesided}. (More generally, it suffices that \eqref{eq:slackcondition} hold, at the cost of a constant $B = \bigO(1)$.)
\end{assumption}

The two filters thus differ in two respects: the two-sided filter has $B=1$ unconditionally but spends two Krylov vectors per unit of degree and converges linearly in the gap, $\Xi_D^{\mathrm{2s}}\le8(1+\delta/\Omega)^{-2\lfloor(D-1)/2\rfloor}$; the one-sided filter uses the full degree and converges with the square root of the gap, $\Xi_D^{\mathrm{1s}}\le8(1+2\sqrt{\delta/\Omega})^{-2(D-1)}$, but is bounded only under Assumption~\ref{ass:edge}. In both lemmas the suppression $\varepsilon_D$ is the exact reciprocal of a Chebyshev polynomial evaluated outside $[-1,1]$, hence strictly less than $1$ as Definition~\ref{def:filter} requires; the closed-form exponentials are upper bounds on it, and they exceed $1$---i.e.\ carry no information---until $D-1\gtrsim(2\ln2)\,\Omega/\delta$ for the two-sided filter and $D-1\gtrsim\ln2\,\sqrt{\Omega/\delta}$ for the one-sided one. This is the regime in which the exponential convergence of the method sets in, and it is the only regime in which the bounds below are of interest. Since every statement below is monotone in $\varepsilon_D$, the closed forms may be substituted for the exact values throughout, which is how the convergence factors are written in Theorems~\ref{thm:general} and~\ref{thm:edge}. Everything below is stated for an arbitrary admissible filter with parameters $(\varepsilon_D,B)$ and then specialized to the two.

\subsection{An approximate ground state in the effective Krylov space}\label{app:projector}

We now prove the Chebyshev analogue of \cite[Thm.~2]{Kirby24a}. The proof follows \cite[App.~A]{Kirby24a} step by step; the differences are confined to (i) the substitution of an admissible filter (Definition~\ref{def:filter}) for the trigonometric filter and (ii) the propagation of the coefficient bound $\norm{c}^2\le 2B^2$ in place of $\norm{c}\le 1$.

\begin{theorem}[Approximate ground-state projector; cf.\ {\cite[Thm.~2]{Kirby24a}}]\label{thm:projector}
Instate the setup of Section~\ref{app:setup}, fix $\delta>0$ as in Section~\ref{app:filter}, and assume $\eta_S \le \epsilon$. Let $c$ be the coefficient vector of an admissible filter $q$ with parameters $(\varepsilon_D,B)$, and define the ansatz coordinates and state
\begin{subequations}\label{eq:ansatz}
\begin{align}
\tilde c &:= \Pi' c, \\
c' &:= \tilde c\,\sqrt{\frac{\tilde c^\dagger \mS \tilde c}{\tilde c^\dagger \mS' \tilde c}}, \\
\ket{\psi} &:= \mVeff{} c' = \mV \tilde c ,
\end{align}
\end{subequations}
with $\mVeff{}$ as in \eqref{eq:effective}. Then $\ket{\psi} = P\ket{\psi_0}$ for an operator $P$ diagonal in the eigenbasis of $H$, $P\ket{E_k} = \beta_k'\ket{E_k}$, with
\begin{equation}\label{eq:betabounds}
\abs{\beta_k'}^2 \;\le\;
\begin{cases}
\;2 + \alpha_k, & E_k - E_0 < \delta,\\[2pt]
\;\Xi_D + \alpha_k, & E_k - E_0 \ge \delta,
\end{cases}
\end{equation}
where $\Xi_D = 2\varepsilon_D^2$ and the $\alpha_k \ge 0$ satisfy
\begin{equation}\label{eq:zeta}
\sum_{k=0}^{N-1}\abs{\gamma_k}^2 \alpha_k \;\le\; \zeta \;:=\; 4B^2 D\,\bigl(\epsilon + \eta_S\bigr) .
\end{equation}
Moreover,
\begin{subequations}\label{eq:normbounds}
\begin{align}
\braket{\psi|\psi} &\;\ge\; \frac{\norm{c'}^2}{2B^2}\Bigl(\abs{\gamma_0}^2 - 2B^2(\epsilon+\eta_S)\Bigr), \\
\braket{\psi|\psi} &\;\ge\; \abs{\gamma_0}^2 - 4B^2(\epsilon+\eta_S) \;=:\; \abs{\gamma_0'}^2 .
\end{align}
\end{subequations}
\end{theorem}

\begin{proof}
\emph{Diagonal action and pointwise bounds.}
Using \eqref{eq:factorization} and $\ket{\psi}=\mV\tilde c$,
\begin{equation}
\ket{\psi} \;=\; \sum_{k=0}^{N-1}\gamma_k \Bigl(\textstyle\sum_{j}\tilde c_j\, T_j(x_k)\Bigr)\ket{E_k}
\;=\; \sum_k \gamma_k\,\beta_k'\,\ket{E_k},
\end{equation}
with $\beta_k' = \bigl[\Phi\,\tilde c\,\bigr]_k$, which defines $P$. Writing $\tilde c = c + (\Pi'-\id)c$ and $\beta_k := [\Phi c]_k = q(x_k)$,
\begin{equation}\label{eq:betaprime}
\begin{split}
\abs{\beta_k'}^2
&= \bigl|\beta_k + [\Phi\,\Pi'^{\perp} c]_k\bigr|^2 \\
&\;\le\; 2\abs{\beta_k}^2 + 2\,\norm{c}^2 \sum_{j=0}^{D-1}\bigl|[\Phi\,\Pi'^{\perp}]_{kj}\bigr|^2 \\
&\;=\; 2\abs{\beta_k}^2 + \alpha_k ,
\end{split}
\end{equation}
with $\Pi'^\perp := \id-\Pi'$, by Cauchy--Schwarz; this defines $\alpha_k := 2\norm{c}^2\sum_j\abs{[\Phi\Pi'^\perp]_{kj}}^2$. By \ref{F1}--\ref{F3}, $\abs{\beta_k}\le 1$ when $E_k-E_0<\delta$ and $\abs{\beta_k}\le \varepsilon_D$ when $E_k-E_0\ge\delta$; inserting these into \eqref{eq:betaprime} yields \eqref{eq:betabounds}. [This is \cite[Eqs.~(96)--(103)]{Kirby24a} with the phase matrix $\Phi_{kj}=e^{ijE_k\,dt}$ replaced by the Chebyshev--Vandermonde matrix and with the extra factor $\norm{c}^2\le 2B^2$ retained.]

\emph{Collective bound on the $\alpha_k$.}
Exactly as in \cite[Eqs.~(104)--(111)]{Kirby24a},
\begin{equation}
\begin{split}
\sum_k \abs{\gamma_k}^2\alpha_k
&\;=\; 2\norm{c}^2\, \Tr\bigl(\Pi'^\perp\,\Phi^\dagger\Gamma^\dagger\Gamma\Phi\,\Pi'^\perp\bigr) \\
&\;=\; 2\norm{c}^2\, \Tr\bigl(\Pi'^\perp \mS\,\Pi'^\perp\bigr) \\
&\;\le\; 2\norm{c}^2\, D\bigl(\epsilon+\eta_S\bigr)
\;\le\; \zeta ,
\end{split}
\end{equation}
where the middle equality uses $\mS = \Phi^\dagger\Gamma^\dagger\Gamma\Phi$ from \eqref{eq:factorization} (the step at which \cite{Kirby24a} uses $\Psi^\dagger\Psi=\id$, valid verbatim here), and the trace bound is \cite[Eq.~(111)]{Kirby24a}: it relies only on the spectral decomposition of $\mS$, Weyl's inequality, and the Davis--Kahan theorem~\eqref{eq:daviskahan}, applied to $\mS$ vs.\ $\mS'$, and is independent of the Krylov basis.

\emph{Norm lower bounds.}
Following \cite[Eqs.~(91)--(93)]{Kirby24a}: first, since $\norm{\tilde c}\le\norm{c}$ and $\tilde c^\dagger\mS'\tilde c = c^\dagger\mS'' c$,
\begin{equation}
\begin{split}
\frac{\braket{\psi|\psi}}{\norm{c'}^2}
&= \frac{\tilde c^\dagger \mS'\tilde c}{\norm{\tilde c}^2}
\ \ge\ \frac{c^\dagger \mS'' c}{\norm{c}^2} \\
&\ge\ \frac{c^\dagger \mS c - \norm{c}^2\norm{\mS''-\mS}}{\norm{c}^2} \\
&\ge\ \frac{\abs{\gamma_0}^2}{2B^2} - (\epsilon+\eta_S),
\end{split}
\end{equation}
using $\norm{\mS''-\mS}\le\norm{\mS''-\mS'}+\norm{\mS'-\mS}\le \epsilon+\eta_S$ (the eigenvalues of $\mS'$ on the null space of $\mS''$ lie in $[-\eta_S,\epsilon]$ because $\mS\succeq0$, exactly as argued after \cite[Eq.~(91)]{Kirby24a}), $\norm{c}^2\le 2B^2$, and
\begin{equation}\label{eq:overlapfloor}
\begin{split}
c^\dagger \mS c \;=\; \norm{\mV c}^2 &\;=\; \sum_k \abs{\gamma_k}^2\, q(x_k)^2 \\
&\;\ge\; \abs{\gamma_0}^2\,q(x_0)^2 \;=\; \abs{\gamma_0}^2
\end{split}
\end{equation}
by \ref{F1}. Second, without reference to $c'$,
\begin{equation}
\begin{split}
\braket{\psi|\psi} &= \tilde c^\dagger \mS \tilde c \\
&= c^\dagger \mS c + c^\dagger(\mS''-\mS)c - c^\dagger\Pi'(\mS''-\mS)\Pi'c \\
&\ge\ \abs{\gamma_0}^2 - 2\norm{c}^2\norm{\mS''-\mS}
\ \ge\ \abs{\gamma_0'}^2 .
\end{split}
\qedhere
\end{equation}
\end{proof}

Two remarks on the comparison with \cite[Thm.~2]{Kirby24a}, whose exponential suppression factor is $8(1+\pi\delta/\Omega)^{-2d}$ for a real-time space of dimension $2d+1$. First, with the two-sided filter the factor is $\Xi_D^{\mathrm{2s}}\le8(1+\delta/\Omega)^{-2\lfloor(D-1)/2\rfloor}$: the same linear-in-gap exponent per Krylov vector, up to the constant $\pi$, as it must be, since in the angular variable $\theta=\arccos x$ the two spaces resolve the same angular gap (Appendix~\ref{app:RescalingFactor}). With the one-sided filter the factor is $8(1+2\sqrt\gamma)^{-2(D-1)}$ with $\gamma\ge\delta/\Omega$: per unit of Krylov dimension the exponent improves from $\bigO(\delta/\Omega)$ to $\bigO(\sqrt{\delta/\Omega})$, the Chebyshev acceleration \cite{Saad80a}, available only at the edge. Second, $\zeta = 4B^2D(\epsilon+\eta_S)$ versus $2D(\epsilon+\eta_S)$ there, and $\abs{\gamma_0'}^2 = \abs{\gamma_0}^2-4B^2(\epsilon+\eta_S)$ versus $\abs{\gamma_0}^2-2(\epsilon+\eta_S)$: constant-factor losses traceable to $\norm{c}^2\le2B^2$ in place of $\norm{c}\le1$.

\subsection{Energy error bounds}\label{app:main}

The remaining steps of \cite{Kirby24a} use only the conclusions of Theorem~\ref{thm:projector}, Weyl's inequality, and the Davis--Kahan theorem, and are basis-independent. We state them for an arbitrary admissible filter and then specialize; the proofs are those of \cite[Thm.~3, Thm.~4]{Kirby24a} with the indicated substitutions, which we condense for self-containedness.

\begin{theorem}[cf.\ {\cite[Thm.~3]{Kirby24a}}]\label{thm:expectation}
Let $\heff$ be any Hermitian operator, let $\ket{\psi}$ be the state of Theorem~\ref{thm:projector}, and let $\widetilde\Delta := E_1' - E_0$, where $E_1'$ is the second-lowest eigenvalue of $\heff$. Then, for any $0<\delta<\delta'<\norm{H}$ with $\norm{\heff-H} < \delta'-\delta$,
\begin{multline}\label{eq:expectation}
\frac{\bra{\psi}(\heff - E_0)\ket{\psi}}{\braket{\psi|\psi}}
\;\le\;
\delta'\,\mathbbm{1}\bigl(\delta' > \widetilde\Delta\bigr)
+ \norm{\heff-H} \\
+ 6\norm{H}\biggl(
\frac{\norm{\heff-H}}{\delta'-\delta}
+ \frac{\zeta}{\braket{\psi|\psi}}
+ \frac{\Xi_D}{\braket{\psi|\psi}}
\biggr).
\end{multline}
\end{theorem}

\begin{proof}[Proof (condensed; following {\cite[Eqs.~(118)--(128)]{Kirby24a}})]
Decompose $\bra{\psi}(\heff-E_0)\ket{\psi}/\braket{\psi|\psi}$ over the spectral projectors of $\heff$, splitting at $E_0+\delta'$. The low part contributes at most $\delta'\mathbbm{1}(\delta'>\widetilde\Delta)+\norm{\heff-H}$ (Weyl's inequality). For the high part, split $\ket{\psi}=\ket{\psi_{\le E_0+\delta}}+\ket{\psi_{>E_0+\delta}}$ in the eigenbasis of $H$. Theorem~\ref{thm:projector} gives $\norm{\ket{\psi_{>E_0+\delta}}}^2 \le \Xi_D+\zeta$, and the Davis--Kahan theorem~\eqref{eq:daviskahan}, with $Q'$ the spectral projector of $\heff$ onto $[E_0+\delta',\infty)$, gives $\norm{Q'\ket{\psi_{\le E_0+\delta}}}^2 \le \braket{\psi|\psi}\,\bigl(\norm{\heff-H}/(\delta'-\delta)\bigr)^2\le\braket{\psi|\psi}\,\norm{\heff-H}/(\delta'-\delta)$, the last step because $\norm{\heff-H}<\delta'-\delta$ by hypothesis; Young's inequality combines these into a bound on the weight of $\ket{\psi}$ above $E_0+\delta'$ in the spectrum of $\heff$, which multiplies an energy prefactor, the spectral extent of $\heff$ above $E_0$, which is less than $3\norm{H}$ (using $\norm{\heff-H}<\delta'-\delta<\norm{H}$). Collecting constants yields \eqref{eq:expectation}.
\end{proof}

\begin{theorem}[Error bound for an admissible filter; cf.\ {\cite[Thm.~4]{Kirby24a}}]\label{thm:generic}
Instate the setup of Sections~\ref{app:setup}--\ref{app:filter}, let $q$ be an admissible filter with parameters $(\varepsilon_D,B)$, and define
\begin{subequations}\label{eq:constants}
\begin{align}
\chi &:= \eta_H + \eta_S\norm{H}, \\
\abs{\gamma_0'}^2 &:= \abs{\gamma_0}^2 - 4B^2(\epsilon+\eta_S), \\
\Delta' &:= \Delta - \frac{2B^2\chi}{\abs{\gamma_0'}^2}, \\
\zeta &:= 4B^2 D(\epsilon+\eta_S) .
\end{align}
\end{subequations}
Then the ground-state energy estimate $\widetilde E_0$ of the noisy, thresholded Chebyshev--Krylov method satisfies
\begin{multline}\label{eq:generic}
\widetilde{E}_0 - E_0
\;\le\;
\delta'\,\mathbbm{1}\bigl(\delta' > \Delta'\bigr)
+ \frac{2B^2\chi}{\abs{\gamma_0'}^2} \\
+ \frac{6\norm{H}}{\abs{\gamma_0'}^2}
\left(
\frac{2B^2\chi}{\delta'-\delta}
+ \zeta
+ \Xi_D
\right)
\end{multline}
for any parameters $0<\delta<\delta'<\norm{H}$, provided
\emph{(i)} $2B^2\chi/\abs{\gamma_0'}^2 < \delta'-\delta$,
\emph{(ii)} $\epsilon \ge \eta_S$, and
\emph{(iii)} $\abs{\gamma_0'}^2 > 0$.
\end{theorem}

\begin{proof}
Take $\heff$ to be the effective Hamiltonian \eqref{eq:effective} built at the ansatz point $c'$ of Theorem~\ref{thm:projector}. By \eqref{eq:matchpoint} and the variational principle, $\widetilde E_0 - E_0 \le \bra{\psi}(\heff-E_0)\ket{\psi}/\braket{\psi|\psi}$. Apply Theorem~\ref{thm:expectation}; substitute $\norm{\heff-H}\le 2B^2\chi/\abs{\gamma_0'}^2$ (Lemma~\ref{lem:effH}, using (i) to satisfy the hypothesis of Theorem~\ref{thm:expectation}), $\braket{\psi|\psi}\ge\abs{\gamma_0'}^2$ (second bound in \eqref{eq:normbounds}), and $E_1'-E_0 \ge \Delta - \norm{\heff-H} \ge \Delta'$ (Weyl's inequality), which permits replacing $\widetilde\Delta$ by $\Delta'$ in the indicator.
\end{proof}

Inserting the two filters of Section~\ref{app:filter} gives the two results of this appendix. The first is the main one: it holds for \emph{any} rescaling of the Hamiltonian, in particular for the block-encoding normalization used in practice.

\begin{theorem}[Main bound, any placement of the ground state; cf.\ {\cite[Thm.~4]{Kirby24a}}, {\cite[Thm.~1]{Kirby23a}}]\label{thm:general}
Instate the setup of Section~\ref{app:setup} with any a priori bounds $E_{\min}'\le E_0$, $E_{\max}'\ge E_{\max}$ (no assumption on the edge slack $\nu$), and define $\chi$, $\abs{\gamma_0'}^2$, $\Delta'$, $\zeta$ by \eqref{eq:constants} with $B=1$. Then
\begin{multline}\label{eq:general}
  \widetilde E_0-E_0\ \le\
  \delta'\,\mathbbm 1(\delta'>\Delta')
  +\frac{2\chi}{\abs{\gamma_0'}^2} \\
  +\frac{6\norm{H}}{\abs{\gamma_0'}^2}\biggl(
    \frac{2\chi}{\delta'-\delta}+\zeta \\
    +8\Bigl(1+\frac{\delta}{\Omega}\Bigr)^{-2\lfloor(D-1)/2\rfloor}\biggr)
\end{multline}
for any $0<\delta<\delta'<\norm{H}$, provided (i) $2\chi/\abs{\gamma_0'}^2<\delta'-\delta$, (ii) $\epsilon\ge\eta_S$, and (iii) $\abs{\gamma_0'}^2>0$.
\end{theorem}

\begin{proof}
Theorem~\ref{thm:generic} with the two-sided filter of Lemma~\ref{lem:twosided}, $B=1$ and $\Xi_D=\Xi_D^{\mathrm{2s}}$.
\end{proof}

\begin{theorem}[Edge-aligned bound]\label{thm:edge}
Under Assumption~\ref{ass:edge} (more generally, with the constant $B$ of Lemma~\ref{lem:onesided}), and with $\chi$, $\abs{\gamma_0'}^2$, $\Delta'$, $\zeta$ as in \eqref{eq:constants},
\begin{multline}\label{eq:edge}
\widetilde{E}_0 - E_0
\;\le\;
\delta'\,\mathbbm{1}\bigl(\delta' > \Delta'\bigr)
+ \frac{2B^2\chi}{\abs{\gamma_0'}^2} \\
+ \frac{6\norm{H}}{\abs{\gamma_0'}^2}
\biggl(
\frac{2B^2\chi}{\delta'-\delta}
+ \zeta \\
+ 8\Bigl(1+2\sqrt{\delta/\Omega}\Bigr)^{-2(D-1)}
\biggr)
\end{multline}
under the same provisos (i)--(iii).
\end{theorem}

\begin{proof}
Theorem~\ref{thm:generic} with the one-sided filter of Lemma~\ref{lem:onesided} and $\gamma\ge\delta/\Omega$ from \eqref{eq:filter}.
\end{proof}

The two bounds have identical noise structure---linear in $\chi$, with the same threshold term $\zeta$---and differ only in the convergence factor and in the conditions under which the coefficient bound $B$ is controlled. Theorem~\ref{thm:general} is the exact analogue, in the Chebyshev--Krylov space, of the real-time bound \cite[Thm.~4]{Kirby24a}, and it improves the noise scaling of \cite[Thm.~1]{Kirby23a}, which is built on the same two-sided filter, from $\eta^{1/(1+\alpha)}$ to linear; Theorem~\ref{thm:edge} adds the square-root acceleration that edge alignment makes possible. For the corollaries below we write
\begin{equation}\label{eq:rho}
\rho_D :=
\begin{cases}
\bigl(1+\Delta/(2\Omega)\bigr)^{-2\lfloor(D-1)/2\rfloor}, & \text{(Theorem~\ref{thm:general})},\\[3pt]
\bigl(1+\sqrt{2\Delta/\Omega}\bigr)^{-2(D-1)}, & \text{(Theorem~\ref{thm:edge})},
\end{cases}
\end{equation}
for the convergence factor at the window $\delta=\Delta/2$.

\begin{corollary}[Noiseless case]\label{cor:noiseless}
With $\eta_H=\eta_S=0$ and $\epsilon\to0^+$, choosing $\delta = \Delta/2$ and $\delta'\uparrow\Delta$,
\begin{equation}
0 \;\le\; \widetilde{E}_0 - E_0 \;\le\; \frac{48\,\norm{H}}{\abs{\gamma_0}^2}\,\rho_D ,
\end{equation}
the analogue for this method of the Kaniel--Paige--Saad bound for the Lanczos process \cite{Saad80a}; the edge-aligned case carries the characteristic $\sqrt{\Delta/\Omega}$ exponent, the general case the $\Delta/\Omega$ exponent of the classical power-basis analysis.
\end{corollary}

\begin{corollary}[Simplified noisy bound]\label{cor:asymptotic}
With $B=1$ (Theorem~\ref{thm:general}, or Theorem~\ref{thm:edge} under Assumption~\ref{ass:edge}), choose $\epsilon = \Theta(\eta_S)$, $\delta = \Delta/2$, and $\delta' = 3\Delta/4$. Condition (i) of Theorem~\ref{thm:generic} then reads $8\chi < \Delta\abs{\gamma_0'}^2$, and it implies $\delta' \le \Delta'$, so that the indicator term vanishes. Define the dimensionless noise rate $\eta := \max\bigl(\eta_S,\,\eta_H/\norm{H}\bigr)$, so that $\chi \le 2\norm{H}\eta$. Then
\begin{equation}\label{eq:simplified}
\widetilde{E}_0 - E_0
\;\le\;
\bigO\!\left(
\frac{\norm{H}}{\abs{\gamma_0'}^2}
\left[
\Bigl(\frac{\norm{H}}{\Delta} + D\Bigr)\eta
\;+\;
\rho_D
\right]
\right),
\end{equation}
which is linear in the noise rate $\eta$, with the overall energy scale $\norm{H}$ factored out so that the bracket is dimensionless.
\end{corollary}

\begin{proof}
With $\delta'-\delta = \Delta/4$ the $\chi$ terms of \eqref{eq:generic} are $2\chi/\abs{\gamma_0'}^2 + 48\norm{H}\chi/(\Delta\abs{\gamma_0'}^2) = \bigO(\norm{H}^2\eta/(\Delta\abs{\gamma_0'}^2))$ (using $\Delta \le \norm{H}$ to absorb the first), $\zeta = 4D(\epsilon+\eta_S) = \Theta(D\eta_S)$ contributes $\bigO(\norm{H}D\eta/\abs{\gamma_0'}^2)$, and the convergence term $6\norm{H}\,\Xi_D/\abs{\gamma_0'}^2$ at the window $\delta = \Delta/2$ is $\bigO(\norm{H}\rho_D/\abs{\gamma_0'}^2)$ for either filter. The window is fixed at $\delta = \Delta/2$, rather than tied to $\delta'$, so that the convergence factor is exactly $\rho_D$: a window $\delta = c\Delta/2$ with $c<1$ would change the base of the exponential, a discrepancy that no constant in the $\bigO(\cdot)$ absorbs uniformly in $D$.
\end{proof}

Unlike the real-time method, where the entries of $\mH$ and $\mS$ are estimated from two distinct families of estimands ($\langle e^{ik\,\Delta t H}\rangle$ and $\langle H e^{ik\,\Delta t H}\rangle$), here \emph{every} entry of both matrices is a fixed linear functional \eqref{eq:pair}--\eqref{eq:Hmoments} of the single family $\mu_0,\dots,\mu_{2D-1}$. A single machine-level precision parameter therefore fixes both noise levels at once (in particular, the maximum in the definition of $\eta$ is automatically balanced), and it is natural to state the bound directly in terms of that parameter:

\begin{corollary}[Machine-level bound]\label{cor:machine}
Suppose each Chebyshev moment is estimated with accuracy
\begin{equation}
\abs{\widetilde\mu_\ell - \mu_\ell} \;\le\; \varsigma, \qquad \ell = 0,\dots,2D-1,
\end{equation}
where $\varsigma$ is dimensionless (as are the $\mu_\ell \in [-1,1]$); e.g.\ $\varsigma = \bigO(N_s^{-1/2})$ for $N_s$ shots per moment. Let
\begin{equation}
\bar\lambda := \max\bigl(E_{\mathrm{max}}',\,-E_{\mathrm{min}}'\bigr)
= \frac{\Omega+\abs{E_{\mathrm{min}}'+E_{\mathrm{max}}'}}{2} \ge \norm{H},
\end{equation}
which equals the subnormalization $\lambda$ for the symmetric block-encoding window $[-\lambda,\lambda]$. Then, entrywise in \eqref{eq:pair} and \eqref{eq:Hmoments},
\begin{subequations}\label{eq:xipropagation}
\begin{align}
\eta_S &\;\le\; D\,\varsigma, \\
\eta_H &\;\le\; D\,\bar\lambda\,\varsigma, \\
\chi &\;\le\; 2D\,\bar\lambda\,\varsigma ,
\end{align}
\end{subequations}
and choosing $\epsilon = D\varsigma$, $\delta = \Delta/2$, $\delta' = 3\Delta/4$, with $B=1$ as in Corollary~\ref{cor:asymptotic} (admissible under the same condition $8\chi < \Delta\abs{\gamma_0'}^2$, which holds in particular when $16D\bar\lambda\varsigma < \Delta\abs{\gamma_0''}^2$),
\begin{equation}\label{eq:machine}
\frac{\widetilde{E}_0 - E_0}{\bar\lambda}
\;\le\;
\bigO\!\left(
\frac{1}{\abs{\gamma_0''}^2}
\left[
\Bigl(\frac{\bar\lambda}{\Delta} + D\Bigr) D\,\varsigma
\;+\;
\rho_D
\right]
\right),
\end{equation}
where $\abs{\gamma_0''}^2 := \abs{\gamma_0}^2 - 8D\varsigma$.
\end{corollary}

\begin{proof}
Each entry of $\mS$ is the average of two moments, so $\abs{\Delta\mS_{jk}}\le\varsigma$; each entry of $\mH$ is $\tfrac{\Omega}{8}$ times a sum of four moments plus $\tfrac{E_{\mathrm{min}}'+E_{\mathrm{max}}'}{2}\,\mS_{jk}$, so $\abs{\Delta\mH_{jk}} \le \bigl(\tfrac{\Omega}{2} + \tfrac{\abs{E_{\mathrm{min}}'+E_{\mathrm{max}}'}}{2}\bigr)\varsigma = \bar\lambda\varsigma$, where the stated identity for $\bar\lambda$ follows because $\tfrac{E_{\mathrm{min}}'+E_{\mathrm{max}}'}{2} \pm \tfrac{\Omega}{2}$ equal $E_{\mathrm{max}}'$ and $E_{\mathrm{min}}'$. The conversion $\norm{A} \le D\max_{jk}\abs{A_{jk}}$ gives \eqref{eq:xipropagation}, and $\epsilon = D\varsigma \ge \eta_S$ satisfies assumption (ii) of Theorem~\ref{thm:generic} deterministically. Insert into \eqref{eq:general} or \eqref{eq:edge}: the $\chi$ terms contribute $\bigO(\norm{H}\bar\lambda D\varsigma/\Delta)/\abs{\gamma_0''}^2$ (using $\bar\lambda\ge\norm{H}$ and $\Delta\le\norm{H}$ to absorb the bare $\chi$ term), $\zeta = 4D(\epsilon+\eta_S)\le 8D^2\varsigma$ contributes $\bigO(\norm{H}D^2\varsigma)/\abs{\gamma_0''}^2$, and $\abs{\gamma_0'}^2 = \abs{\gamma_0}^2 - 4(\epsilon+\eta_S) \ge \abs{\gamma_0''}^2$. Finally divide both sides by $\bar\lambda$ and drop the factor $\norm{H}/\bar\lambda \le 1$.
\end{proof}

\begin{remark}[Substitution route from Corollary~\ref{cor:asymptotic}]\label{rem:substitution}
Eq.~\eqref{eq:machine} can equivalently be read off from Corollary~\ref{cor:asymptotic} by pure substitution: by \eqref{eq:xipropagation}, $\eta = \max(\eta_S,\eta_H/\norm{H}) \le (\bar\lambda/\norm{H})\,D\varsigma$, and inserting this into \eqref{eq:simplified} and dividing both sides by $\bar\lambda$ gives
\begin{equation}
\frac{\widetilde{E}_0-E_0}{\bar\lambda}
\;\le\;
\bigO\!\left(
\frac{\norm{H}/\bar\lambda}{\abs{\gamma_0''}^2}
\left[
\Bigl(\frac{\norm{H}}{\Delta}+D\Bigr)\frac{\bar\lambda}{\norm{H}}\,D\varsigma
+ \rho_D
\right]
\right),
\end{equation}
which reduces to \eqref{eq:machine} upon dropping $\norm{H}/\bar\lambda\le1$. In this substitution, only the \emph{combination} $\norm{H}\eta \le \bar\lambda D\varsigma$ may be traded for noise in machine units; the remaining explicit factor of $\norm{H}$ in \eqref{eq:simplified} is a genuine energy scale of the problem---it descends from the bound $3\norm{H}$ on the spectral extent of the effective Hamiltonian above $E_0$ in the proof of Theorem~\ref{thm:expectation}---and must not be replaced by a noise quantity. In particular, the Krylov convergence term carries this energy prefactor and no factor of $\varsigma$: it survives the noiseless limit, consistently with Corollary~\ref{cor:noiseless}.
\end{remark}

Both sides of \eqref{eq:machine} are dimensionless: energies are measured in units of $\bar\lambda$, which is precisely the scale at which the qubitization walk represents the Hamiltonian ($h = H/\lambda$ for the symmetric window), and $D\varsigma$ is the entrywise noise floor of the matrix pair itself ($\eta_S \le D\varsigma$ by \eqref{eq:xipropagation}). The entrywise-to-spectral conversion is responsible for that factor of $D$; even so, in the same machine units the result compares favorably with the $(D^4\eta)^{1/(1+\alpha)}$ of \cite[Thm.~1]{Kirby23a}. The parameter $\varsigma$ is the quantity directly connected to device characteristics---shot budgets, readout fidelity---which makes \eqref{eq:machine} the form most convenient for resource estimation. In the general case the convergence term $\rho_D$ is small only once $D\gtrsim\Omega/\Delta$; for the chains of Section~\ref{sec:FoldedQSD} at the block-encoding normalization this is several hundred, so the simulated range $D\le60$ lies in the regime where thresholding, not the bound's noise term, decides the error (Section~\ref{sec:fold-noise}).

\subsection{Discussion}\label{app:discussion}

\paragraph{Comparison of guarantees.}
The analysis of \cite{Epperly22a}, adapted to the Chebyshev space in \cite[Thm.~1]{Kirby23a}, yields an energy error $\bigO\bigl((D^4\eta)^{1/(1+\alpha)}\bigr)$ with $\alpha\in[0,1/2]$, plus a Krylov term $\bigO\bigl(\abs{\gamma_0}^{-2}(1+\delta/2)^{-D}\bigr)$ (for a spectrum normalized to $[-1,1]$). Theorem~\ref{thm:general}, built on the same two-sided filter, improves the noise term to $\bigO(D\norm{H}\eta)$, i.e.\ $\bigO(D\eta)$ at the normalization $\norm{H}\le1$ of \cite{Kirby23a}---linear in $\eta$, matching the real-time result of \cite{Kirby24a}---while keeping the linear-in-gap convergence term; Theorem~\ref{thm:edge} additionally improves, per unit Krylov dimension, the convergence exponent from $\bigO(\delta/\Omega)$ to $\bigO(\sqrt{\delta/\Omega})$ when the ground state is edge-aligned. To reach a target suppression $e^{-L}$ of the excited-state contamination, the real-time analysis \cite{Kirby24a} requires Krylov dimension $\sim L\Omega/(\pi\delta)$, whereas Theorem~\ref{thm:edge} requires $\sim \tfrac{1}{4}L\sqrt{\Omega/\delta}$: a quadratic improvement in $\Omega/\delta$, which is significant precisely in the small-gap regime where these methods are most stressed. We emphasize that relative to \cite{Kirby23a} this is an improvement of \emph{analysis}, not of algorithm: the quantum circuits and classical post-processing are exactly those of \cite{Kirby23a}.

\paragraph{Three separable ingredients; depth versus dimension.}
The comparison between the exponential factors $8\bigl(1+\pi\delta/(2\norm{H})\bigr)^{-2d}$ of \cite[Eq.~(46)]{Kirby24a} and $8\bigl(1+2\sqrt{\delta/\Omega}\bigr)^{-2(D-1)}$ in \eqref{eq:edge} decomposes into three independent effects, which should be distinguished.
\emph{(i) Base.} At matched normalization ($\Omega\approx 2\norm{H}$), the base improves from $\pi\delta/\Omega$ to $2\sqrt{\delta/\Omega}$; since the factor has the form $(1+\beta)^{-(\cdot)}$, the larger base means faster decay whenever $\delta/\Omega < 4/\pi^2\approx0.40$, i.e.\ in every case of practical relevance. This quadratic acceleration is the principal improvement.
\emph{(ii) Exponent per matrix dimension.} The real-time Krylov basis $e^{ij\,\Delta t H}\ket{\psi_0}$, $j=-d,\dots,d$, spends $D_{\mathrm{rt}}=2d+1$ vectors to span trigonometric polynomials of degree $d$ (positive and negative frequencies are both needed), so its filter has degree $d$ and squared-amplitude exponent $2d = D_{\mathrm{rt}}-1$. The Chebyshev--Krylov basis spans algebraic polynomials of degree $D-1$ with $D$ vectors---each dimension contributes a full degree---giving exponent $2(D-1)$: a further factor of $2$ at equal matrix dimension, multiplicative with (i). This factor benefits every cost tied to the matrix dimension: the number of estimands ($2D$ moments), the size of the thresholded generalized eigenvalue problem, and the factor $D$ in $\zeta$.
\emph{(iii) Depth accounting.} Ingredient (ii) is largely cancelled, however, if one fixes the maximum coherent circuit depth rather than the dimension. The real-time pair is Toeplitz, with entries depending on $j-k\in[-2d,2d]$, so its deepest circuit evolves for time $2d\,\Delta t$: depth $\propto 2d$, equal to its exponent. Our deepest circuits estimate moments up to $\mu_{2D-1}$, which the measurement scheme of \cite{Kirby23a} realizes with only $\lfloor\ell/2\rfloor\le D-1$ walk steps for $\mu_\ell$ (unitarity of the walk lets the bra and the ket each absorb half of the polynomial): depth $\propto D-1$, \emph{half} our exponent. Per unit of maximum depth the two accounting factors therefore coincide, and what survives asymptotically is ingredient (i): $2\ln\bigl(1+2\sqrt{\delta/\Omega}\bigr)$ of suppression per walk step versus $\ln\bigl(1+\pi\delta/\Omega\bigr)$ per time step $\Delta t=\pi/\Omega$, a $\Theta(\sqrt{\Omega/\delta})$ advantage in query units.
Finally, the noise-term prefactors ($2B^2$ vs.\ $1$ in the $\chi$ terms, $4B^2D$ vs.\ $2D$ in $\zeta$) are a third, subdominant ingredient, and the apparent factor-$2B^2$ disadvantage is an artifact of the proof of \ref{F5}: replacing the crude step $\int q^2w \le B^2\!\int w$ by the exact weighted mass of the filter gives, for $\nu=0$,
\begin{equation}\label{eq:refinedcoeff}
\norm{c}^2 \;\le\; \frac{2}{\pi}\arccos\!\Bigl(1-\frac{2\delta}{\Omega}\Bigr) + 2\varepsilon_D^2
\;\le\; 2\sqrt{\delta/\Omega} + 2\varepsilon_D^2 \;\ll\; 1,
\end{equation}
and propagating this constant through Theorems~\ref{thm:projector}--\ref{thm:edge} in place of $2B^2$ makes the constants at least as tight as the real-time ones. (The trigonometric filter admits the analogous refinement with $\norm{c}^2\lesssim \delta/\Omega$: the extra edge weight of the Chebyshev measure, responsible for ingredient (i), reappears here with the opposite sign---$\sqrt{\delta/\Omega}$ vs.\ $\delta/\Omega$---but only in these subdominant additive terms.)

\paragraph{The cost of the square root.}
Two requirements. First, the access model: the Chebyshev space requires a block encoding of $H$ and coherent applications of the qubitization walk operator, a more fault-tolerant-oriented primitive than the low-depth Trotterized evolutions available to the real-time method---this trade-off is discussed at length in \cite{Kirby23a,Kirby24a} and is unchanged here. Second, Assumption~\ref{ass:edge}: the rescaling must use a lower bound $E_{\mathrm{min}}'\le E_0$ that is tight to accuracy $\sim\sqrt{\delta\Omega}/D$ (Remark~\ref{rem:slack}); otherwise the filter's supremum $B$ on the spectrally empty edge interval inflates the coefficient bound \ref{F5}, and with it the threshold and noise terms, exponentially in the slack. The real-time/trigonometric filter needs no such assumption because the complex exponentials are unimodular on the whole real line---there is no spectrally empty region on which the filter is uncontrolled. This is the essential structural difference between the two settings, and it is invisible at the level of noiseless span arguments (both spaces equal $\mathcal{K}_D(H,\ket{\psi_0})$ or its unitary analogue): it resides entirely in the conditioning of the representation.

\paragraph{The role of the Chebyshev--Krylov basis.}
Property \ref{F5} fails in the monomial Krylov basis $h^j\ket{\psi_0}$: expanding the very same filter state $q(h)\ket{\psi_0}$ in powers $h^j$ produces coefficients of size exponential in $D$, so the thresholding term $\zeta \propto \norm{c}^2 D(\epsilon+\eta_S)$ would be exponentially amplified. The Chebyshev--Krylov basis is thus not merely implementable via qubitization; it is the basis in which the noise analysis closes with $\bigO(1)$ constants. This is a precise linear-algebraic formulation of the well-known observation that Chebyshev polynomials mitigate the ill-conditioning of the power basis.

\paragraph{Costs.}
All $D^2$ entries of $(\mH,\mS)$ are functions of the $2D$ Chebyshev moments \eqref{eq:moments}; the moment $\mu_\ell$ is estimated from circuits containing $\lfloor\ell/2\rfloor \le D-1$ walk steps (plus one additional block-encoding application for odd $\ell$) \cite{Kirby23a}; the Toeplitz-plus-Hankel structure \eqref{eq:Hmoments} is the exact analogue of the Toeplitz measurement economy of the real-time method. Statistical errors enter Theorem~\ref{thm:edge} only through $\eta_H,\eta_S$, so standard entrywise-to-spectral conversions apply.

\paragraph{Open items.}
(1) A lower bound on the signed error (variationality control), as in \cite[Cor.~1.1]{Kirby24a}, should transfer with the same substitutions; we have not carried this out. (2) The interplay between the threshold choice $\epsilon=\Theta(\eta_S)$ vs.\ $\Theta(\sqrt{\eta_S})$ discussed in \cite{Kirby24a} is unchanged in structure here. (3) Whether an admissible filter (Definition~\ref{def:filter}) can retain a square-root rate for interior $x_0$ while remaining uniformly bounded on $[-1,1]$---a two-interval, Zolotarev-type construction---is open; degree-splitting arguments suggest at most a factor of $2$ is at stake, but we have not pursued the optimal construction. The angular-gap analysis of Appendix~\ref{app:RescalingFactor} suggests the answer is negative for the resolving power of the moments themselves, which would make Theorem~\ref{thm:general} essentially tight in the interior. (4) Extension to excited states, and numerical experiments in the style of \cite[Sec.~5]{Kirby24a} tracking the tightness of \eqref{eq:general} and \eqref{eq:edge}, are natural next steps.

\section{Impacts of the Rescaling Factor}
\label{app:RescalingFactor}

The Hamiltonian rescaling factor---$\Omega$ in the general affine form
\eqref{eq:qsd-rescale}, or $\lambda$ for the symmetric window $h=H/\lambda$---of
Section~\ref{sec:ChebyshevQSD} affects both the Chebyshev-based QSD
(Section~\ref{sec:fold-normalization}) and the subsequent ground-state
preparation stage (Section~\ref{sec:prep-normalization}), for two distinct
reasons. This appendix collects the details: what is and is not invariant
under the rescaling, the angular-gap mechanism behind the energy dependence
and its scaling with system size, its consistency with the published bounds,
the second mechanism behind the preparation cost, and why the operational
constraint cannot be circumvented by a spectral shift.

\subsection{What is, and what is not, invariant}
\label{app:rescale-invariance}

Since $T_k$ has degree exactly $k$, the Chebyshev--Krylov space coincides with
the power Krylov space,
\begin{equation}
\begin{split}
  \operatorname{span}\{T_k(h)\ket{\psi_0}\}_{k<D}
  &=\operatorname{span}\{H^k\ket{\psi_0}\}_{k<D} \\
  &=\mathcal K_D(H,\ket{\psi_0}),
\end{split}
\end{equation}
for \emph{every} affine rescaling $h=(2H-(E_{\min}'+E_{\max}')\id)/\Omega$. In
exact arithmetic the Rayleigh--Ritz values at a given $D$ are therefore
independent of $(E_{\min}',E_{\max}')$, and so is the noiseless energy estimate
obtained without thresholding. Everything else depends on the rescaling: the
conditioning of the Gram matrix $\mS$ in the Chebyshev--Krylov basis, the set of
directions that survive a finite threshold $\epsilon$, the way moment noise of
a given width $\sigma$ propagates into the energy, the coefficient vector $c$
that represents the Ritz state and hence its preparation cost, and the
admissibility of a given filter polynomial for QSVT. The single parameter that
organizes all of these is the position of the ground state in $[-1,1]$, i.e.\
the edge slack $\nu=2(E_0-E_{\min}')/\Omega$ of \eqref{eq:qsd-rescale}. For the
symmetric window, $\Omega=2\lambda$ and
\begin{subequations}\label{eq:app-slack}
\begin{align}
  \nu&=1-\frac{\abs{E_0}}{\lambda}, \\
  x_0&=-1+\nu=-\frac{\abs{E_0}}{\lambda}.
\end{align}
\end{subequations}
The two conventions of Section~\ref{sec:fold-setup} are two points on this
dial: $\lambda_{\mathrm{edge}}=1.000001\abs{E_0}$ gives $\nu=10^{-6}$, whereas
$\lambda_{\mathrm{BE}}=(2L-1)JS(S+1)$ gives $\nu=\Theta(1)$ ($0.39$ and $0.29$
for the two models of Section~\ref{sec:FoldedQSD}). Assumption~\ref{ass:edge}
of the error analysis is the statement $\nu\approx0$; Remark~\ref{rem:slack}
quantifies how much slack is tolerable, $E_0-E_{\min}'\lesssim
\sqrt{\delta\Omega}/D$, a condition that $\lambda_{\mathrm{edge}}$ satisfies at
all $D\le60$ used here and that $\lambda_{\mathrm{BE}}$ violates by two orders
of magnitude.

\subsection{The angular gap and the energy}
\label{app:rescale-angle}

Substituting $x_i=\cos\theta_i$ in $\mu_k=\sum_i\abs{\gamma_i}^2T_k(x_i)$ turns
the Chebyshev moments into trigonometric moments,
\begin{equation}
  \mu_k=\sum_i\abs{\gamma_i}^2\cos(k\theta_i),
  \label{eq:app-trigmoments}
\end{equation}
with $\theta_i=\arccos x_i\in[0,\pi]$. Correspondingly, the
Chebyshev--Krylov space is the even trigonometric span in the operator
$\Theta:=\arccos h$, and the Chebyshev QSD is, in these variables, exactly the
real-time QSD of \cite{Epperly22a,Kirby24a} with ``time step'' one and
``phases'' $\theta_i$. Both the Rayleigh--Ritz convergence rate and the
resolvability of the ground state against noise are therefore governed by gaps
in $\theta$, not in the energy: the trigonometric filter of
\cite[Lemma~3.3]{Epperly22a} suppresses the excited states at the rate
$2(1+a)^{-k}$ with $a$ the \emph{angular} gap, and the phase matrix $\Phi$ of
\eqref{eq:qsd-factorization} is a Vandermonde matrix in $\cos\theta_i$.

\begin{figure*}[t]
\centering
\includegraphics[width=\textwidth]{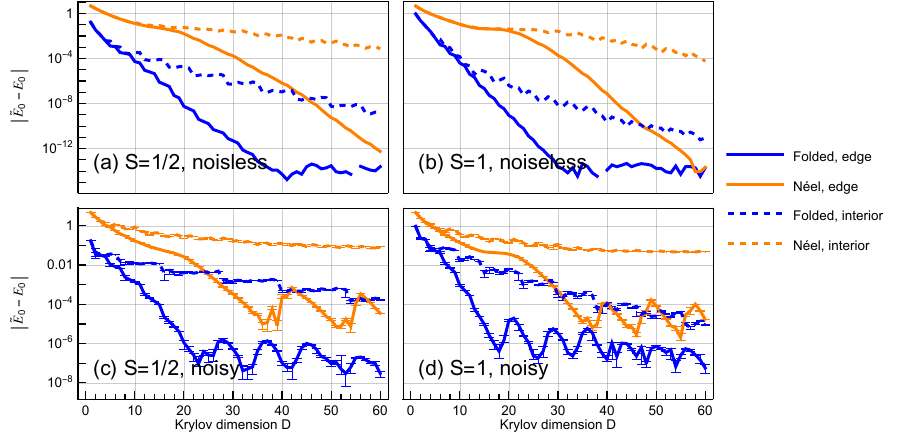}
\caption{Effect of the block-encoding normalization. Energy error versus Krylov
dimension for both reference states under the edge-aligned normalization
$\lambda_{\mathrm{edge}}$ \eqref{eq:fold-lambdaedge} (solid) and under the
physical LCU normalization $\lambda_{\mathrm{BE}}=(2L-1)JS(S+1)$
\eqref{eq:fold-lambdaBE} (dashed; the convention of
Figures~\ref{fig:fold-noiseless} and~\ref{fig:fold-noise}). (a, b) Noiseless.
(c, d) Moment noise $\sigma=10^{-5}$ (medians over 40 realizations, with
error bars marking the interquartile range). (a, c) $S=1/2$, $2L=24$; (b, d)
$S=1$, $2L=12$.}
\label{fig:app-scale}
\end{figure*}

The map from energies to angles has Jacobian $d\theta/dE=-[\lambda\sin\theta]^{-1}$,
which diverges at the edges $\theta\to0,\pi$. Writing $\hat\Delta:=\Delta/\lambda$
and expanding $\theta_0-\theta_1=\arccos x_0-\arccos(x_0+\hat\Delta)$ around $x_0$,
\begin{equation}
  \theta_0-\theta_1\simeq
  \begin{cases}
    \sqrt{2\hat\Delta}\,, & x_0=-1\ \text{(edge)},\\[3pt]
    \hat\Delta/\sin\theta_0\,, & \abs{x_0}<1\ \text{(interior)},
  \end{cases}
  \label{eq:app-thetagap}
\end{equation}
with $\sin\theta_0=\sqrt{1-x_0^2}$, where the edge case uses $1-\cos\varphi\simeq\varphi^2/2$ for
$\varphi=\pi-\theta_1$; this is \eqref{eq:fold-thetagap} of the main text. The angular gap---the quantity that a Krylov step
actually resolves---is thus the \emph{square root} of the normalized energy gap
at the edge but only \emph{linear} in it in the interior. This is the same
mechanism that produces the exponent $(1+2\sqrt{\delta/\Omega})^{-2(D-1)}$ of
Theorem~\ref{thm:edge} through the one-sided Chebyshev filter of
Lemma~\ref{lem:onesided}, and the linear-in-gap exponent
$(1+\delta/\Omega)^{-2\lfloor(D-1)/2\rfloor}$ of Theorem~\ref{thm:general} (and
of \cite[Thm.~1]{Kirby23a}) through the two-sided filter of
Lemma~\ref{lem:twosided} (Section~\ref{app:rescale-bounds}).
For the chains of Section~\ref{sec:FoldedQSD}: with $\Delta=0.153J$ ($S=\half$,
$2L=24$), \eqref{eq:app-thetagap} gives $\theta_0-\theta_1=0.171$ at
$\lambda_{\mathrm{edge}}=10.45J$ and $0.011$ at $\lambda_{\mathrm{BE}}=17.25J$
($x_0=-0.61$); with $\Delta=0.097J$ ($S=1$, $2L=12$), $0.111$ at
$\lambda_{\mathrm{edge}}=15.67J$ and $0.006$ at $\lambda_{\mathrm{BE}}=22J$
($x_0=-0.71$)---losses of $15\times$ and $18\times$ in angular resolution per
Krylov step, matching the flattening of the noiseless slopes in
Figure~\ref{fig:app-scale}.

Noise is affected through the inverse map. A moment uncertainty of width
$\sigma$ translates into an uncertainty $\delta\theta$ of the resolved ground
angle, which converts into an energy error
\begin{equation}
  \delta E=\lambda\,\abs{\cos(\theta_0+\delta\theta)-\cos\theta_0}\simeq
  \begin{cases}
    \tfrac12\lambda\,(\delta\theta)^2, & \text{edge},\\[3pt]
    \lambda\sin\theta_0\,\delta\theta, & \text{interior}.
  \end{cases}
  \label{eq:app-noisejacobian}
\end{equation}
At the edge the first-order term vanishes, so a given angular uncertainty
costs only a second-order energy error; in the interior the full Jacobian
$\lambda\sin\theta_0$ applies. Together with the loss of angular resolution
\eqref{eq:app-thetagap}, this is why the same $\sigma$ costs two to three
orders of magnitude more energy error at $\lambda_{\mathrm{BE}}$ than at
$\lambda_{\mathrm{edge}}$ (Figure~\ref{fig:app-scale}(c, d)), and why
the floors are linear in $\sigma$ when edge-aligned but sublinear in the
interior (Figure~\ref{fig:fold-noise}), where the thresholding bias
$\bigO(\epsilon^{1/(1+\alpha)})$ of \cite{Epperly22a,Kirby23a} dominates.

\subsection{Numerical comparison of the two normalizations}
\label{app:rescale-numerics}

Figure~\ref{fig:app-scale} compares the two conventions directly, for the
same chains, reference states, threshold rule, and noise model as in
Section~\ref{sec:FoldedQSD}. Noiseless (panels a, b): under $\lambda_{\mathrm{BE}}$
the exponential slopes flatten by roughly the ratio \eqref{eq:app-thetagap};
for $S=\half$, $2L=24$ the folded state reaches only $2\times10^{-9}$ at $D=60$
(instead of machine precision by $D\approx37$), and the \Neel{} state does not
converge at all on this depth budget ($7\times10^{-4}$ at $D=60$, versus
$8\times10^{-13}$ when edge-aligned). With noise (panels c, d): the same $\sigma$
costs two to three orders of magnitude more energy error, because the
$\theta\to E$ conversion of the moment uncertainty carries the Jacobian
$\lambda\sin\theta_0$ without the edge enhancement \eqref{eq:app-noisejacobian};
at $\sigma=10^{-5}$ the folded floor rises from $\sim10^{-7}$ to
$\sim2\times10^{-4}$ and the \Neel{} floor from $\sim4\times10^{-5}$ to
$\sim10^{-1}$ ($S=\half$, $2L=24$). Edge-aligned, the floors are linear in
$\sigma$ with an offset equal to the overlap ratio, as predicted by
\eqref{eq:qsd-simplified}; interior, they are the sublinear floors of
Figure~\ref{fig:fold-noise}. Two remarks are in order. First, the folded-state
advantage is robust---indeed amplified---under the physical normalization: at
$\sigma=10^{-5}$ the folded state is a factor $\sim500$ more accurate than the
\Neel{} state for both models, compared with $\sim10^2$ when edge-aligned.
Second, any classical benchmark of Chebyshev-based quantum Krylov methods must
state its $\lambda$ convention explicitly: edge-aligned figures answer the
idealized question, LCU-normalized figures describe the physical algorithm, and
the two differ by orders of magnitude.

\subsection{Scaling with system size}
\label{app:rescale-scaling}

\begin{figure*}[t]
\centering
\includegraphics[width=\textwidth]{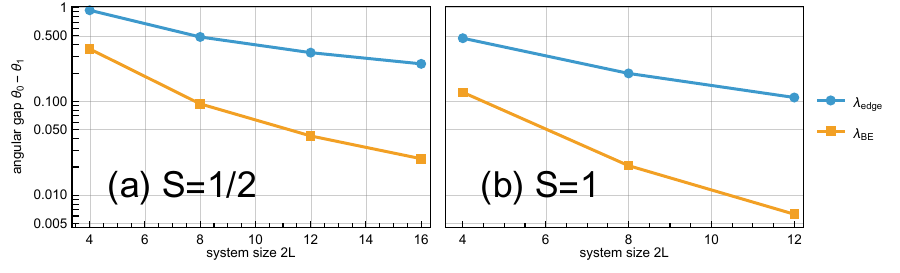}
\caption{Angular gap $\theta_0-\theta_1$ between the two lowest levels versus
system size under the edge-aligned ($\lambda_{\mathrm{edge}}$) and
block-encoding ($\lambda_{\mathrm{BE}}$) normalizations. (a) Gapless $S=1/2$
chain: $\Delta\propto(2L)^{-1}$ while both normalizations grow $\propto2L$, so
$\theta_{\mathrm{edge}}\propto(2L)^{-1}$ but $\theta_{\mathrm{BE}}\propto(2L)^{-2}$,
and the edge-to-interior ratio grows linearly with system size (from
$\approx2.5$ at $2L=4$ to $\approx10$ at $2L=16$). (b) Open $S=1$ chain:
$E_1-E_0$ is the exponentially decaying Kennedy splitting, and the ratio grows
even faster ($\approx4$ at $2L=4$ to $\approx18$ at $2L=12$).}
\label{fig:app-thetagap}
\end{figure*}

Both normalizations are extensive, $\lambda_{\mathrm{edge}}\simeq\abs{E_0}\propto
2L$ and $\lambda_{\mathrm{BE}}=(2L-1)JS(S+1)$, with a size-independent ratio
$\lambda_{\mathrm{BE}}/\lambda_{\mathrm{edge}}\to JS(S+1)/\abs{e_0}$ ($1.65$ for
$S=\half$, $2L=24$, and $1.40$ for $S=1$, $2L=12$), where $e_0$ is the
ground-state energy per site. Combining the two cases of \eqref{eq:app-thetagap}, the ratio of
angular gaps is
\begin{equation}
  \frac{(\theta_0-\theta_1)_{\mathrm{edge}}}{(\theta_0-\theta_1)_{\mathrm{BE}}}
  =\sin\theta_0\,\sqrt{\frac{2\lambda_{\mathrm{BE}}^2}{\Delta\,\lambda_{\mathrm{edge}}}}
  \ \propto\ \sqrt{\frac{2L}{\hat\Delta}},
  \label{eq:app-ratio}
\end{equation}
so the interior penalty \emph{worsens} with system size for any gap that does
not grow. For the gapless $S=\half$ chain, $\Delta\propto(2L)^{-1}$ gives
$\theta_{\mathrm{edge}}\propto(2L)^{-1}$, $\theta_{\mathrm{BE}}\propto(2L)^{-2}$,
and a ratio growing linearly in $2L$: measured $\approx5\times$ at $2L=8$ and
$\approx10\times$ at $2L=16$ (Figure~\ref{fig:app-thetagap}(a)), and
$15\times$ at $2L=24$ by \eqref{eq:app-thetagap} (Section~\ref{app:rescale-angle}). For the open
$S=1$ chain the lowest gap is the exponentially small Kennedy splitting between
the singlet and the edge triplet, and the ratio grows faster still
($\approx4\times$ at $2L=4$, $\approx18\times$ at $2L=12$;
Figure~\ref{fig:app-thetagap}(b)). The effective gaps relevant to the
singlet folded state are larger (Section~\ref{sec:fold-state}) but obey the same
square-root-versus-linear dichotomy. For the folding ladder of
Section~\ref{sec:prep-ladder} the consequence is the one recorded there: with
$\lambda_{\mathrm{BE}}\propto L$ the ground state sits at a fixed interior point,
and the Krylov dimension needed to resolve a gap scales as $\hat\Delta^{-1}$
rather than $\hat\Delta^{-1/2}$.

\subsection{Consistency with the published bounds}
\label{app:rescale-bounds}

Theorems~\ref{thm:general} and~\ref{thm:edge} of Appendix~\ref{app:ChebyshevQSD}
make the dichotomy \eqref{eq:app-thetagap} precise on the side of rigorous
guarantees: identical noise structure, a convergence exponent linear in
$\delta/\Omega$ for any placement of the ground state and proportional to
$\sqrt{\delta/\Omega}$ at the edge. Edge alignment improves the exponent to
its square-root form, and nothing else changes.

None of this contradicts the analyses of \cite{Epperly22a,Kirby23a,Kirby24a};
it sharpens their reading. Theorem~1 of \cite{Kirby23a} assumes only that the
spectrum of $h$ lies in $[-1,1]$, so it holds under either normalization, and
its convergence term $(1+\delta/2)^{-\lfloor D/2\rfloor}$ is linear in the
normalized energy tolerance $\delta$. Its proof \cite[App.~B]{Kirby23a} builds
the spectral filter $f(x)=p^*\bigl((x-x_0)^2/4\bigr)$ from the minimax residual
polynomial $p^*$ of its Lemma~B.2---a rescaled Chebyshev polynomial on $[a,b]$
with $\beta\le2(1+\sqrt{a/b})^{-d}$: squaring the distance to $x_0$ yields a
\emph{two-sided} suppressor, valid wherever $x_0$ sits, at the cost of halving
the degree and shrinking the window to $[\delta^2/4,1]$, whence the linear rate.
This is precisely the interior case of \eqref{eq:app-thetagap} (up to the
$\bigO(1)$ factor $1/\sin\theta_0$), so at $\lambda_{\mathrm{BE}}$ the guarantee
is essentially tight, and the flattened slopes of Figure~\ref{fig:app-scale}
are exactly the behavior the theorem predicts. At the edge, suppression is
needed on one side only, and the same Lemma~B.2 applied at full degree to
$y=(x-x_0)/2\in[\delta/2,1]$ gives $\beta\le2(1+\sqrt{\delta/2})^{-D}$---the
square-root enhancement, which the two-sided construction cannot capture. This
one-sided construction is exactly the filter $q$ of Lemma~\ref{lem:onesided},
whose defining property is that no spectrum lies below $x_0$; the coefficient
bound \eqref{eq:parseval}, which it needs for the noise analysis, is where the
slack $\nu$ enters and where $\lambda_{\mathrm{BE}}$ fails. The two-sided
construction, by contrast, is bounded on all of $[-1,1]$, which is why it
yields the position-agnostic Theorem~\ref{thm:general}.
Equivalently, in the trigonometric picture of \cite{Epperly22a} the convergence
is always linear in the angular gap, and \eqref{eq:app-thetagap} is nothing but
the arccos Jacobian applied to it. The noise analysis of \cite{Kirby24a} is
stated for real-time Krylov pairs, where the angle $\theta=E\,dt$ is linear in
the energy and the spectral range $E_{\max}-E_0$ enters through
$\bigl(1+\pi\delta/(E_{\max}-E_0)\bigr)^{-2d}$ \cite[Eq.~(34)]{Kirby24a};
applied to the Chebyshev pair through the substitution $\Theta=\arccos h$, it
remains valid with the angular gap \eqref{eq:app-thetagap} in the role of
$\pi\delta/(E_{\max}-E_0)$. In short: the interior case saturates the normalization-agnostic
upper bounds (Theorem~\ref{thm:general}), while the edge-aligned case
\emph{beats} them, its faster, square-root convergence being guaranteed
separately by Theorem~\ref{thm:edge} under Assumption~\ref{ass:edge}. An upper
bound can be beaten, not broken.

\subsection{The second mechanism: preparation cost}
\label{app:rescale-preparation}

The preparation cost of Section~\ref{sec:StatePreparation} depends on the
rescaling through an independent mechanism, because it is a property of the
\emph{representation} $c$ of the Ritz state rather than of the state. Three
effects follow from the position of $x_0$.
\emph{(i) Coherence of the coefficients.} At the edge, $T_k(-1)=(-1)^k$, so a
filter peaked at $x_0=-1$ has alternating-sign coherent coefficients with
$\norm{c}_1=\abs{p(-1)}=\norm{p}_\infty$: the saturation case of
Lemma~\ref{lem:prep-norms}, hence $\pQSVT=\pLCU$ exactly and an inert $\ell_1$
gauge fixing. In the interior the coefficients are incoherent and the modest
QSVT advantage and $\ell_1$ slack of Section~\ref{sec:prep-numerics} appear.
\emph{(ii) Protection of the raw pipeline.} With no room between the ground
state and the interval edge, a spurious Ritz value can displace the true one
only by falling below $-1$, a discrete per-realization event; the raw
(unthresholded) pipeline therefore retains a ``good'' mode with $\bigO(1)$
probability at the edge, whereas in the interior it fails uniformly.
\emph{(iii) Admissibility of the certified filter.} QSVT requires
$\abs{p}\le1$ on all of $[-1,1]$. The one-sided shifted-Chebyshev amplifier
$T_D(\ell(x))/T_D(\ell(x_0))$, which achieves the square-root acceleration
$D=\bigO(\hat\Delta^{-1/2}\log\tfrac1\epsilon)$, grows without bound below $x_0$
and is admissible only at the edge; in the interior one must use a two-sided
step approximation at degree $\Theta(\hat\Delta^{-1}\log\tfrac1\epsilon)$
(Section~\ref{sec:prep-ceiling}). This is the state-preparation face of the
same arccos Jacobian: the square-root acceleration is an edge-only luxury on
both fronts.

\subsection{Spectral shifts do not help}
\label{app:rescale-noshift}

Can edge alignment be restored on hardware by shifting the Hamiltonian? Block
encoding $H-s\id$ with the same LCU decomposition raises the subnormalization
to $\lambda_{\mathrm{BE}}+\abs s$ (the identity is one more term in the linear
combination), so the encodable window becomes $[s-\lambda_{\mathrm{BE}}-\abs s,\,
s+\lambda_{\mathrm{BE}}+\abs s]$, whose lower edge $s-\lambda_{\mathrm{BE}}-\abs s
\le-\lambda_{\mathrm{BE}}$ never rises above $-\lambda_{\mathrm{BE}}$. The ground
state can therefore never be brought closer to the lower edge than it is at
$s=0$; a numerical scan over all $s$ confirms that the angular gap
\eqref{eq:app-thetagap} is maximized at $s=0$. Genuine remedies must reduce
the normalization itself toward the spectral radius: tighter LCU
decompositions (e.g.\ grouping anticommuting terms), or uniform spectral
amplification of the block encoding \cite{Low17a}, which trades additional
queries for a block encoding of $H/\lambda'$ with $\lambda'$ closer to
$\norm{H}$. Absent such remedies, the interior, linear-in-gap convergence of
\eqref{eq:app-thetagap} is the physically relevant one for the Chebyshev
method, the edge-aligned convention remains the natural but idealized default
of classical benchmarking, and any reported accuracy must state which
convention it uses---they differ by orders of magnitude.

\section{Affleck--Kennedy--Lieb--Tasaki (AKLT) Model}
\label{app:AKLT}

In this appendix we consider the spin-$1$ AKLT chain and derive in closed form the fidelity between the singlet
ground state of an open chain of $2L$ sites and the tensor product of the
singlet ground states of its two halves, each of length $L$. Working entirely
with the matrix-product-state (MPS) transfer matrix, we obtain the fidelity in
\emph{exact closed form} for every $L>1$,
\begin{equation}
  F(L)=\frac{\big(1+3(-1/3)^{L}\big)^{2}}{4\big(1+3\,(1/9)^{L}\big)},
\end{equation}
which approaches $\tfrac14$ as $L\to\infty$ with a sign-alternating correction
of order $(1/3)^{L}$. We explain the origin of every factor, emphasize that a
frequently quoted two-term expression is only the leading asymptotics of this
result, and note the role of the $SU(2)$ gauge carried by the AKLT matrices in
identifying the singlet with the identity boundary matrix.

\subsection{Matrix-Product Representation}

The AKLT model is the spin-$1$ chain with the Hamiltonian that projects every
neighbouring pair onto total spin $2$~\cite{Affleck87a,Affleck88a}. Its ground
state is the prototypical valence-bond solid and admits an exact MPS
representation with \emph{bond dimension} $2$~\cite{Fannes92a,Schollwock11a}. With
$\sigma^{\pm},\sigma^{z}$ the Pauli operators, the site tensors are
\begin{equation}
  A^{+}=\sqrt{\tfrac{2}{3}}\,\sigma^{+},\qquad
  A^{0}=-\sqrt{\tfrac{1}{3}}\,\sigma^{z},\qquad
  A^{-}=-\sqrt{\tfrac{2}{3}}\,\sigma^{-},
  \label{eq:Amats}
\end{equation}
one tensor for each physical value $m\in\{+,0,-\}$.

An open MPS has two free bond indices. 
We close them with a \emph{boundary matrix} $B\in\mathbb{C}^{2\times 2}$,
\begin{equation}
  \ket{\Psi_{B}}
  =\sum_{m_{1}\cdots m_{L}}
   \tr\!\big[\,B\,A^{m_{1}}A^{m_{2}}\cdots A^{m_{L}}\big]\,
   \ket{m_{1}\cdots m_{L}} .
  \label{eq:PsiB}
\end{equation}
A rank-one $B=\beta\alpha^{\dagger}$ reproduces the usual boundary-vector state
$\sum \alpha^{\dagger}A^{m_{1}}\cdots A^{m_{L}}\beta\,\ket{\bm m}$; a general $B$
spans the four-dimensional space of edge states, matching the four-fold ground
degeneracy of the open AKLT chain. Physically, $B$ is the joint state of the two
unpaired edge spin-$\tfrac12$'s, reshaped into a matrix.

We first note that the singlet corresponds to $B=\Id$.
The AKLT tensors are $SU(2)$ covariant: for the spin-$1$ Wigner matrix
$D^{1}(g)$ and the spin-$\tfrac12$ matrix $u(g)$,
\begin{equation}
  \sum_{m'}D^{1}(g)_{mm'}\,A^{m'} = u(g)^{\dagger}\,A^{m}\,u(g).
  \label{eq:cov}
\end{equation}
Threading \eqref{eq:cov} through \eqref{eq:PsiB} shows that a global rotation
acts purely on the boundary, $U_{g}^{\otimes L}\ket{\Psi_{B}}=\ket{\Psi_{u\,B\,u^{\dagger}}}$,
so $\ket{\Psi_{B}}$ is rotationally invariant ($S=0$) precisely when $B$ commutes
with all $u(g)$, i.e.\ $B\propto\Id$ by Schur's lemma. Hence the singlet ground
state of a chain of $L$ sites is the trace closure,
\begin{equation}
  \ket{s_{L}}=\sum_{m_{1}\cdots m_{L}}
  \tr\!\big[A^{m_{1}}\cdots A^{m_{L}}\big]\,\ket{m_{1}\cdots m_{L}} .
  \label{eq:singlet}
\end{equation}
This is identical to the ground state of the periodic ring of $L$ sites (see
the remark at the end of Section~\ref{app:AKLT-fidelity}). Note that the spin singlet
of the two edge spin-$\tfrac12$'s is the antisymmetric tensor $\epsilon$ in spin
space, yet it corresponds to $B=\Id$ (a $\delta$ in bond space): the AKLT tensors
carry an internal $\epsilon$ gauge, so the bond-space identity encodes the
spin-space singlet. Closing instead with $B\propto\sigma^{x,y,z}$ yields the
three triplet states.


Define the single-layer transfer operator on $2\times2$ matrices,
\begin{equation}
  \Emap(X)=\sum_{m}A^{m}\,X\,(A^{m})^{\dagger},
\end{equation}
and its matrix representation
\begin{equation}
  E=\sum_{m}A^{m}\otimes (A^{m})^{*}\in\mathbb{C}^{2^{2}\times2^{2}},
\end{equation}
related by vectorization, $E\,\vecop(X)=\vecop(\Emap(X))$. For the tensors
\eqref{eq:Amats} one finds the standard spectrum
\begin{subequations}
\begin{align}
  \Emap(\Id)&=\Id, \\
  \Emap(\sigma^{a})&=-\tfrac13\,\sigma^{a}\qquad(a=x,y,z),
\end{align}
\end{subequations}
so $E$ has eigenvalues $\{1,-\tfrac13,-\tfrac13,-\tfrac13\}$. The AKLT tensors
additionally satisfy $(A^{\pm})^{\dagger}=-A^{\mp}$ and $(A^{0})^{\dagger}=A^{0}$,
which makes $\Emap$ self-adjoint.

The singlet norm is a single closed loop of $L$ transfer steps,
\begin{equation}
  \braket{s_{L}|s_{L}}=\Tr E^{L}=1+3\left(-\tfrac13\right)^{L}.
  \label{eq:norm}
\end{equation}


We next compute $\braket{s_{2L}|s_{L}\otimes s_{L}}$, the overlap amplitude of
the full-chain singlet on $2L$ sites with the product of two half-chain singlets,
all defined by \eqref{eq:singlet}.
In the doubled (ket\,$\otimes$\,bra) bond space, each half contributes $L$
transfer steps. The ket consists of two independent traces (one per half), while
the bra is a single trace threading both halves. Closing each half's ket bond
with $B=\Id$ amounts to a partial trace over the ket index of $E^{L}$, leaving a
$2\times2$ operator on the bra bond,
\begin{equation}
  M_{L} := \Tr_{\mathrm{ket}}\!\big[E^{L}\big],
\end{equation}
with components
\begin{equation}
  (M_{L})_{jk}=\sum_{i}\big(E^{L}\big)_{(ij),(ik)} .
\end{equation}
The bra then joins the two halves into one loop, so
\begin{equation}
  \braket{s_{2L}|s_{L}\otimes s_{L}} = \Tr\!\big[M_{L}^{2}\big].
  \label{eq:Otrace}
\end{equation}
Using the eigendecomposition of $E$ and $\tr(\sigma^{a})=0$, only the identity
channel survives the ket-trace, giving
\begin{equation}
  M_{L}=\tfrac12\,\big(\Tr E^{L}\big)\,\Id
       =\tfrac12\big(1+3(-\tfrac13)^{L}\big)\Id,
\end{equation}
and therefore, from \eqref{eq:Otrace},
\begin{equation}
\begin{split}
  \braket{s_{2L}|s_{L}\otimes s_{L}}
  &=\tfrac12\big(1+3(-\tfrac13)^{L}\big)^{2} \\
  &=\tfrac12+3\left(-\tfrac13\right)^{L}+\tfrac92\left(\tfrac19\right)^{L}.
\end{split}
  \label{eq:overlap}
\end{equation}

\subsection{The fidelity}
\label{app:AKLT-fidelity}

The fidelity is the squared overlap of the normalized states. Combining
\eqref{eq:overlap} with the norms \eqref{eq:norm}, and noting that the
denominator carries one full-chain norm and the square of the half-chain norm,
\begin{equation}
\begin{split}
  F(L)
  &=\frac{\abs{\braket{s_{2L}|s_{L}\otimes s_{L}}}^{2}}{\braket{s_{2L}|s_{2L}}\;
  \braket{s_{L}|s_{L}}^{2}} \\
  &=\frac{\tfrac14\big(1+3(-\tfrac13)^{L}\big)^{4}}
            {\big(1+3(-\tfrac13)^{L}\big)^{2}\big(1+3(\tfrac13)^{2L}\big)} ,
\end{split}
\end{equation}
the square of the half-chain norm cancels, leaving the exact result
\begin{equation}
  F(L)=\frac{\big(1+3\left(-\tfrac13\right)^{L}\big)^{2}}
            {4\big(1+3\left(\tfrac19\right)^{L}\big)}
      =\frac{\big(3^{L}+3(-1)^{L}\big)^{2}}{4\,(3^{2L}+3)} .
  \label{eq:exact}
\end{equation}
(The overlap amplitude itself, $\sqrt{F}=\bigl(1+3(-\tfrac13)^L\bigr)/
\bigl(2\sqrt{1+3(\tfrac19)^L}\bigr)$, tends to $\tfrac12$.)
This holds for all integer $L>1$. (At $L=1$ the half chain has no bond and
\eqref{eq:singlet} gives a zero-norm state, so $F$ is undefined; this is a
single-site artifact, not a parity effect---the state \eqref{eq:singlet} is a
genuine $S=0$ singlet for every $L\ge 2$, of either parity.)

Expanding \eqref{eq:exact},
\begin{equation}
  F(L)=\tfrac14+\tfrac32\left(-\tfrac13\right)^{L}
      +\tfrac32\left(\tfrac19\right)^{L}-\tfrac92\left(-\tfrac1{27}\right)^{L}+\cdots ,
  \label{eq:expand}
\end{equation}
so $F\to\tfrac14$ with a sign-alternating leading correction
$\tfrac32(-\tfrac13)^{L}$ governed by the AKLT correlation length
$\xi=1/\ln 3\approx 0.91$. Even $L$ approaches from above, odd $L$ from below.
We stress that the two-term truncation $\tfrac14+\tfrac32(-\tfrac13)^{L}$---or
the squared two-term amplitude $\bigl(\tfrac12+\tfrac32(-\tfrac13)^{L}\bigr)^2$
sometimes quoted---is \emph{only} the leading asymptotics: it reproduces the
limit and the first correction but not the $(\tfrac19)^{L}$ term, and is not
exact at any finite $L$. The exact object is \eqref{eq:exact}.

The limit is the square of a recoupling amplitude. The product
$\ket{s_{L}}\otimes\ket{s_{L}}$ pairs the four edge spin-$\tfrac12$'s within
each half, whereas $\ket{s_{2L}}$ pairs the cut bond internally and the two
outer edges together. The overlap amplitude of these two distinct singlet
pairings of four spin-$\tfrac12$'s is $\tfrac12$, hence $F_\infty=\tfrac14$.
Equivalently, gluing two rings into one requires re-pairing at \emph{two} cut
points, each contributing a factor $1/\sqrt2$ to the amplitude, hence
$(1/\sqrt2)^{2}=\tfrac12$ in amplitude and $\tfrac14$ in fidelity. An
amplitude $1/\sqrt2$ (fidelity $\tfrac12$) would arise only with a single cut,
e.g.\ if the outer edges were left free rather than projected onto the global
singlet.

\paragraph{Periodic boundary conditions.}
Everything above was phrased for open chains, but it covers the closed chain
verbatim. On a ring of $L$ sites the AKLT Hamiltonian has a \emph{unique}
ground state, the translation-invariant valence-bond solid, whose MPS
representation is the periodic contraction of the same site tensors
\eqref{eq:Amats}, i.e.\ precisely the trace closure \eqref{eq:singlet}. The
singlet ground state of the open chain and the ground state of the ring are
therefore the \emph{same vector} $\ket{s_L}$ for every $L\ge2$, and the
periodic fold fidelity
\begin{equation}
  F^{\mathrm{PBC}}(L)
  =\frac{
    \abs{\braket{g^{\mathrm{PBC}}_{2L}|g^{\mathrm{PBC}}_{L}\otimes g^{\mathrm{PBC}}_{L}}}^{2}
  }{
    \norm{g^{\mathrm{PBC}}_{2L}}^{2}\,\norm{g^{\mathrm{PBC}}_{L}}^{4}
  } = F(L)
\end{equation}
coincides identically---not merely asymptotically---with the singlet--singlet
fidelity \eqref{eq:exact} of the open chain, including the norms
\eqref{eq:norm}, which are the ring partition functions $\Tr E^{L}$. 
This is the coincidence illustrated in Figure~\ref{fig:fid-aklt}. The two
geometries differ only in how the state is selected and in the energy cost of
the fold: on the open chain $\ket{s_L}$ must be isolated from the four-fold
degenerate edge manifold (sector-resolved folding,
Section~\ref{sec:fid-aklt}), whereas on the ring it is the unique ground state
and no resolution is needed; and the folded state of the ring differs from
its target through \emph{two} cut bonds rather than one, which is why the
interface energy is $20/9$ instead of $10/9$ while the fidelity---set by the
recoupling of valence bonds at two points in either geometry---is the same.
For a generic gapped chain such as the spin-$1$ Heisenberg model the two
fidelities do not coincide at finite $L$ (Figure~\ref{fig:fid-haldane}); the
exact identity here is a consequence of the AKLT ground state being a pure
valence-bond solid, whose edge modes are exactly decoupled from the bulk.

\subsection{Numerical illustration}

Figure~\ref{fig:fid-aklt} shows $F(L)$ and the deviation
$\lvert F-\tfrac14\rvert$, the latter falling on $\tfrac32(1/3)^{L}$ over more
than ten orders of magnitude. Such overlaps and their finite-size scaling are the basic
objects of the ground-state-fidelity approach to quantum matter
\cite{Zanardi06a,Cozzini07a}.

\bibliography{References}

\end{document}